\documentclass{aa}  

\usepackage{graphicx}
\usepackage{txfonts}
\usepackage{subcaption}
\usepackage{booktabs}
\usepackage{xspace}
\usepackage{xcolor}
\usepackage[hidelinks]{hyperref}

\newcommand{\numax}{\ensuremath{\nu_{\rm max}}\xspace}
\newcommand{\dnu}{\ensuremath{\Delta\nu}\xspace}
\newcommand{\dpi}{\ensuremath{\Delta\Pi_1}\xspace}
\newcommand{\q}{\ensuremath{q}\xspace}

\begin{document}

\title{Period spacings and global seismic parameters for K2 red giants using deep learning}

\titlerunning{Period spacings for K2 red giants using deep learning}
\authorrunning{N. Ghanghas et al.}


\author{Nipun Ghanghas\inst{1}\corrauth{nipun@tifr.res.in}
    \and Siddharth Dhanpal\inst{1}\email{}
    \and Shravan Hanasoge\inst{1,2}\email{}
    \and Praneeth Netrapalli\inst{3}\email{}
    \and Karthikeyan Shanmugam \inst{3}\email{}
    }
\institute{Department of Astronomy and Astrophysics, Tata Institute of Fundamental Research, Mumbai, 400005, India 
\and Center for Space Science, NYUAD Institute, New York University Abu Dhabi, PO Box 129188, Abu Dhabi, UAE 
\and Google Research India, Bengaluru, 560016, India}

\abstract
{Gravity-mode period spacings ($\Delta\Pi_1$) of red giants probe the stellar core directly, constraining its structure, mass and evolutionary state. Their measurement requires resolving narrow, densely spaced mixed modes and has so far relied on the four-year baseline of Kepler. Recovering $\Delta\Pi_1$ from the much shorter ($\sim$80-day) baselines typical of K2 remains largely unexplored at the ensemble scale.}
{We develop an automated machine-learning technique to measure global asteroseismic parameters and gravity-mode period spacings for red giants from single-campaign K2 photometry.}
{Two deep residual neural networks take the full-resolution power spectrum of an $\sim$80-day light curve as input, without background fitting or mode identification, and return a probability distribution for each parameter, yielding a point estimate and asymmetric uncertainties. They are trained on $\sim$8~million synthetic spectra and evaluated on held-out synthetics, on Kepler data degraded to K2-like resolution, and on K2 observations.}
{On Kepler data at K2-like resolution, $\nu_{\rm max}$ and $\Delta\nu$ are recovered for 96\% and 91\% of stars with robust dispersions of 3.6\% and 1.2\%; $\Delta\Pi_1$ for 22\%, with a dispersion of 0.9\%. Applied to 18,704 K2 red giants, the inferred $\nu_{\rm max}$ and $\Delta\nu$ agree with catalogue values to 5.8\% and 1.9\%, the additional scatter arising from known K2 artefacts. Among the 2,059 young red giants we obtain $\Delta\Pi_1$ for 232 stars with a median fractional uncertainty of 1.3\%, following the same $\Delta\nu$--$\Delta\Pi_1$ sequence as the Kepler sample although the training set carries no imprint of that relation.}
{We show that $\nu_{\rm max}$, $\Delta\nu$ and -- for a subset of young red giants -- $\Delta\Pi_1$ can be recovered from a single $\sim$80-day K2 campaign within an automated, probabilistic machine-learning framework. This approach can also be extended to short-baseline samples from TESS and, in future, Roman and PLATO.}


\maketitle
\nolinenumbers

\section{Introduction}
Asteroseismology has transformed the study of stellar structure and evolution by providing direct access to stellar interiors \citep{Aerts_JCD_book, Garcia_2019_review}. In solar-like oscillators, where modes are stochastically excited by near-surface convection, the observed power excess and the regular frequency pattern within it yield the global seismic parameters \numax\ and \dnu, from which masses and radii follow through scaling relations. Red giants occupy a special place in this picture because their oscillation spectra contain mixed modes, which behave as pressure modes in the envelope and as gravity modes in the radiative core \citep{Beck2011, Bedding_2011Natur.471..608B, Mosser_2011A&A...532A..86M}. These modes carry information that no envelope diagnostic can provide, and their asymptotic period spacing \dpi\ is tied directly to the buoyancy cavity: it separates hydrogen-shell-burning stars from core-helium-burning ones far more cleanly than any classical indicator \citep{Bedding_2011Natur.471..608B}, constrains the core mass and the extent of mixing beyond the convective core during the main sequence \citep{Montalban_2013ApJ...766..118M}, and, together with the coupling factor $q$, probes the evanescent region separating the two cavities \citep{Mosser_2017_coupling}.

Extracting \dpi is nonetheless demanding. Mixed modes are narrow, closely spaced in frequency, and further complicated by rotational splitting, so their ensemble-scale measurement has traditionally required the long, uninterrupted four-year baseline of the nominal Kepler mission \citep{Kepler_mission}. That dataset has supported a succession of increasingly automated analyses: \citet{Mosser_2012_evolution} first measured period spacings for several hundred giants by fitting the asymptotic mixed-mode pattern; \citet{Mosser_and_Belkacem_I_2015} showed that the pattern becomes regular in a stretched period variable, which \citet{vrard_et_al} exploited to measure \dpi automatically for some 6,100 Kepler red giants; \citet{Mosser_2017_coupling} extended the same framework to the coupling factor, and \citet{Gehan_2018A&A...616A..24G} to core rotation rates. More recent work has emphasised the joint treatment of the full mixed-mode parameter set: \citet{kuszlewicz2023MixedmodeEnsembleAsteroseismologya} applied Bayesian optimisation to obtain \dpi, $q$, the g-mode phase offset and the rotational splitting for 1,074 low-luminosity giants, while \citet{Gang_Li_2024} fitted asymptotic expressions accounting for near-degeneracy effects to 2,495 stars, doubling the number of measured core rotation rates. These catalogues define the current state of the art, and provide the reference against which any shorter-baseline method must be tested.

They are also confined to a single 100~square-degree field. The surveys that have followed Kepler trade baseline for sky coverage: K2 observed a succession of fields along the ecliptic in campaigns of $\sim$80~days \citep{k2_mission}, TESS surveys almost the entire sky in $\sim$27-day sectors \citep{Tess_mission_ricker_2015}, and PLATO \citep{plato_mission_2024} and the Roman Galactic Bulge Time-Domain Survey \citep{Spergel_2015_Roman, Huber_2023_Roman} will extend this approach to long-duration fields and to the Galactic bulge respectively. The scientific return of this trade is substantial. The K2 Galactic Archaeology Program has delivered seismic parameters for $\sim$19,000 red giants spanning a far wider range of Galactic environment and metallicity than the Kepler field affords \citep{Stello2017, k2_dr3}, and TESS has extended detections to $\sim$158,000 giants across the sky \citep{quick_look_Hon_el_al}. The core diagnostics have largely not followed. The seismic content recovered from short baselines has been almost exclusively \numax, which requires only that the power excess be localised, and in favourable cases \dnu\ \citep{quick_look_Hon_el_al}; measuring \dpi\ at these resolutions has remained out of reach for ensembles. \citet{davies2016AsteroseismologyRedGiantsa} demonstrated that \dpi\ can in principle be recovered from as little as $\sim$70~days of data for a limited set of stars, but no automated method delivering validated period spacings for a large short-baseline sample has been available.

The difficulty is one of resolution rather than of signal-to-noise alone. The spacing between adjacent mixed modes scales as $\Delta\Pi_1 \nu^2$, which reduces sharply as a star evolves and \numax\ decreases; and for an evolved giant it falls below the Rayleigh resolution of an $\sim$80-day time series, so that the g-mode pattern is smeared into an unresolved forest. The problem is compounded as stars ascend the giant branch: mixed-mode inertia rises and suppresses the visibility of dipole modes \citep{Grosjean_2014A&A...572A..11G}. The regime in which \dpi\ remains accessible from a single campaign is therefore the less-evolved end of the red-giant branch, and we find that our networks return confident period spacings only for stars with $\Delta\nu \gtrsim 9~\mu$Hz, which we refer to throughout as ``young'' red giants.

Machine learning has proved well suited to seismic inference at scale, and has already been applied to the short-baseline regime where classical analyses become difficult. Convolutional networks operating on power spectra have been used to classify evolutionary state \citep{Hon2017}, to detect oscillations and estimate \numax\ with expert-level reliability \citep{Hon_2018}, and to compile the all-sky TESS red-giant catalogue \citep{quick_look_Hon_el_al}. Closer to the present work, \citet{dhanpal2022, dhanpal2023} showed that a neural network can infer \numax, \dnu, \dpi\ and $q$ directly from a Kepler power spectrum, without background fitting or mode identification, by recasting the regression as a classification over discretised parameter bins and thereby returning a full predictive distribution for each parameter. That formulation is useful here because the measurement of \dpi\ at coarse resolution is frequently ambiguous, admitting several candidate spacings, and a distributional output represents such ambiguity explicitly rather than collapsing it into a single value with an inadequate error bar.

These networks were nonetheless trained and applied on fully resolved four-year spectra. Whether the same approach survives a nearly twenty-fold degradation in frequency resolution, where the mixed-mode pattern is only partially resolved and the information content of the spectrum is qualitatively different, is not something that can be settled by extrapolation. That is the question addressed here. We train two networks on synthetic spectra generated at K2-like resolution -- one for the pressure-mode parameters across the full red-giant range, one for the gravity-mode parameters in the young red-giant regime -- and validate them in three stages of increasing realism: on held-out synthetics, on Kepler light curves degraded to K2-like baselines and compared against four-year reference values, and finally on K2 observations themselves. We then apply the method to the K2 red-giant sample and report period spacings for those stars where the measurement is confident.

This paper is organised as follows. Sect.~\ref{sec:data} describes the K2 and Kepler observations and their processing; Sect.~\ref{sec:methods} presents the construction of the synthetic training sets, the network architecture, and the training procedure; Sect.~\ref{sec:results} reports the validation and the application to the K2 red giants; and Sect.~\ref{sec:summary} summarises our conclusions.

\section{Data}
\label{sec:data}
\subsection{K2 Observations}
\label{sec:data_k2}
We use detrended light curves from the \texttt{K2SFF} pipeline \citep{Vanderburg_2014}, which corrects for flux trends correlated with spacecraft pointing drift and other systematics. The light curves are retrieved from the Mikulski Archive for Space Telescopes (MAST) using the \texttt{lightkurve} package \citep{Lightkurve_2018}, taking the long-cadence ($\sim$30~min) data for the red giants listed in K2~GAP~DR3 \citep{k2_dr3}. We high-pass filter each lightcurve with a 4-day boxcar filter to suppress low frequency trends, and apply 5$\sigma$ clipping to reject outlier flux measurements. Following \cite{stello2015OscillatingRedGiants}, we fill gaps shorter than 1.5 hours using linear interpolation to reduce the impact of window function \citep{garcia2014ImpactGapsKepler}. We then compute Lomb--Scargle periodograms \citep{Lomb_1976,Scargle_1982,Press_Rybicki_1989} on a fixed frequency grid corresponding to an $\sim$88-day baseline. Fixing the power spectral density (PSD) length for every star is required so that the spectra can be processed by the convolutional layers of our network (Sect.~\ref{sec:model_architecture}). In total our sample comprises 18,704 K2 red giants with $\Delta\nu$ in the range $1$--$19~\mu$Hz, the interval over which our models are designed to operate. Of these, 2,059 stars fall in the range $\Delta\nu = 9$--$19~\mu$Hz, which we refer to as ``young red giants'' throughout this work.

\subsection{Kepler data at K2-like resolution}
\label{sec:data_ekic}

To benchmark the models against the literature we construct a set of Kepler light curves degraded to K2-like baselines, allowing our inferences from short segments to be compared directly with values derived by classical techniques from the full four-year time series. We start from the 21,144 red giants identified by \citet{dhanpal2022} and retrieve the corresponding long-cadence PDCSAP light curves via \texttt{lightkurve} \citep{Lightkurve_2018}, stitched across all available quarters. From each light curve we extract a single contiguous segment, with a length drawn at random from the empirical distribution of K2 campaign durations spanning 67--88 days, and a randomly chosen starting quarter. Because mode excitation is stochastic and the granulation background and photon noise differ from one epoch to another, such a segment is an independent realisation of the oscillation signal rather than a coarser view of the full-length spectrum, and is therefore representative of what a genuine single-campaign observation would provide. Segments are rejected if they contain a single contiguous gap exceeding 40\% of the nominal length, if the actual time span falls below 60\% of that length, or if fewer than half of the expected cadences survive.

The retained segments are processed exactly as the K2 light curves in Sect.~\ref{sec:data_k2}: a 4-day boxcar high-pass filter, 5$\sigma$ clipping of outlying flux measurements, and linear interpolation across gaps of up to three consecutive cadences ($\sim$1.5~h). Periodograms are computed on the same fixed frequency grid as the K2 spectra, 2,088 bins spanning 3.02--277.77~$\mu$Hz at a spacing of 0.132~$\mu$Hz set by the $\sim$88-day Rayleigh resolution, so that the two datasets are directly interchangeable as network inputs.

\section{Methods}\label{sec:methods}
We train two networks with identical architecture: a \texttt{p-mode model}, which infers the pressure-mode--dominated global parameters \numax\ and \dnu, and a \texttt{g-mode model}, which infers the gravity-mode period spacing \dpi\ and the coupling factor $q$ for young red giants (Sect. \ref{sec:training}). The input to each model is a 1-D array of power from the full PSD, normalised by its maximum value, with no background correction or hand-crafted feature extraction; the networks operate on the complete spectrum and learn the relevant features directly. The construction of the synthetic training data is detailed in Sect. \ref{sec:training_set}, and the network architecture in Sect. \ref{sec:model_architecture}.

\subsection{Training set}
\label{sec:training_set}
We generate realistic sets of synthetic spectra based on asymptotic theory \citep{Aerts_JCD_book, Garcia_2019_review} to train the model. This approach is necessary because observed $\Delta\Pi_{1}$ measurements for red giants are scarce, which makes training directly on observed (Kepler) data infeasible. We use the simulator framework developed by \cite{othman_benomar_2023_spectra_simulator} and implement changes needed in calculations of mode frequencies, heights, and widths to account for unresolved modes at K2-like resolution. The full theoretical framework and detailed methodology for generating these synthetic datasets are described in Appendix \ref{appendix:generating_synthetics}.

We construct two separate datasets to train two distinct models (see Section \ref{sec:training} for details), each comprising approximately 10 million synthetic red giant samples. Of these, 82.5\% are used for training, 15\% for validation, and 2.5\% for testing. The two datasets share identical ranges in all parameters, except for $\Delta\nu$ and $\Delta\Pi_{1}$. The first dataset spans a $\Delta\nu$ range of $1$--$19~\mu\mathrm{Hz}$ and $\Delta\Pi_{1}$ from $40$ to $500$ seconds. The second consists entirely of young red giants, restricted to $\Delta\nu$ between $9$--$19~\mu\mathrm{Hz}$ and $\Delta\Pi_{1}$ between $40$--$150$ seconds.

\begin{figure*}
    \centering
    \includegraphics[width=0.98\textwidth]{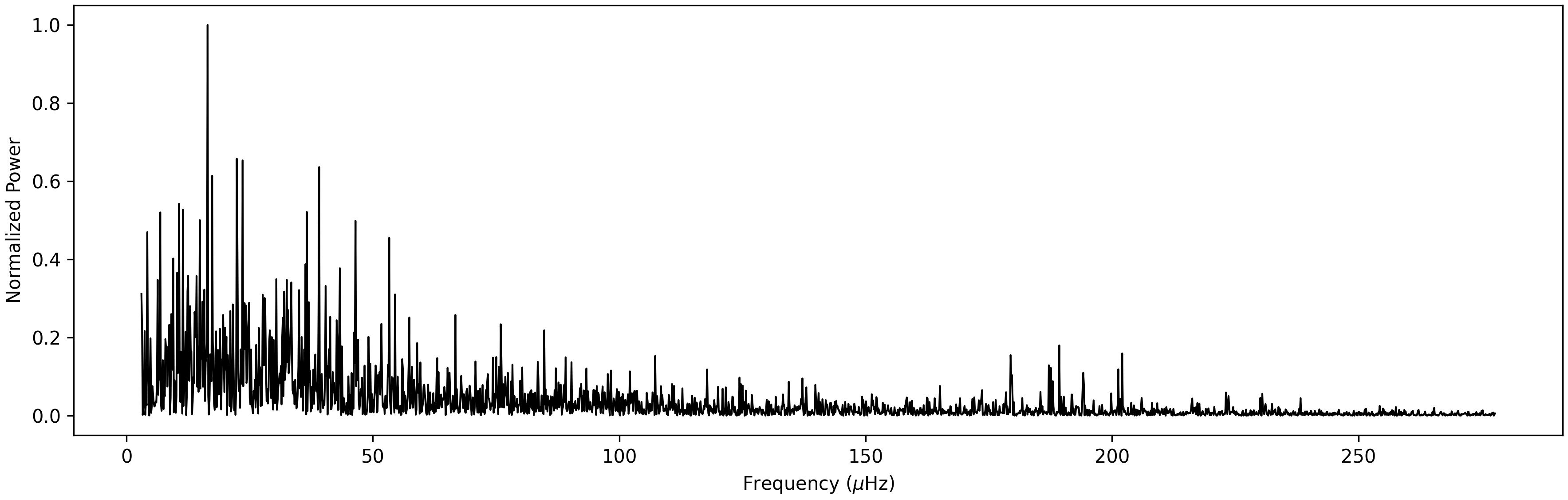}
    \caption{Example of a synthetic PSD at K2-like resolution, with $\nu_{\mathrm{max}} = 190.34~\mu$Hz, $\Delta\nu = 14.52~\mu$Hz, $\Delta\Pi_{1} = 89.55$ s, and $q = 0.15$.}
    \label{fig:k2_synthetics_example}
\end{figure*}

Each parameter is sampled uniformly over its range, except the inclination angle $\iota$, which is drawn from an isotropic distribution, $P(\iota) \propto \sin \iota$. Crucially, we sample $\Delta\Pi_{1}$ and $q$ independently of $\Delta\nu$. Although this produces some combinations not realised in nature, it is a deliberate choice: because the training set contains no relationship between \dpi and the other parameters, the network cannot recover \dpi as a by-product of a learned correlation and must infer it directly from the spectral morphology. Table \ref{tab:parameter_ranges} lists the full parameter ranges used to generate the synthetic datasets, and Figure \ref{fig:k2_synthetics_example} shows an example synthetic PSD.

\begin{table*}
\centering
\caption{Ranges of seismic parameters used to create the synthetics dataset}
\hspace*{-1.2cm}
\resizebox{0.55\textwidth}{!}{%
\begin{tabular}{l c c c}
\hline
\hline
\textbf{Parameter} & \textbf{Dataset 1} & \textbf{Dataset 2} & \textbf{Distribution} \\ \hline
$\Delta\nu$ ($\mu$Hz) & 1--19 & 9--19 &Uniform \\
$\nu_{\rm{max}}/\nu_{\rm{max,scaled}}$ & 0.9--1.1 & Same &Uniform\\
$\Delta\Pi_{1}$ (s) & 40--150 (if $\Delta\nu>$ 9)& 40--150 &Uniform\\
                    & 40--500 (if $\Delta\nu<$ 9)&        &Uniform\\
$q$ & 0--0.5 & Same &Uniform \\
$\epsilon_{p}$ & 0--1  & Same &Uniform \\
$\epsilon_{g}$ & 0--1  & Same &Uniform \\
$\beta_p ~(\equiv \alpha_l~n_{\rm{max}}) $ & 0.0--0.1  & Same &Uniform \\
$d_{01}$ & ($-0.6$ to $0.6$)/$\Delta\nu$ & Same &Uniform \\
$d_{02}/d_{02,\rm{scaled}}$ & (0.8 to 1.2)/$\Delta\nu$ & Same &Uniform \\
$d_{03}/d_{03,\rm{scaled}}$ & (0.8 to 1.2)/$\Delta\nu$ & Same &Uniform \\
Core rotation ($\mu$Hz) & 0.005--2.8  & Same &Uniform \\
Envelope rotation ($\mu$Hz) & 0.005--0.4  & Same &Uniform \\
Inclination $\iota$ (deg) & 0--90  & Same &Isotropic \\
$A_g$ & 0.8--1.2  & Same &Uniform \\
$B_g$ & $-2.2$ to $-1.8$ & Same &Uniform \\
$C_g$ & 0--1.0 & Same &Uniform \\
$A_\tau$ & 0.8--1.2 & Same &Uniform \\
$B_\tau$ & $-1.0$ to $-0.9$ & Same &Uniform \\
$C_\tau$ & 0--1.0 & Same &Uniform \\
$p$ & 1.8--2.4 & Same &Uniform \\
$N_0$ & 0.1--29500 & Same &Log-Uniform \\
Noise Realisations & 1--3 & Same &Random \\
Frequency range \\for ML training ($\mu$Hz) & 3.02--277.77 & Same &2088 bins\\
Observation time (days) & 88.0  & Same &Fixed value \\
SNR & 10--20 & Same &Uniform \\ \hline
\end{tabular}%
}
\label{tab:parameter_ranges}
\end{table*}

\subsection{Model architecture}
\label{sec:model_architecture}
\begin{figure*}
    \centering
    \subfloat[\label{fig:wide_resnet_layers}]{
        \includegraphics[width=0.84\textwidth]{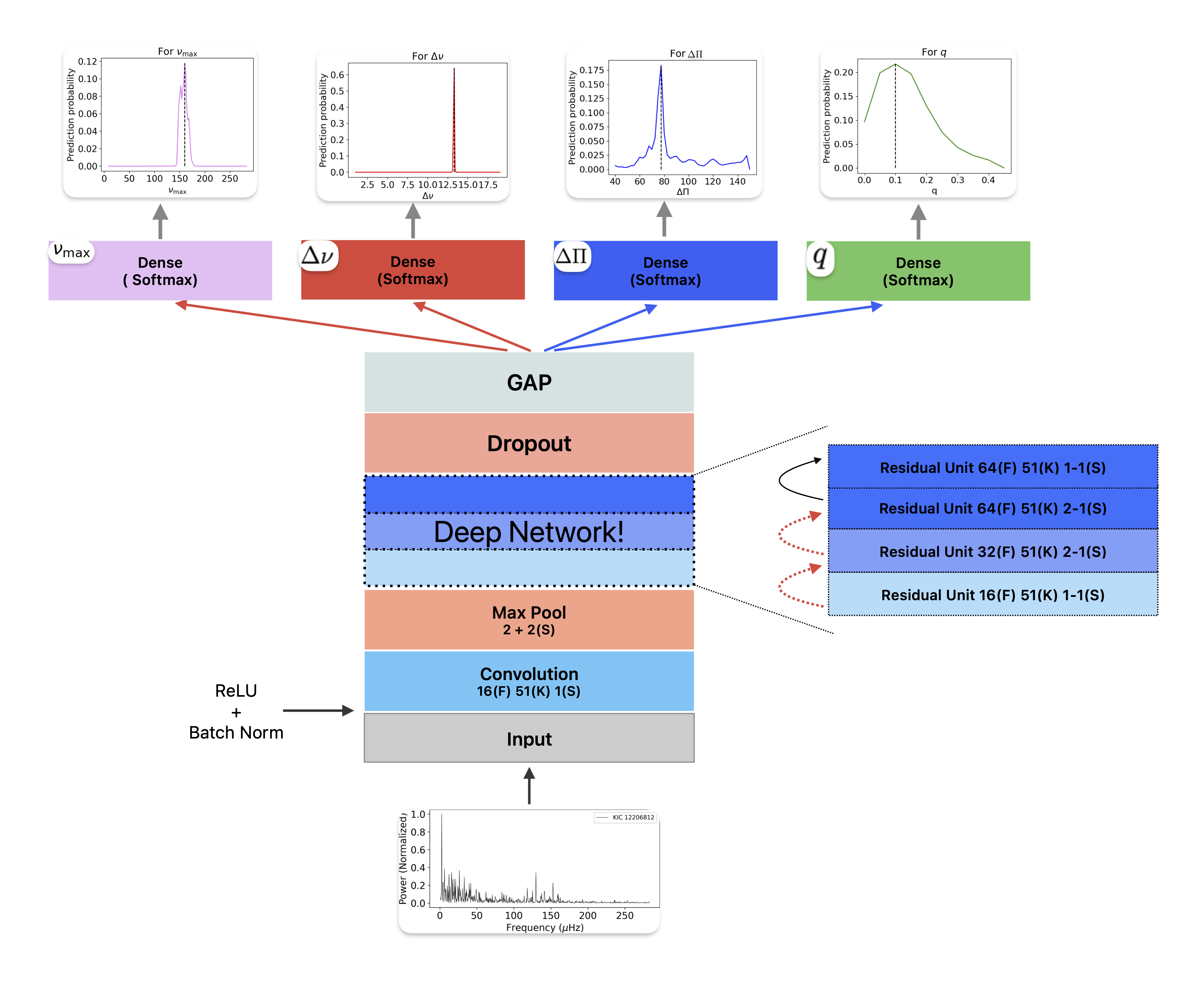}
    } \\
    \subfloat[\label{fig:wide_resnet_skip_connection}]{
    \includegraphics[width=0.6\textwidth]{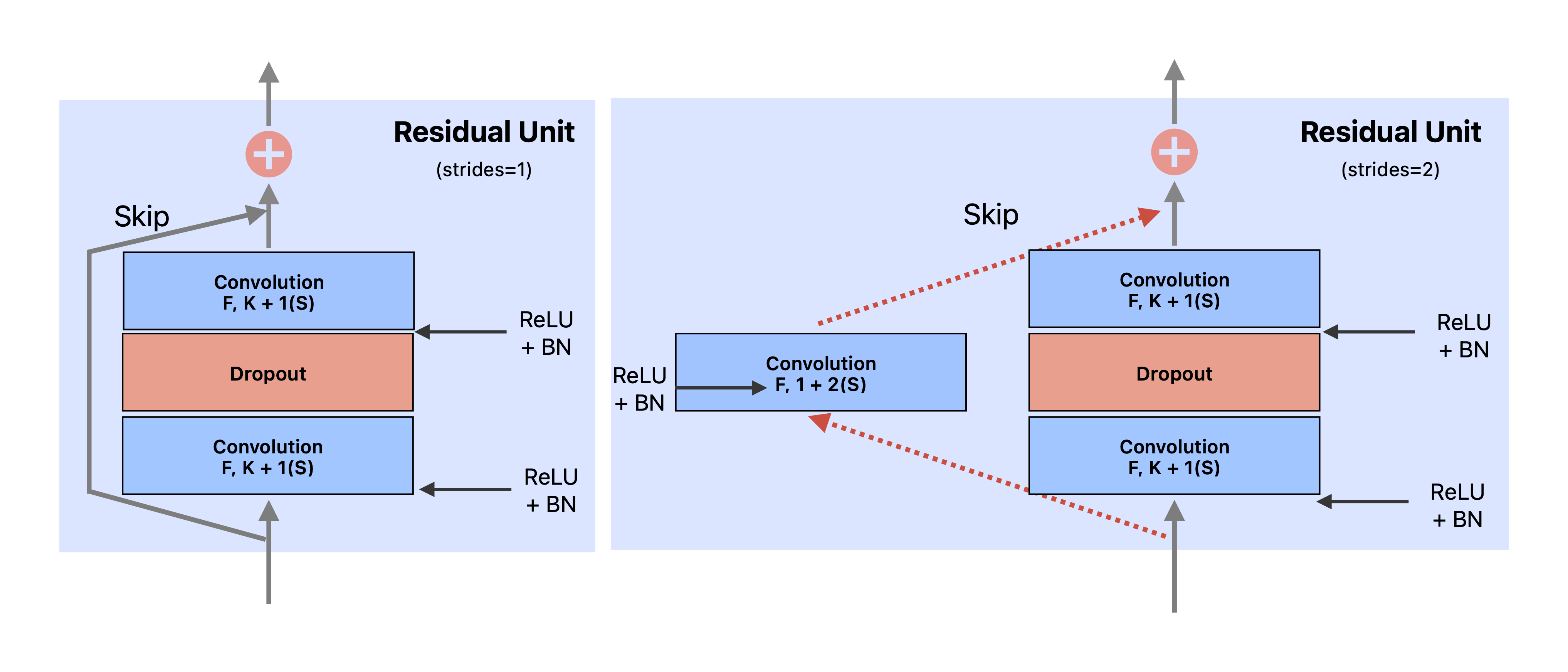}
    }
    \caption{ResNet-based architecture adopted for both models: \textbf{(a)} the overall architecture and \textbf{(b)} the architecture of the residual unit blocks, for a stride of 1 (left) and a stride of 2 (right).}
    \label{fig:wide_resnet}
\end{figure*}

In our initial exploration, we tried several architectures -- a standard CNN, a CNN-LSTM, a ResNet, and a Vision Transformer -- spanning both convolution-based and attention-based approaches. In these exploratory tests we found the ResNet, which augments a CNN with skip (residual) connections, to give the most stable training and the best validation performance, and we adopted it as the backbone for both K2 models. This is consistent with the ability of residual connections to ease gradient flow in deeper networks and to capture features across the multiple frequency scales present in the PSD. For each target parameter, the final layer applies a softmax over a set of discrete bins spanning the parameter's range, so that the network outputs a probability distribution over the bins rather than a single point estimate (see Sect.~\ref{sec:regression_to_classification}). The adopted architecture, including the number and composition of the residual blocks, is shown in Fig.~\ref{fig:wide_resnet}.

\subsection{Recasting regression as classification}
\label{sec:regression_to_classification}
Identifying the parameters corresponding to the PSD of a given star is naturally a regression problem. A well-established alternative, however, is to recast regression as classification by discretizing the continuous target range into a fixed number of bins and training the network to predict the correct bin. This approach has been applied successfully across a range of domains, from estimating a person's age from a photograph \citep{Rothe_2015} to deriving photometric redshifts for large galaxy surveys \citep{Pasquet_2019,Shuntov_2020}, where the binned output is valued not only for its accuracy but because it exposes the full shape of the inferred distribution rather than collapsing it to a single number. Most directly relevant here, the same formulation has been used to infer global seismic parameters for Kepler red giants from long-baseline data \citep{dhanpal2022,dhanpal2023}. We adopt it but apply it to the substantially shorter, lower-resolution K2 spectra. Each parameter's range is divided into bins, and the model is trained to classify the bin corresponding to a given PSD. The network then outputs an array of probabilities, one per bin, giving the probability that the parameter falls in each bin.

There are several scientific and practical reasons for this choice. First, classification can improve predictive accuracy, particularly when the relationship between inputs and outputs is highly non-linear \citep{Stewart_2022arXiv221105641S} or when the output parameters span a wide range (as is often the case in asteroseismology). Second, classification inherently respects physical constraints -- for example, it avoids predicting unphysical negative values for quantities like $\Delta\nu$ or $\Delta\Pi_1$, which sometimes arise in unconstrained regression models. Lastly, and importantly, formulating the inference as a classification problem allows us to recover a probabilistic distribution over the parameter space, giving us access not just to a point prediction but to a meaningful measure of the uncertainty and possible multimodality in the inferred values.

As established by \cite{richard1991_nn_bayesian_dist}, neural network classifiers trained with one-hot encoded target vectors and a cross-entropy loss function yield outputs that can be interpreted as Bayesian posterior probabilities. In this context, the output probability distribution over the bins may be interpreted as a quantized approximation to the posterior distribution of the parameter, conditioned on the observed data and the learned model.

We note, however, that the network output serves only as an approximation to the underlying posterior. Each parameter has its own softmax output, so the network returns marginal distributions rather than a joint distribution over all four, and the covariance between parameters is not accessible. The fidelity of the approximation depends on the network capacity, the amount of training data, and how well the training set reflects the true likelihood and prior distributions: our uniform sampling corresponds to a flat prior over each parameter's range (Sect.~\ref{sec:training_set}), and the realism of the synthetic spectra is assessed against observations in Sect.~\ref{sec:results_ekic}. We compare the network output directly with a Markov-chain Monte Carlo (MCMC) posterior for a Kepler red giant at K2 resolution in Sect.~\ref{sec:mcmc_comparison}.

Nonetheless, the binned output provides an empirical distribution over each parameter's range, from which we extract a point estimate and an asymmetric uncertainty interval (Sect.~\ref{sec:point_estimates}).

This offers a significant advantage over standard regression-based neural networks, which typically do not provide any probability distribution directly. Predictive uncertainties can be recovered for such networks through deep ensembles \citep{lakshminarayanan2017simplescalablepredictiveuncertainty} or Monte Carlo dropout \citep{gal2016dropoutbayesianapproximationrepresenting}, but the former requires training multiple instances of the model, substantially increasing training time, and neither reliably captures complex multi-modal behaviour. This presents a significant limitation in our specific application because the asteroseismic parameters, especially \dpi, can exhibit multimodal distributions. Alternatively, mixture density networks \citep{bishop1994_mixture_density_networks,hon2020AsteroseismicInferenceSubgiant} can be used to obtain probability distributions, but these necessitate a predefined number of modes and are susceptible to mode collapse. A comprehensive comparison including all these different methods is beyond the scope of this work.

\subsection{Hyperparameter tuning}
\label{sec:hyperparameter_tuning}
We tuned the network's hyperparameters -- including the kernel size, number of residual blocks, dropout rate, and learning rate -- on the validation set. The kernel size was selected via a grid search over values from 5 to 91: the validation loss decreases with increasing kernel size and saturates at $\approx$51, with negligible improvement beyond, so we adopt 51 as the smallest kernel in this saturated regime, minimizing the model's memory footprint and training time. At the K2-like frequency resolution this spans $\sim$6.7~$\mu$Hz, representing the frequency context seen by the first-layer filters; the PSD is supplied at its native resolution, so finer structure is preserved in the input.
The remaining hyperparameters were fixed after exploratory tuning on the validation set, retaining the configuration that gave stable convergence and low validation loss. The adopted network uses four residual blocks with [16, 32, 64, 64] filters, a dropout rate of 0.1, and a base learning rate of $1.5\times10^{-4}$, giving 987,940 trainable parameters. The output of each model is discretised into bins of 1~$\mu$Hz for $\nu_{\rm max}$, 0.05~$\mu$Hz for $\Delta\nu$, 1.25~s for $\Delta\Pi_1$, and 0.025 for $q$; the procedure for choosing these bin sizes is described in Appendix~\ref{appendix:choosing_bin_size}. The same configuration is used for both models.

\subsection{Training}\label{sec:training}
The input to each network is the normalized power spectrum, and the targets are the one-hot-encoded bin indices for each parameter (\numax, \dnu, \dpi, and $q$). The networks are trained as supervised classifiers, minimizing the categorical cross-entropy loss summed over all output parameters.

Measuring \dpi\ from single-campaign K2 data is challenging, and this motivated our use of two specialized networks rather than one. Although \cite{dhanpal2023} recovered all four parameters with a single network from four-year Kepler spectra, the coarser frequency resolution of K2 makes the period spacing far harder to extract, particularly for more evolved stars: the spacing between successive radial orders shrinks, granulation noise rises at low frequencies, the increasing mixed-mode inertia suppresses mode visibility \citep{Grosjean_2014A&A...572A..11G}, and the density of mixed modes drops once helium burning begins \citep{Gehan_2018A&A...616A..24G}. Consistent with this, when we trained a single network to infer \dpi\ across the full \dnu\ range, we obtained useful results only for \dnu\ $\gtrsim 9~\mu$Hz. We therefore split the problem: a p-mode model covering the full range for \numax\ and \dnu, and a dedicated g-mode model restricted to the young--red-giant regime (\dnu\ $> 9~\mu$Hz), where a network optimized specifically for the g-mode--dominated parameters recovers \dpi\ more reliably than a single general-purpose network.

Both networks use an identical architecture but are tailored to different parameters. The p-mode model assigns weights of 0.4 to each of \numax\ and \dnu\ and 0.1 to each of \dpi\ and $q$; the g-mode model reverses this to $(0.1, 0.1, 0.4, 0.4)$, prioritizing \dpi\ and $q$. In both cases we retain all four parameters as outputs, rather than dropping the down-weighted ones, as we found this reduces catastrophic outliers -- predictions that fall far from the true values -- in our synthetic and Kepler-at-K2-resolution test sets. A plausible explanation is that the additional parameters help the model distinguish the dipole mixed modes from the surrounding pressure modes in ambiguous cases. For all results presented here, we take \numax\ and \dnu\ from the p-mode model and \dpi\ and $q$ (for young red giants) from the g-mode model. We infer $q$ throughout, but do not report it as a result in its own right: it is used only as a diagnostic for the \dpi\ inferences, since the reliability of a period spacing depends strongly on the coupling strength, and stars with very weak coupling are rejected on that basis (Sect.~\ref{sec:results_ekic_dpi}).

We use the Adam optimizer, which provides per-parameter adaptive learning rates. Furthermore, rather than a monotonically decaying learning rate, we use a cosine-decay-with-restarts schedule: the periodic restarts can push the optimizer out of sharp local minima and encourage it to settle in broader and better-generalizing solutions. Training is distributed over two GPUs with a batch size of 64 per GPU; this larger effective batch yields smoother gradient estimates, and we scale the learning rate (Sect.~\ref{sec:hyperparameter_tuning}) accordingly to keep the effective step size consistent. Finally, to prevent overfitting once the validation loss plateaus, we apply early stopping with a patience of five epochs -- enough for the model to fully converge while halting shortly after it begins to overfit. We adopt the checkpoint with the lowest validation loss as the final model, so that inference uses the best-performing state rather than the most-trained one. Training a single network under this configuration takes approximately 15~hours. Once trained, inference is inexpensive: each network processes $\sim$20,000 spectra in about 12~s, so the entire K2 sample is analysed in a few tens of seconds. This is the practical advantage of the approach over per-star likelihood sampling: the MCMC fits described in Sect.~\ref{sec:mcmc_comparison} require of order a day of computation for a single star.

\subsection{Point estimates, uncertainties, and quality cuts}
\label{sec:point_estimates}
For each parameter, we take the median of the output probability distribution as the point prediction, and use the 16th and 84th percentiles around that median to quantify the uncertainty: $\sigma_{\text{low}}$ is the difference between the median and the 16th percentile, and $\sigma_{\text{high}}$ that between the 84th percentile and the median. Where a single number is required -- in particular for the quality cuts described below -- we use the combined uncertainty $\sigma_{\text{tot}} = \sigma_{\text{low}} + \sigma_{\text{high}}$, that is, the full width of the 16th--84th percentile interval. This combination gives both a sharp point estimate and a measure of the spread and asymmetry of the underlying distribution. We additionally require the median and the mode of the distribution to be mutually consistent, discarding any inference for which the most probable bin falls outside the 16th--84th percentile interval. Such cases arise when the output is strongly multimodal and the median falls in a low-probability region between competing peaks, so that it is not representative of any of them; this condition is applied to every parameter, alongside the uncertainty cuts below. To check whether these reported uncertainties are actually reliable, we compare predictions with reference values using the normalized residual, defined as:
\[
\text{Normalized residual} =
\begin{cases}
\dfrac{\hat{y} - y}{\sqrt{\sigma_{\text{high}}^2+\sigma_{\text{ref}}^2}}, & \text{if } \hat{y} < y \\[1.2ex]
\dfrac{\hat{y} - y}{\sqrt{\sigma_{\text{low}}^2+\sigma_{\text{ref}}^2}}, & \text{if } \hat{y} \geq y
\end{cases}
\]
where \(\hat{y}\) and \(y\) are the predicted and reference values, \(\sigma_{\text{low}}\) and \(\sigma_{\text{high}}\) are the lower and upper uncertainties on the prediction, and \(\sigma_{\text{ref}}\) is the uncertainty on the reference value. If the uncertainties are well calibrated, the fraction of stars with an absolute normalized residual below 1 -- the empirical $1\sigma$ fraction -- should match the 68.3\% expected for Gaussian errors. A fraction above 0.683 means the model's uncertainties are too large (under-confident); a fraction below 0.683 means they are too small (over-confident).

Because our models produce a prediction for every star regardless of data quality, we need a way to isolate confident inferences. We do this with uncertainty cuts, chosen empirically to reject poor predictions: we retain only inferences for which the combined uncertainty $\sigma_{\text{tot}}$ is below 20\% of the predicted value in \numax\ and below 10\% in \dnu\ and \dpi, together with the additional conditions on \q\ described in Sect.~\ref{sec:results_ekic_dpi}. All uncertainty thresholds quoted in this work refer to this combined width. How these choices trade off completeness against reliability is examined in Appendix~\ref{appendix:dpi_selection_cut} for \dpi, which shows how accuracy, coverage, and the outlier rate each depend on where the threshold is set.

\subsection{MCMC and ML Comparison}
\label{sec:mcmc_comparison}
To illustrate the capability of ML models to approximate Bayesian posterior distributions, we compare the network outputs with posteriors obtained using Markov Chain Monte Carlo (MCMC) analysis for one Kepler red giant truncated to a K2-like baseline. We employ the Metropolis-Hastings algorithm \citep{Metropolis1953, Hastings1970} to sample the posterior, assuming a likelihood based on $\chi^2$ with 2 degrees of freedom, given the observed power spectral density (PSD) and a synthetic model. This methodology follows the approach summarized in \cite{dhanpal2023}, based on the formalism described in \cite{benomar_2009_mcmc}. We refer the reader to these works for further details on the MCMC implementation. Rotational splittings are not included in the MCMC model, as their inclusion was found to hinder convergence at this resolution.

\begin{figure*}
    \centering
    \includegraphics[width=0.95\textwidth]{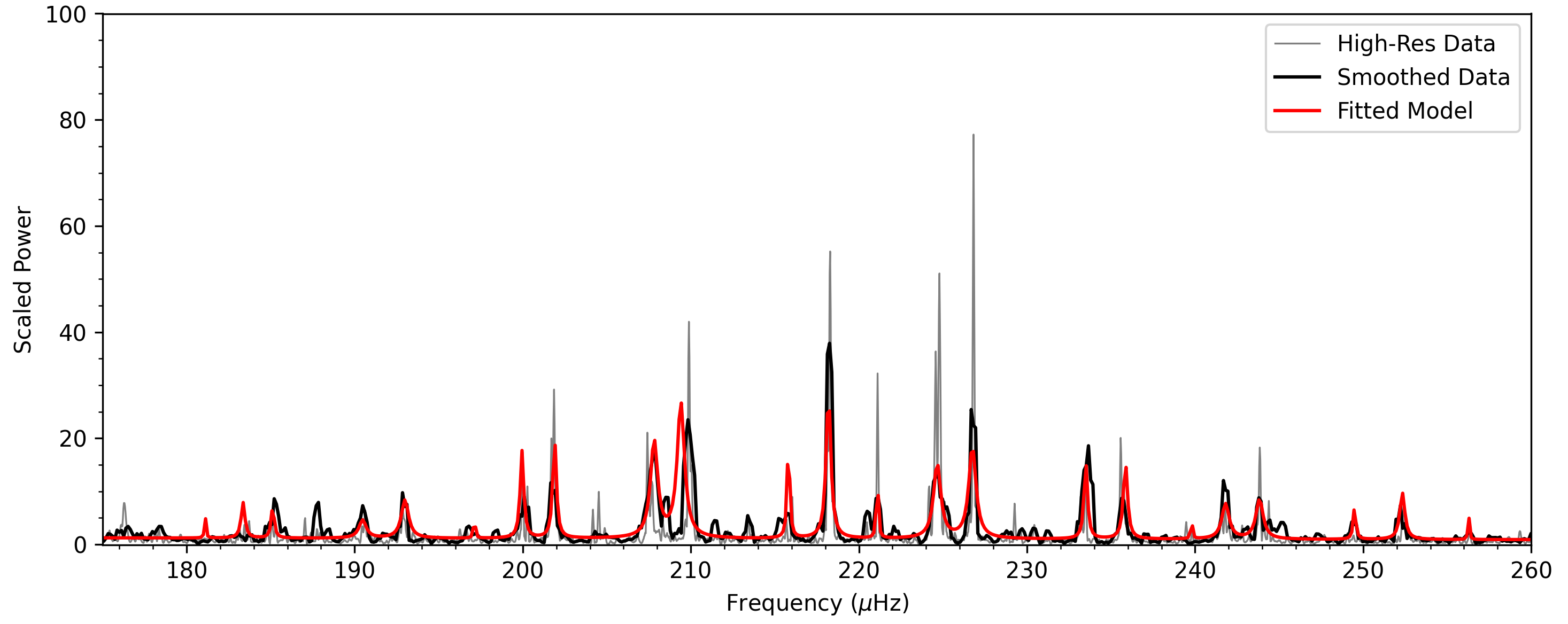}
    \caption{Best-fit MCMC model (red) for KIC 4448777 overlaid on the smoothed PSD data (black) at K2-like resolution. The PSD obtained from the full 4-year Kepler observation is shown in grey for reference.}
    \label{fig:mcmc_fit}
\end{figure*}

\begin{figure*}
    \centering
    \includegraphics[width=0.95\textwidth]{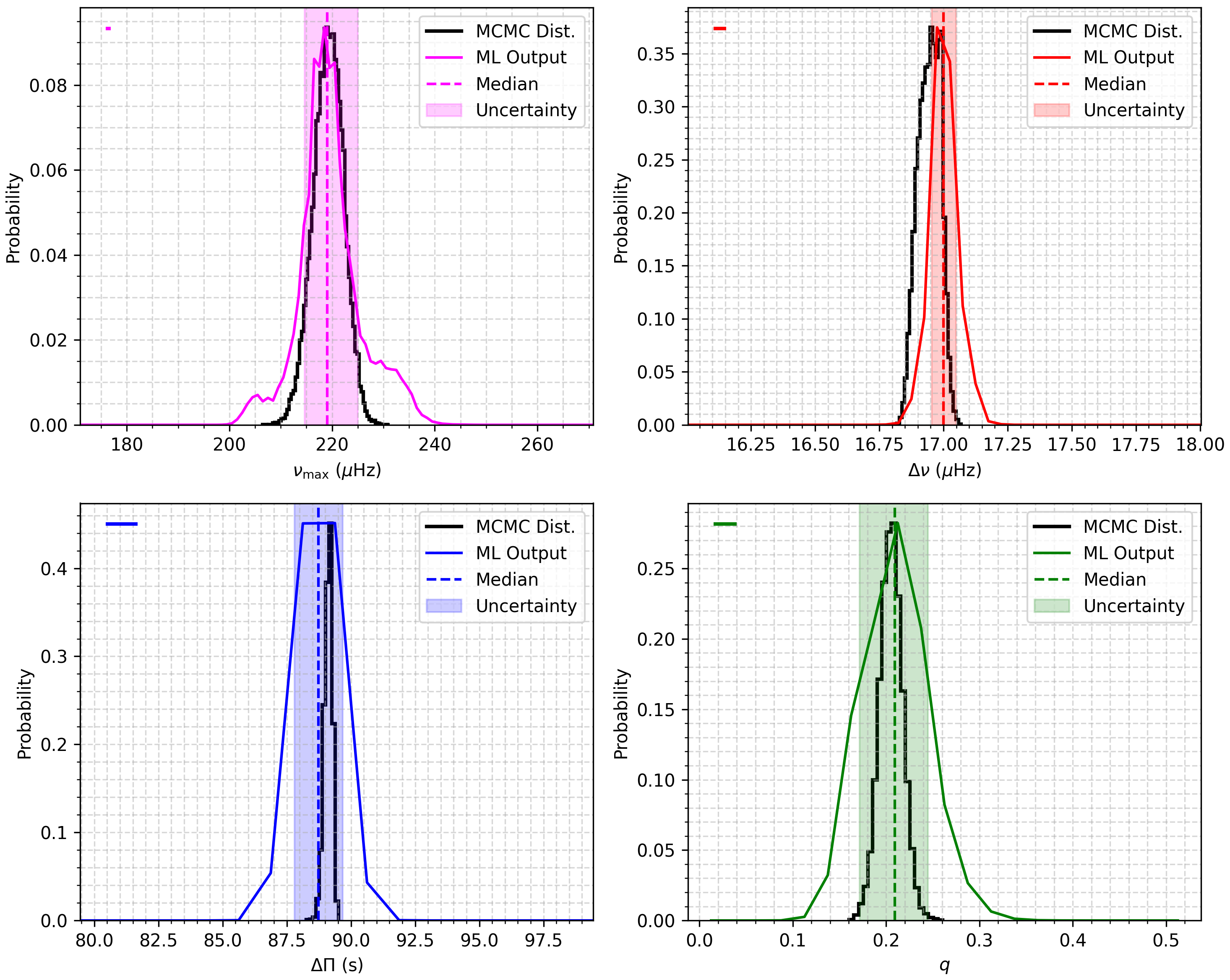}
    \caption{Comparison of output probability distribution for KIC 4448777 at K2-resolution derived from the model (solid curves) and MCMC analysis (histograms). The panels display the posterior distributions for $\nu_{\rm max}$, $\Delta\nu$, $\Delta\Pi_{1}$, and the coupling factor $q$. For visual comparison, the MCMC posteriors have been normalized to match the peak probability density of the ML outputs. The horizontal bar on top left of each panel shows the ML bin size.}
    \label{fig:mcmc_comparison}
\end{figure*}

Fig.~\ref{fig:mcmc_fit} displays the best-fit MCMC model (red) overlaid on the smoothed data (black); the PSD obtained from the full 4-year Kepler observation is shown in grey for reference. Fig.~\ref{fig:mcmc_comparison} presents the comparison between the MCMC posteriors and the model outputs for $\nu_{\rm max}$, $\Delta\nu$,  $\Delta\Pi_{1}$ and $q$. To facilitate visual comparison, the MCMC posteriors have been normalized to match the peak probability of the ML outputs. We observe that the ML outputs and MCMC posteriors overlap and agree well within uncertainties for all parameters. However, the ML distributions for all parameters are broader than the MCMC posteriors, notably so for \dpi and \q. This difference in posterior width is not unexpected. The MCMC algorithm uses significant computational resources (thousands of steps) to map the likelihood surface for a given target, allowing for a highly optimized exploration of the parameter space. In contrast, the ML model provides an instantaneous, generalized inference. The exact difference between the posteriors depends on the output bin sizes and network capacity on one side and the priors and convergence of the chains on the other. This comparison demonstrates that the network output tracks the location and approximate shape of the posterior; the calibration of the uncertainties is assessed from the ensemble comparisons of Sect.~\ref{sec:results_ekic}.

\section{Results}
\label{sec:results}
We validate the models in three stages before applying them to the K2 sample. First, on a held-out set of synthetic stars, which confirms that the networks have not overfitted and that their performance on unseen data matches expectations; for stars passing the quality cuts of Sect.~\ref{sec:point_estimates}, the models recover \numax, \dnu, and \dpi\ to 2.9\%, 0.5\%, and 1.0\% respectively (Appendix~\ref{sec:results_on_synthetics}). Second, on Kepler observations degraded to K2-like resolution, where our inferences can be compared directly against values derived from the full four-year time series. Third, on the K2 observations themselves, against the K2~GAP~DR3 catalogue. Having established the reliability of the inference, we then apply the models to the full K2 red-giant sample -- including \dpi\ for the young red giants -- under the same quality thresholds throughout, and report the resulting high-confidence period spacings in Appendix~\ref{appendix:k2_red_giants_dpi}.

\subsection{Kepler data at K2-like resolution}
\label{sec:results_ekic}
We apply the models to the Kepler segments at K2-like resolution described in Sect.~\ref{sec:data_ekic}, whose baselines approximately match those of a single K2 campaign. Reference values are taken from analyses of the full four-year Kepler time series: \citet{Yu_2018} for \numax and \dnu, obtained with the SYD pipeline and visually verified, and \citet{Gang_Li_2024} for \dpi. Since the inference and the reference derive from the same targets and the same mission, this comparison is largely free of the instrumental systematics that affect the K2 measurements in Sect.~\ref{sec:results_k2gap}, although small scatter is expected due to the reduced frequency resolution and signal-to-noise of the shorter baseline. We note for completeness that the light curves used here and those used in the reference studies were produced by different photometric pipelines, which may contribute to the differences in inferred parameters for a small number of stars.

\subsubsection{\numax and \dnu}
\begin{figure*}
\subfloat[\label{fig:eKIC_numax_comparison_scatter}]{
        \includegraphics[width=0.48\textwidth]{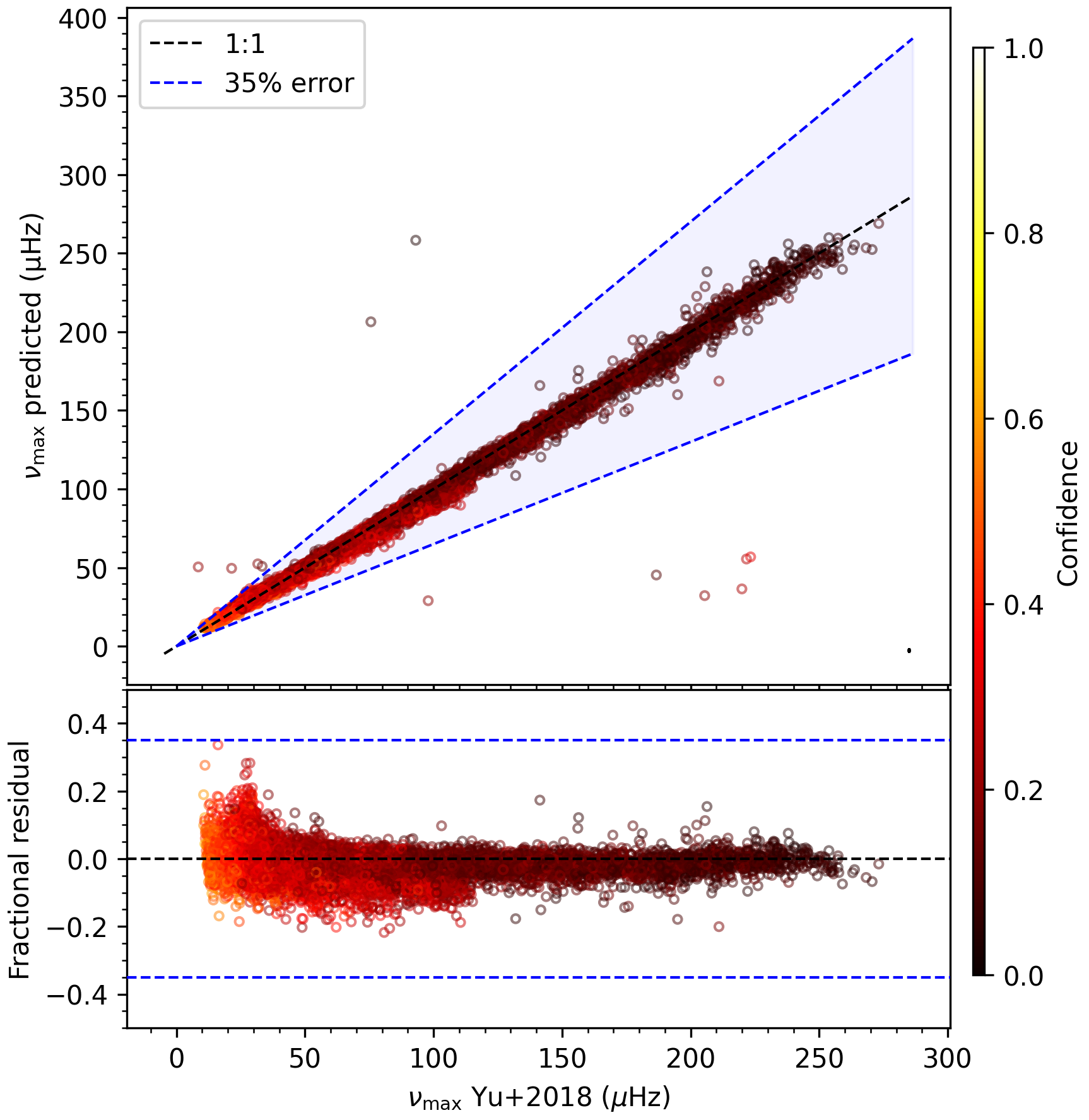}
    }
    \hfill
    \subfloat[\label{fig:eKIC_dnu_comparison_scatter}]{
        \includegraphics[width=0.48\textwidth]{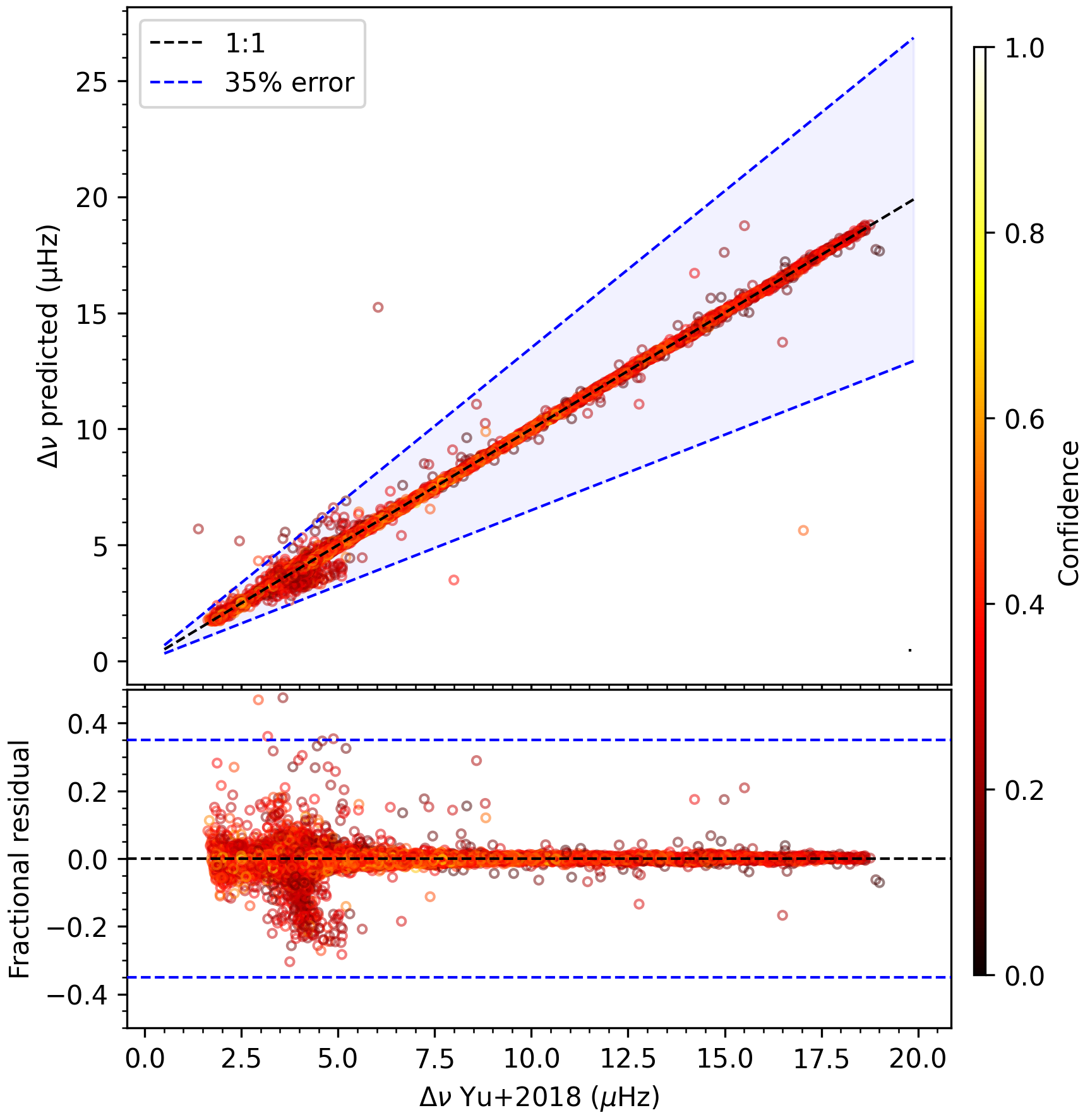}
    } \\ 

    \subfloat[\label{fig:eKIC_numax_comparison_hist}]{
        \includegraphics[width=0.44\textwidth]{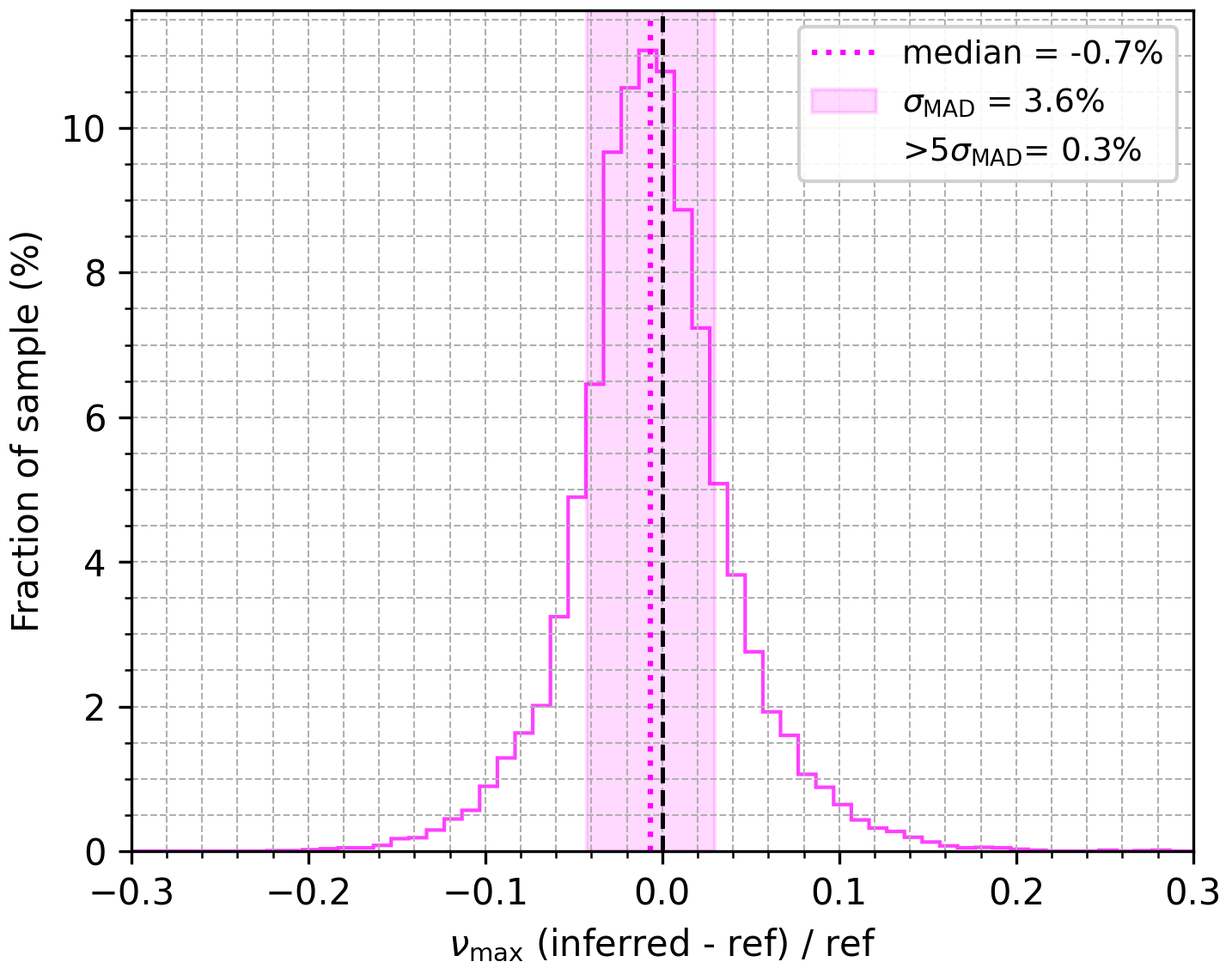}
    }
    \hfill
    \subfloat[\label{fig:eKIC_dnu_comparison_hist}]{
        \includegraphics[width=0.45\textwidth]{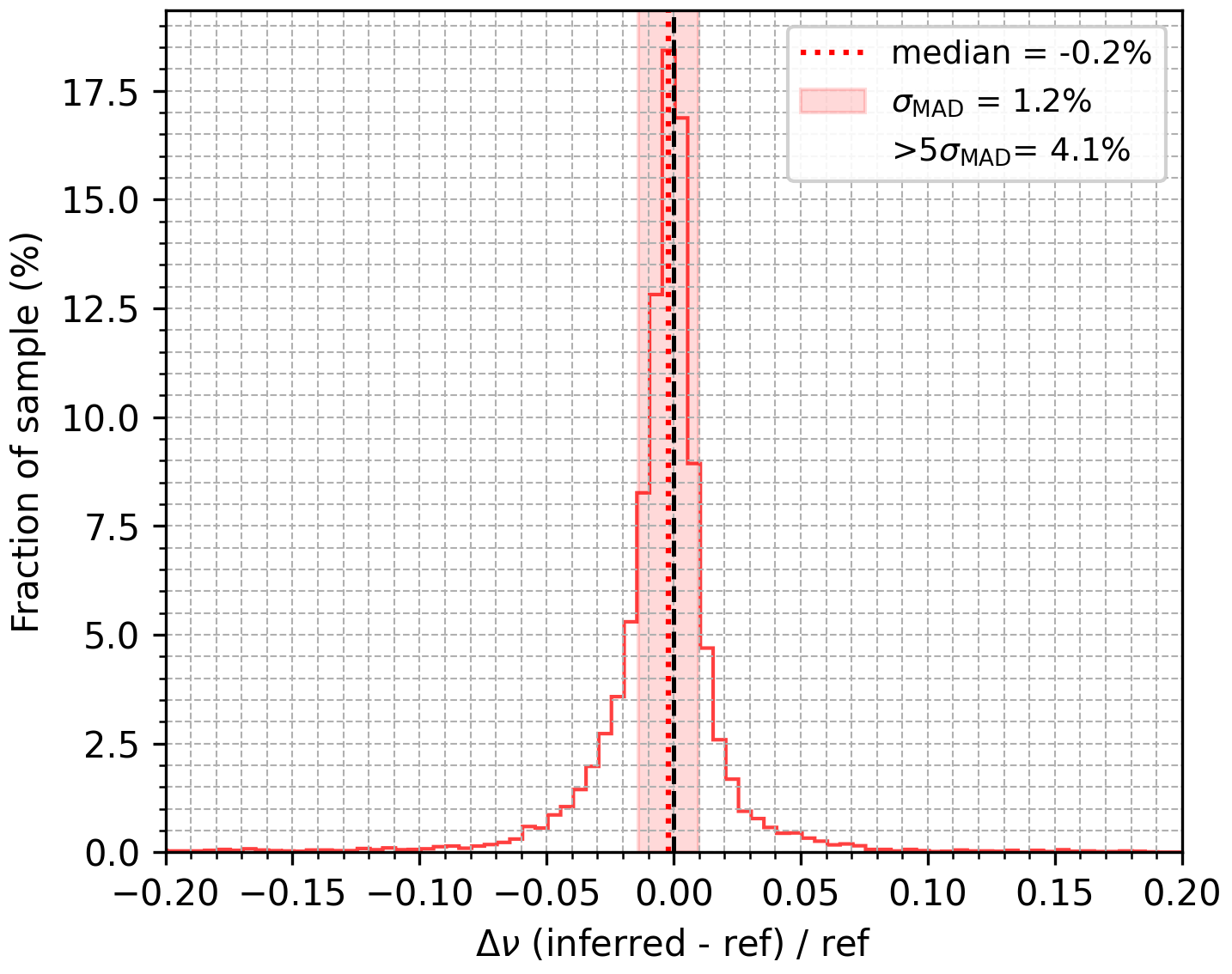}
    }
    \caption{Comparison of model predictions from Kepler data degraded to K2-like resolution with reference values derived from the full-length ($\sim$4-year) Kepler observations of \citet{Yu_2018}: (a) $\nu_{\mathrm{max}}$ and (b) $\Delta\nu$; (c--d) histograms of relative errors in $\nu_{\mathrm{max}}$ and $\Delta\nu$, respectively.}
    \label{fig:eKIC_pmodel_comparison}
\end{figure*}

Of the 15,639 stars in common with \citet{Yu_2018}, 15,054 (96.3\%) satisfy the \numax\ criterion and 14,237 (91.0\%) satisfy the additional criterion on \dnu. The inferred values follow the reference values closely across the full range of both parameters (Figs.~\ref{fig:eKIC_numax_comparison_scatter} and~\ref{fig:eKIC_dnu_comparison_scatter}), and the distributions of fractional differences peak at zero (Figs.~\ref{fig:eKIC_numax_comparison_hist} and~\ref{fig:eKIC_dnu_comparison_hist}). The robust dispersion $\sigma_{\mathrm{MAD}} = 1.4826\,\mathrm{MAD}$, which characterises the core of these distributions without being skewed by the tails discussed below, is $3.6\%$ for \numax\ and $1.2\%$ for \dnu. A small negative bias is present in both parameters, with median fractional offsets of $-0.7\%$ and $-0.2\%$, corresponding to absolute offsets of $-0.32\ \mu$Hz and $-0.013\ \mu$Hz and amounting in each case to less than a fifth of the corresponding dispersion.

We also find a small fraction of stars with larger differences: 0.31\% of the confident \numax sample lies beyond $5\sigma_{\mathrm{MAD}}$, and 4.09\% for \dnu. The higher number for \dnu is not unexpected, as it requires the ridge of consecutive radial orders to be identified, and for a minority of short-baseline spectra that identification can be ambiguous. Outright failures nevertheless remain rare: 99.92\% of confident \numax and 99.93\% of confident \dnu inferences agree with the reference values to better than 35\%.

The uncertainties returned by the models are well calibrated: the empirical $1\sigma$ fractions are 0.698 for \numax and 0.658 for \dnu, bracketing the nominal 0.683 expected for well-calibrated Gaussian uncertainties, such that the reported \numax uncertainties are marginally conservative and those for \dnu marginally optimistic, but in both cases very close to the ideal value.

\subsubsection{\dpi}
\label{sec:results_ekic_dpi}

For \dpi we compare against \citet{Gang_Li_2024}, who measured period spacings from the full four-year Kepler time series using a treatment of rotational splitting that accounts for near-degeneracy effects. To filter out low confidence predictions, we retain inferences with \dpi uncertainties below 10\% and \q uncertainties below 0.12. The uncertainty thresholds were determined empirically to balance coverage and reliability, ensuring that the retained stars exhibit high prediction accuracy and a low fraction of unidentifiable outliers; see Appendix~\ref{appendix:dpi_selection_cut} for details on how this threshold was chosen for \dpi. We further exclude predictions near the upper and lower boundaries -- cases where the model assigns values at the edge of the parameter range, and where the prediction may therefore be truncated by the imposed bounds. Additionally, we discard stars with \q $< 0.05$, as predictions in this regime are particularly prone to failure. This is motivated by the finding of \citet{dhanpal2023}, who showed that both ML and MCMC-based inferences of $\Delta \Pi$ become increasingly unreliable when $q$ is very small.

\begin{figure*}
    \centering
    \subfloat[\label{fig:eKIC_dpi_comparison_scatter}]{
        \includegraphics[width=0.5\textwidth]{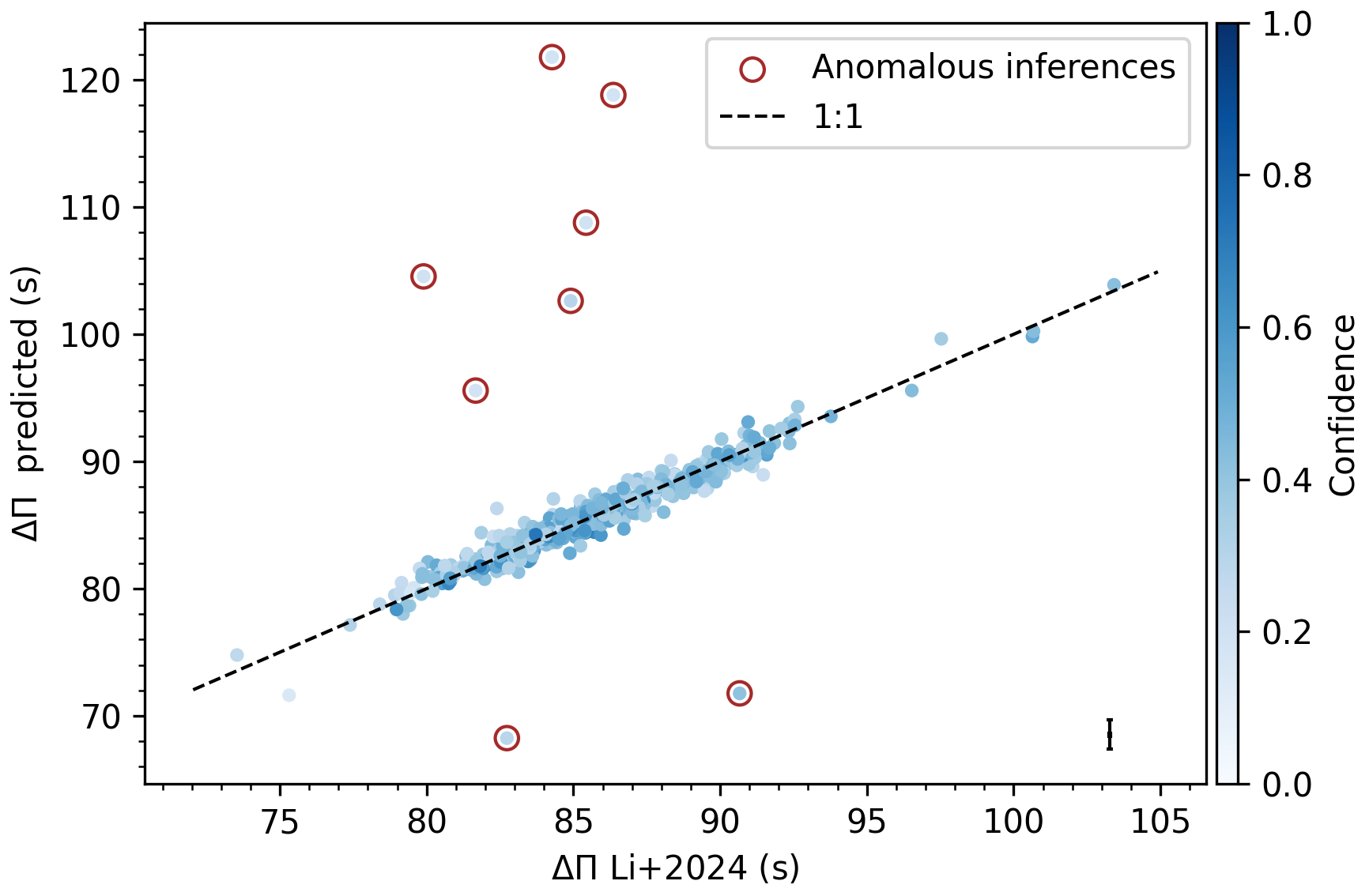}
    }
    \subfloat[\label{fig:eKIC_dpi_comparison_hist}]{
        \includegraphics[width=0.41\textwidth]{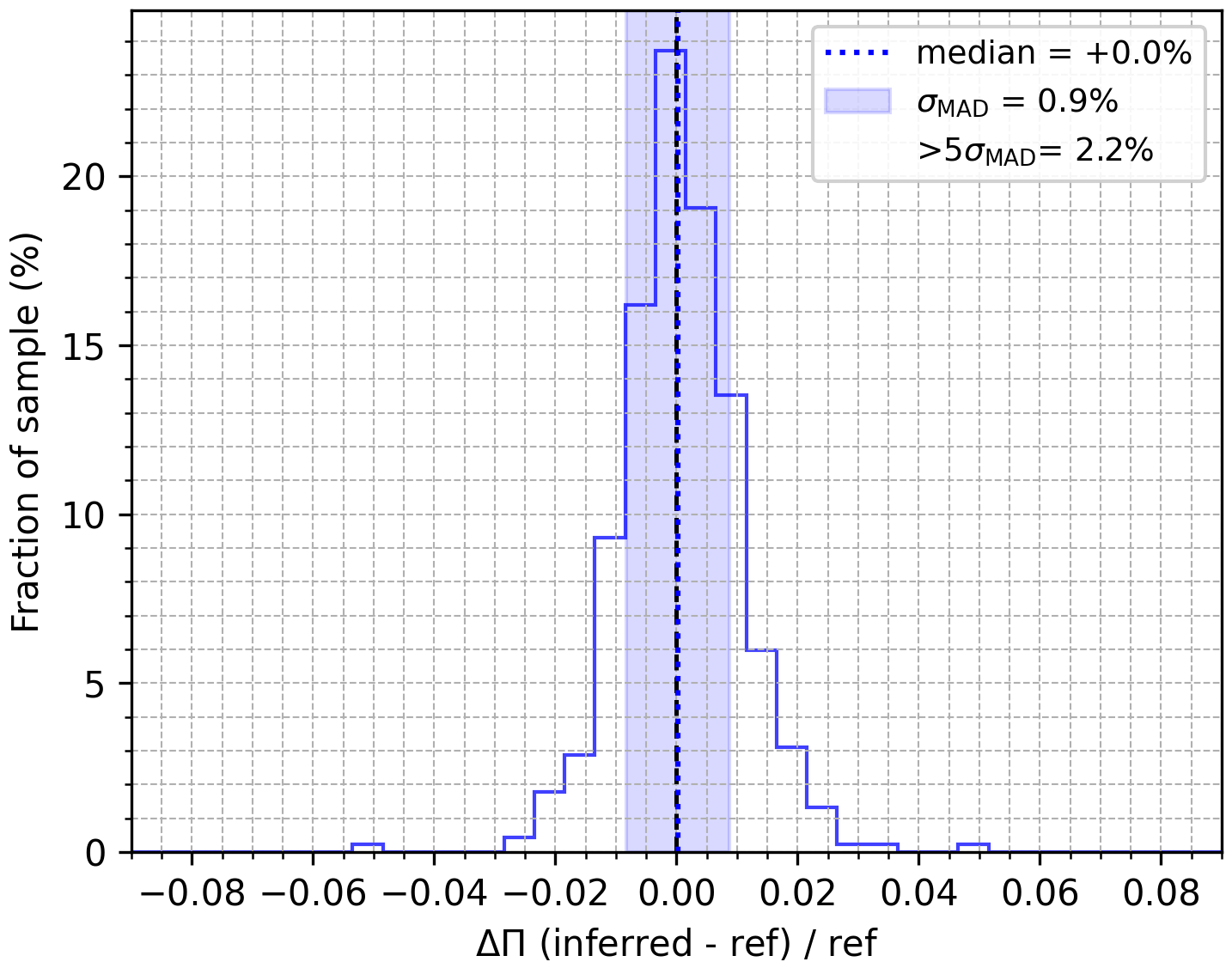}
    }
    \caption{\textbf{(a)} Comparison of our $\Delta\Pi_{1}$ inferences for the 451 confident stars from Kepler data at K2-like resolution with the values derived from the full four-year time series by \citet{Gang_Li_2024}. The black cross in the bottom right corner shows the typical uncertainties in the measurements plotted here. \textbf{(b)} Histogram of relative errors for non-anomalous stars in (a).}
    \label{fig:eKIC_dpi_comparison}
\end{figure*}

\begin{figure*}
    \centering
    \includegraphics[width=0.95\textwidth]{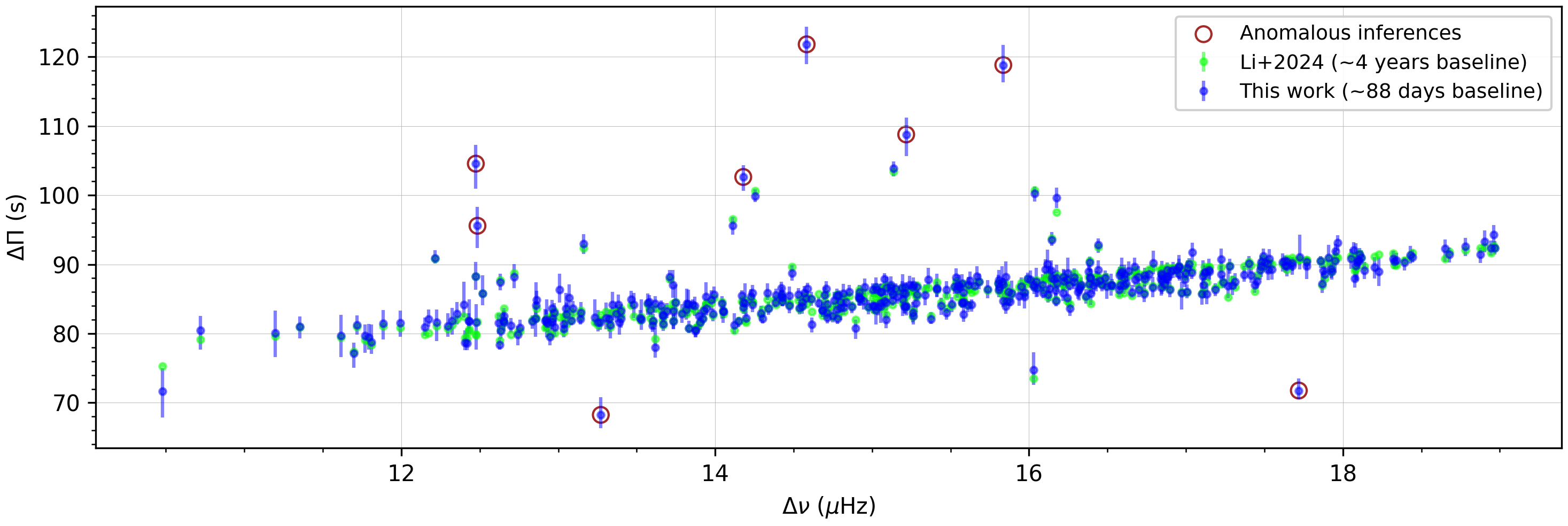}
    \caption{$\Delta\nu$--$\Delta\Pi_{1}$ diagram for the 451 common red giants from Kepler data at K2-like resolution (blue), plotted over the $\Delta\Pi_{1}$ values derived from the full four-year time series by \citet{Gang_Li_2024} (green).}
    \label{fig:eKIC_dnu_dpi_relation}
\end{figure*}

Of the 2,070 stars in common, 451 (21.8\%) satisfy the selection criteria, considerably higher than the 6.8\% yield obtained on the synthetic test set (Appendix~\ref{sec:results_on_synthetics}). This is expected, since any catalogue of measured period spacings is itself restricted to stars whose coupling strengths and mode visibilities permit a determination. Figures~\ref{fig:eKIC_dpi_comparison_scatter} and~\ref{fig:eKIC_dpi_comparison_hist} compare the inferred and reference values and show the distribution of fractional differences. The period spacings agree with \citet{Gang_Li_2024} with a robust dispersion of 0.9\%, with only 2.2\% of the sample lying beyond $5\sigma_{\mathrm{MAD}}$; eight stars show significant discrepancies. Figure \ref{fig:eKIC_dnu_dpi_relation} illustrates the \dnu--\dpi distribution for this sample, where most follow the well-established degenerate sequence: the narrow track in the \dnu--\dpi plane occupied by red giants whose helium cores have become electron degenerate, along which \dpi decreases steadily as the star ascends the giant branch and its core contracts \citep{vrard_et_al,deheuvels2022SeismicSignatureElectron}.

These eight stars, which we flag as anomalous inferences, show significant discrepancies with \citet{Gang_Li_2024} and lie off the degenerate sequence. Upon inspecting the \dpi probability distribution for these stars, we observe the presence of multiple peaks, with a smaller peak at \dpi values lying along the degenerate sequence. Interestingly, this smaller peak aligns closely with the values from \citet{Gang_Li_2024}. This suggests that the model is identifying multiple possible solutions, but with different confidence levels. We label such stars as anomalous inferences precisely because of the presence of this lower-probability peak along the degenerate sequence. These cases warrant further investigation: a preliminary analysis shows that the inferred \q values for these stars differ from those reported by \citet{Gang_Li_2024}, which may be contributing to the ambiguity in their \dpi estimates. A more detailed investigation of this discrepancy is left to future work.

The empirical $1\sigma$ fraction for the high-confidence sample is 0.878, against the 0.683 expected for well-calibrated Gaussian uncertainties, indicating that the reported \dpi\ uncertainties are wider than the residuals warrant; the median quoted uncertainty of 1.32\% exceeds the observed dispersion of 0.9\% by a factor of $\sim$1.5. This behaviour is also seen on synthetic data, where the true values are known exactly and the corresponding figures are 1.40\% and 1.0\% (Appendix~\ref{sec:results_on_synthetics}). It is not a consequence of the discretisation of the network output, since the reported uncertainties do not narrow as the bin width is reduced below 1.25~s and the over-coverage persists essentially unchanged at every bin width tested (Appendix~\ref{appendix:choosing_bin_size}). It appears instead to be a property of the predictive distributions themselves, which are broader than the precision the model ultimately achieves at this resolution. The multimodality described above contributes in the affected cases, although it does not account for the effect in full. The reported \dpi\ uncertainties are therefore conservative by roughly 50\%, and may be treated as upper limits on the true uncertainty.

\subsection{K2 red giants}
Having validated both models on synthetic spectra (Appendix~\ref{sec:results_on_synthetics}) and on Kepler data at K2-like resolution (Sect.~\ref{sec:results_ekic}), we apply them to the K2 observations described in Sect.~\ref{sec:data_k2}, using the same selection criteria throughout.

\subsubsection{Comparison with K2 GAP DR3}
\label{sec:results_k2gap}
We compare the inferred \numax and \dnu values against the K2~GAP~DR3 catalogue \citep{k2_dr3}. Of the 18,704 stars in common, 16,947 (90.6\%) satisfy the \numax criterion and 14,687 (78.5\%) additionally satisfy the criterion on \dnu. The \numax values in K2~GAP~DR3 are obtained by combining measurements from several pipelines, each rescaled by an evolutionary-state--dependent factor chosen to align the asteroseismic radii with Gaia (Sect.~4.2 of \citealt{k2_dr3}). To enable a like-for-like comparison we reverse this calibration using the mean factors of 1.017 for RGB and 1.008 for RC stars, obtaining approximate pre-scaled values. Since the original scaling was pipeline- and star-dependent, this reversal is only approximate and some residual difference is expected.

\begin{figure*}
\subfloat[\label{fig:EPIC_numax_comparison_scatter}]{
        \includegraphics[width=0.48\textwidth]{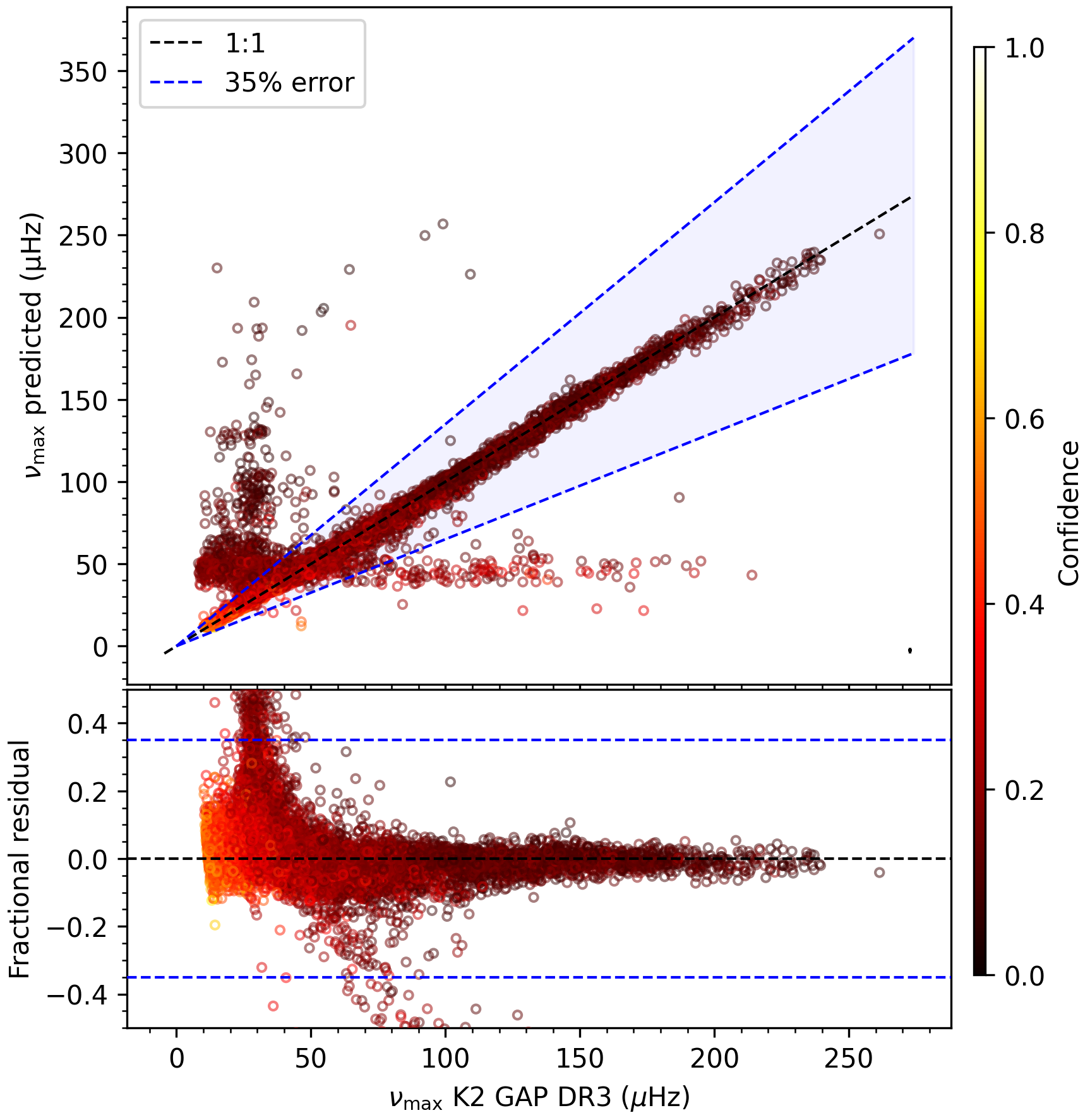}
    }
    \hfill
    \subfloat[\label{fig:EPIC_dnu_comparison_scatter}]{
        \includegraphics[width=0.48\textwidth]{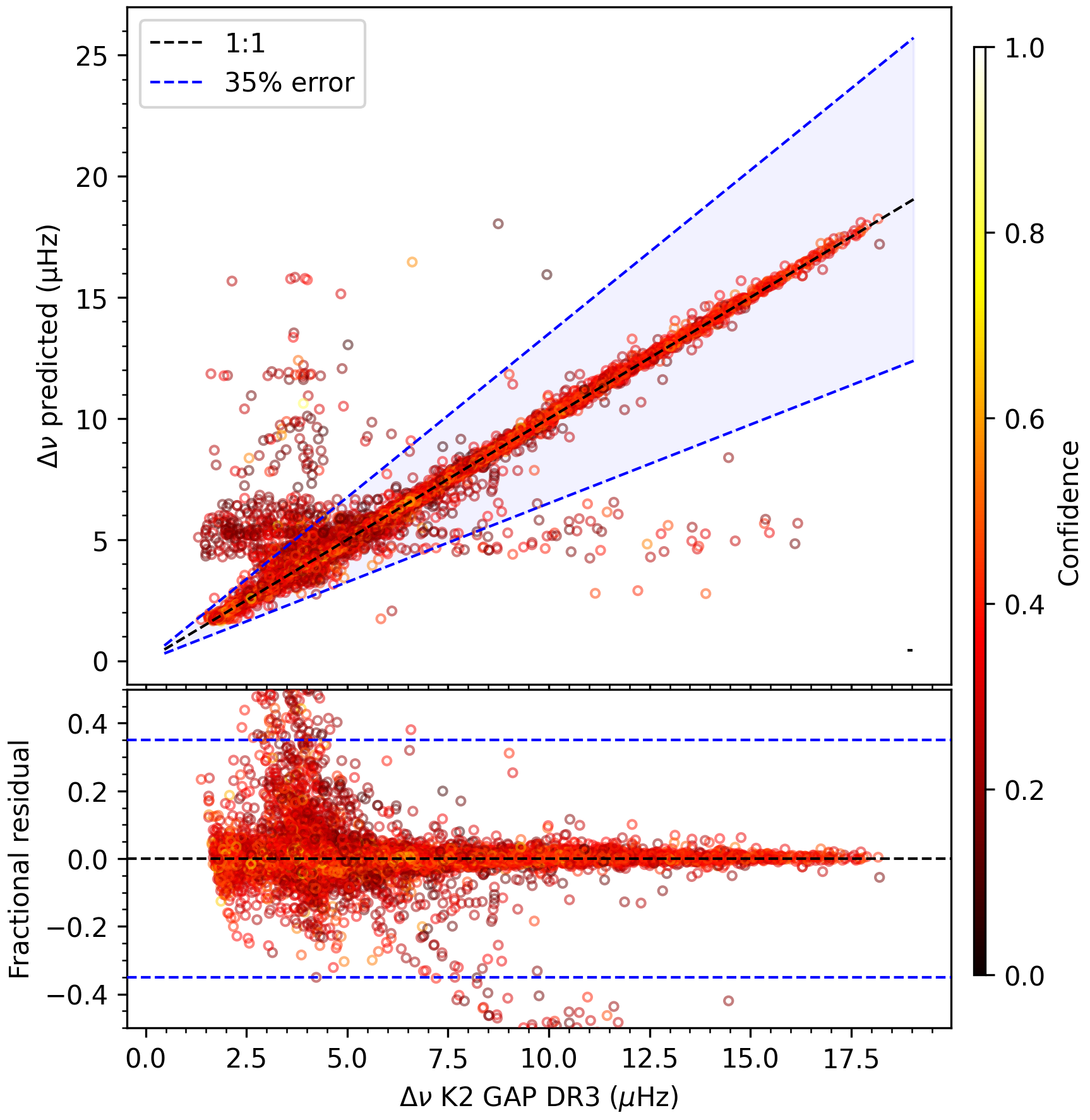}
    } \\ 

    \subfloat[\label{fig:EPIC_numax_comparison_hist}]{
        \includegraphics[width=0.44\textwidth]{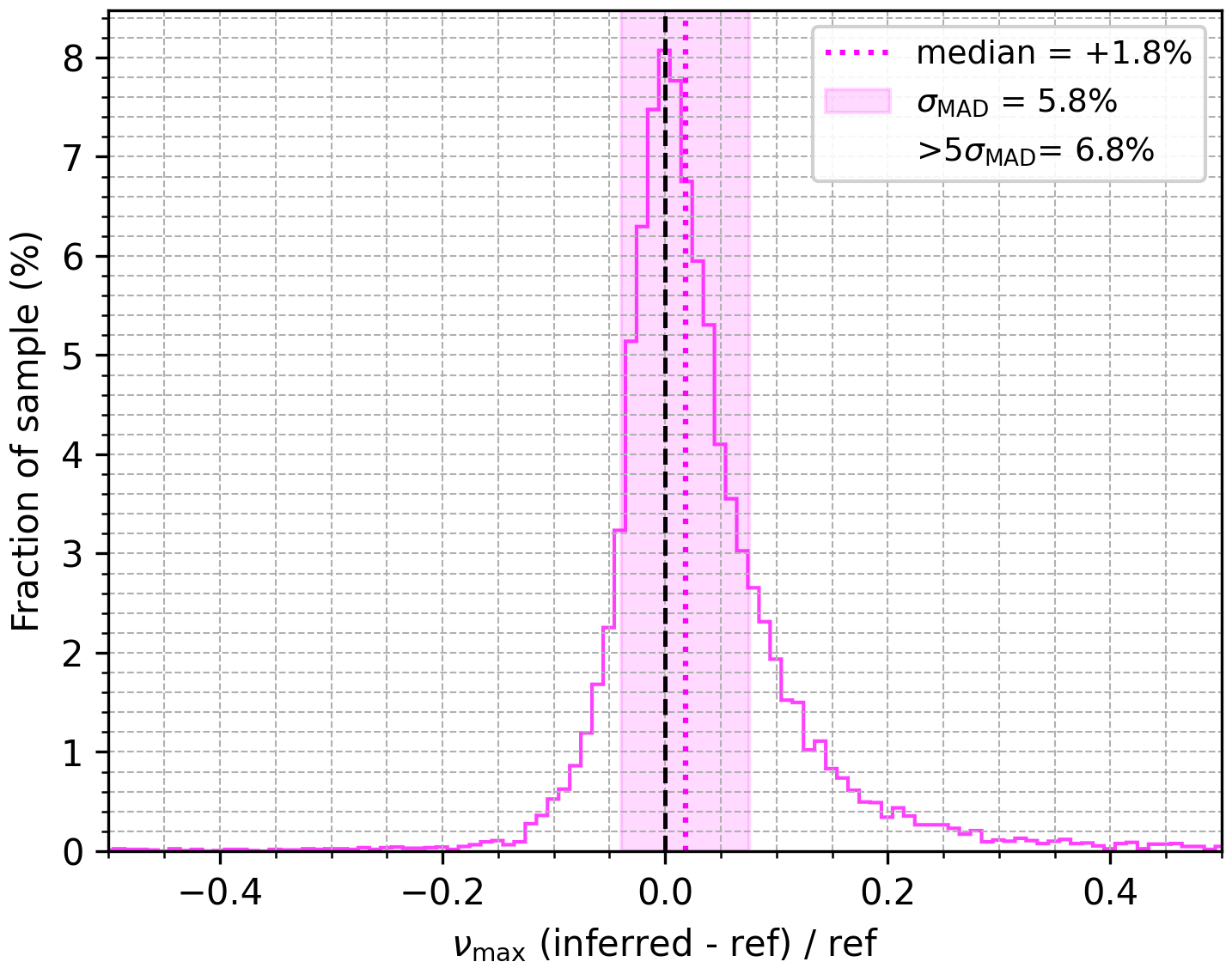}
    }
    \hfill
    \subfloat[\label{fig:EPIC_dnu_comparison_hist}]{
        \includegraphics[width=0.45\textwidth]{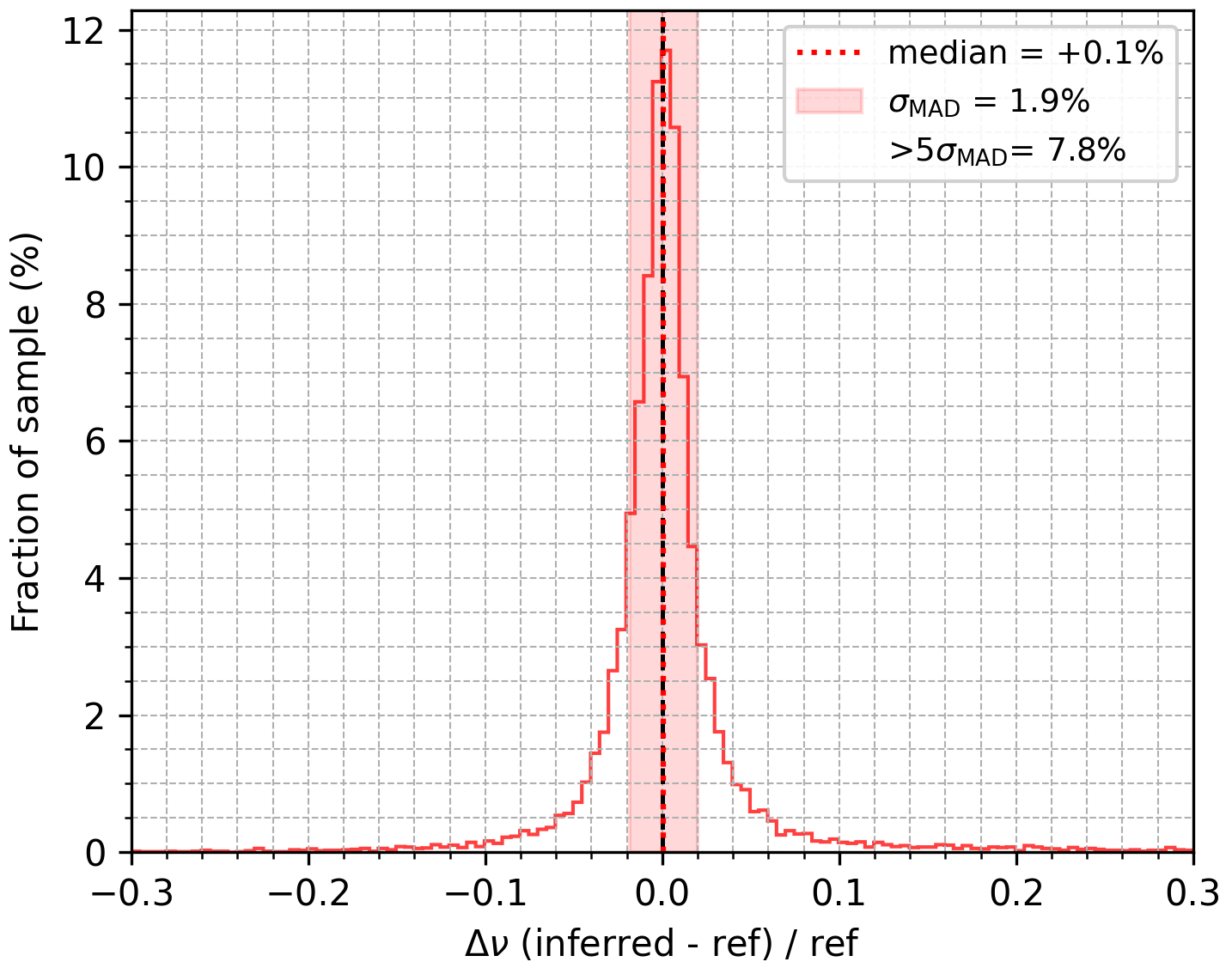}
    }
    \caption{Comparison of the model predictions from K2 observations with reference values from K2~GAP~DR3 \citep{k2_dr3}: (a) $\nu_{\mathrm{max}}$ and (b) $\Delta\nu$; (c--d) histograms of relative errors in $\nu_{\mathrm{max}}$ and $\Delta\nu$, respectively. The approximately unscaled K2~GAP~DR3 values are used, so as not to introduce a bias into the comparison.}
    \label{fig:EPIC_comparison}
\end{figure*}

Figures~\ref{fig:EPIC_numax_comparison_scatter} and~\ref{fig:EPIC_dnu_comparison_scatter} compare the inferred values with the approximately unscaled K2~GAP~DR3 values, and Figs.~\ref{fig:EPIC_numax_comparison_hist} and~\ref{fig:EPIC_dnu_comparison_hist} show the distributions of fractional differences. The robust dispersions are $\sigma_{\mathrm{MAD}} = 5.8\%$ for \numax\ and $1.9\%$ for \dnu, to be compared with $3.6\%$ and $1.2\%$ obtained against Kepler data of the same baseline in Sect.~\ref{sec:results_ekic}. Large discrepancies remain uncommon, with 93.8\% of \numax\ and 97.0\% of \dnu\ inferences agreeing to better than 35\%, but the tails are considerably heavier than in the Kepler comparison: 6.8\% and 7.8\% of the respective samples lie beyond $5\sigma_{\mathrm{MAD}}$, against 0.3\% and 4.1\% previously. A concentration of outliers is apparent near $\nu_{\mathrm{max}} \sim 50~\mu$Hz and $\Delta\nu \sim 5~\mu$Hz, coinciding with the known K2 systematics introduced by spacecraft thruster firings; these occur approximately every six hours and produce spurious peaks near harmonics of $47.19~\mu$Hz in the power spectrum, which can be misidentified as oscillation signal \citep[see][]{K2Handbook}. Neither our synthetic training spectra nor the Kepler validation data contain such features, so the models have no basis for accommodating them. Some scatter is also plausibly associated with the different photometric pipelines, K2~GAP~DR3 being based on EVEREST light curves and this work on K2SFF.

The \dnu inferences show no systematic offset, with a median fractional difference of $+0.1\%$. For \numax\ the median fractional difference is $+1.8\%$, or $0.40\sigma$ in normalized terms. Since the two parameters are inferred from the same spectra by the same network, an offset confined to \numax\ is unlikely to originate in the photometry or in the inference, and is instead comparable in magnitude to the $1.7\%$ calibration factor reversed above; we therefore attribute it principally to the approximate nature of that reversal. The empirical $1\sigma$ fractions are 0.567 for \numax\ and 0.650 for \dnu, so the combined uncertainties do not fully account for the observed differences. Part of this reflects the offset just described, and part the outlier population associated with K2 systematics. 

Overall, these results demonstrate that the models reliably infer $\nu_{\mathrm{max}}$ and $\Delta\nu$ from single-campaign K2 light curves, the difference being dominated by K2-specific instrumental artefacts rather than by any loss of performance intrinsic to the method.

\subsubsection{New \dpi inferences for K2 red-giants}
Having validated both models on Kepler data at K2-like resolution, we now apply the models to the 2,059 K2 red giants with $\Delta\nu$ in the range 9--19~$\mu$Hz. Applying the same selection criteria of Sect.~\ref{sec:results_ekic_dpi} yields a final sample of 232 stars, a fraction of 11.3\%. This lies between the 6.8\% obtained on the synthetic test set, which samples the $\Delta\nu$--$\Delta\Pi_{1}$ plane uniformly and therefore includes regions in which the mixed-mode pattern is intrinsically unresolvable at this baseline, and the 21.8\% obtained in the comparison with \citet{Gang_Li_2024}, whose reference catalogue is itself restricted to stars with measurable period spacings.

\begin{figure*}
    \includegraphics[width=\textwidth]{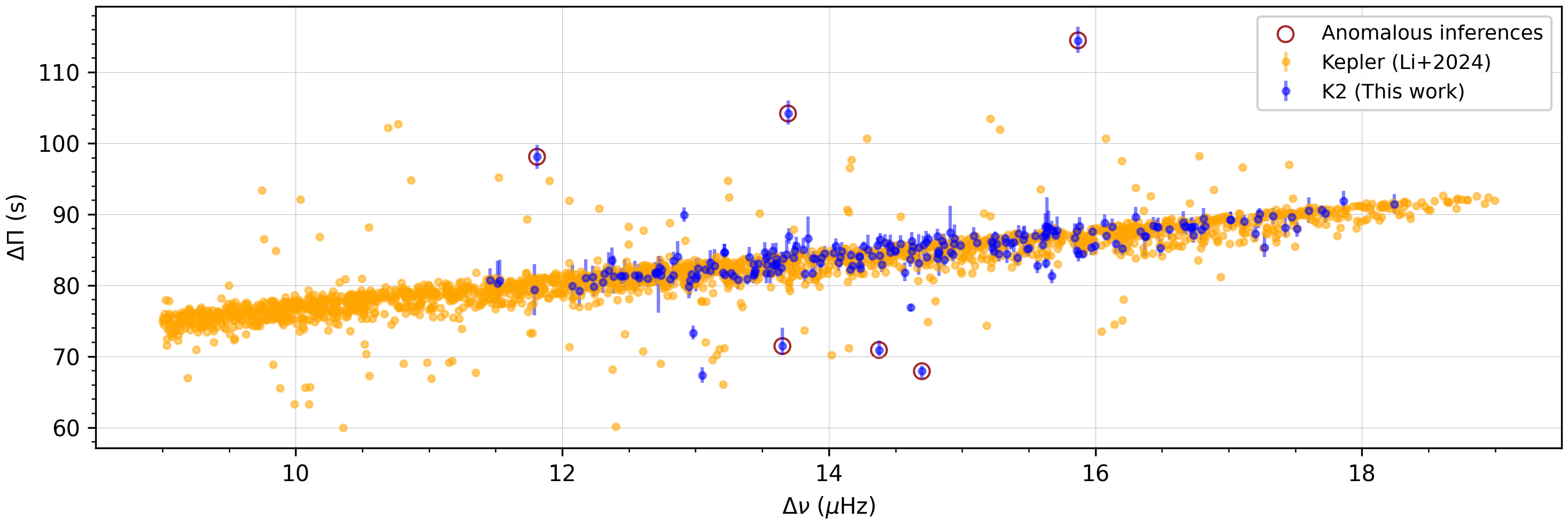}
    \caption{$\Delta\nu$--$\Delta\Pi_{1}$ diagram for the 232 K2 red giants of this work (blue), plotted over the Kepler red giants of \citet{Gang_Li_2024} (lime). Stars flagged as anomalous inferences are marked with open symbols.}
    \label{fig:k2_dnu_dpi_relation}
\end{figure*}

The synthetic training set spans the $\Delta\nu$--$\Delta\Pi_{1}$ plane uniformly and carries no imprint of the observed degenerate sequence, so the networks have no means of preferring on-sequence solutions. Figure~\ref{fig:k2_dnu_dpi_relation} shows the inferred distribution for these 232 stars, overplotted on the Kepler red giants of \citet{Gang_Li_2024}. The K2 inferences trace the same sequence as the Kepler sample. Measured against a fit to the \citet{Gang_Li_2024} sequence, our values are offset by $+0.25 \pm 0.09$~s with a scatter of $1.13$~s, to be compared with a median inferred uncertainty of $1.15$~s (approximately 1.3\%). The offset amounts to $0.3\%$ in $\Delta\Pi_{1}$ and is smaller than the uncertainty on an individual star, and the scatter is consistent with the quoted uncertainties alone. The two scales therefore agree to within the precision of the inferences, despite coming from different missions, different data reduction and entirely different measurement methods. The alignment is a property of the data rather than of the training distribution, and provides a constraint on the absolute $\Delta\Pi_{1}$ scale that the synthetic test set cannot, since the latter shares the forward model used in training.

Nine of the 232 stars lie away from the degenerate sequence. Six are classified as anomalous inferences under the criteria of Sect.~\ref{sec:results_ekic_dpi}, and inspection of their $\Delta\Pi_{1}$ output distributions confirms the expected structure, with a weaker secondary peak coinciding with the degenerate sequence. These are excluded from consideration as candidates for stars lying below the sequence, since a secondary peak on the sequence makes the primary displacement inseparable from a method artefact. The remaining three show well-defined peaks displaced from the relation, and are good candidates for a detailed follow-up. All 232 inferences, including the nine discussed here, are reported in Table~\ref{tab:k2_red_giants_dpi} with anomalous cases flagged.

These measurements constitute the first ensemble set of $\Delta\Pi_{1}$ determinations for K2 red giants, obtained from single-campaign photometry with a median fractional uncertainty of $1.3\%$, and extend ensemble-level core-structure diagnostics beyond the Kepler field to the ecliptic populations sampled by K2.

\section{Summary and Conclusions}
\label{sec:summary}
We have developed a deep-learning-based method to infer key asteroseismic parameters from short-duration, K2-length photometric observations. The model takes a normalized power spectral density as input and outputs a full probability distribution for each parameter, rather than a single point value, learning the relevant features directly from the data without background fitting or manual mode identification. A central advantage of this approach is that each inference yields a percentile-based uncertainty alongside the median as the point prediction. We have assessed these uncertainties throughout this work by comparing the empirical $1\sigma$ fraction against the nominal value expected for well-calibrated Gaussian errors. The model consists of two specialised networks sharing the same architecture: one dedicated to $\nu_{\mathrm{max}}$ and $\Delta\nu$ across the full range of evolutionary states, and a second, restricted to the young-red-giant regime, dedicated to the period spacing $\Delta\Pi_{1}$ and the coupling factor $q$.

We validate the model in three stages.

\begin{itemize}
\item \textit{A held-out synthetic test set} (Appendix~\ref{sec:results_on_synthetics}). The selection criteria retain 96.1\% of stars for $\nu_{\mathrm{max}}$, 93.2\% for $\Delta\nu$, and 6.8\% for $\Delta\Pi_{1}$. The recovered values scatter around the true ones with robust dispersions of 2.9\%, 0.5\%, and 1.0\%, and the reported uncertainties are well calibrated, with empirical $1\sigma$ fractions of 0.681, 0.720, and 0.827. The last of these indicates $\Delta\Pi_{1}$ uncertainties wider than the residuals warrant; these are conservative by roughly 40\%.

\item \textit{Kepler data degraded to K2-like resolution} (Sect.~\ref{sec:results_ekic}), which isolates the effect of the shorter baseline from any instrumental difference between the two missions. Of the 15,639 stars in common with \citet{Yu_2018}, 96.3\% satisfy the $\nu_{\mathrm{max}}$ criterion and 91.0\% the $\Delta\nu$ criterion, with dispersions of 3.6\% and 1.2\%, small negative biases of $-0.7\%$ and $-0.2\%$, and empirical $1\sigma$ fractions of 0.698 and 0.658 -- close to the nominal 0.683 expected for well-calibrated uncertainties. For $\Delta\Pi_{1}$, compared against \citet{Gang_Li_2024}, 21.8\% of the 2,070 common stars pass the selection criteria; their period spacings agree with the reference with a robust dispersion of 0.9\%, only 2.2\% of the sample lying beyond $5\sigma_{\mathrm{MAD}}$, at an empirical $1\sigma$ fraction of 0.878.

\item \textit{The K2 observations themselves}, compared with the K2~GAP~DR3 catalogue \citep{k2_dr3} (Sect.~\ref{sec:results_k2gap}). Of the 18,704 stars in common, 90.6\% satisfy the \numax criterion and 78.5\% the \dnu criterion, with dispersions of 5.8\% and 1.9\%. The larger scatter relative to the Kepler-as-K2 comparison, and the correspondingly lower empirical $1\sigma$ fractions of 0.567 and 0.650, are attributable to K2-specific instrumental artefacts rather than to any loss of the method's intrinsic performance.

\end{itemize}

Applying the model to 2,059 K2 red giants with $\Delta\nu > 9~\mu\mathrm{Hz}$, we obtain confident $\Delta\Pi_{1}$ inferences for 232 stars with a median fractional uncertainty of 1.3\%, an 11.3\% yield. Of these, 223 lie along the well-established $\Delta\nu$--$\Delta\Pi_{1}$ degenerate sequence; six of the remaining nine are flagged as anomalous, with secondary peaks in their $\Delta\Pi_{1}$ probability distributions consistent with the degenerate sequence, while the other three show clean, unambiguous predictions offset from the trend and are good candidates for follow-up. These 232 measurements, reported in Table~\ref{tab:k2_red_giants_dpi} (Appendix~\ref{appendix:k2_red_giants_dpi}), constitute the first ensemble set of $\Delta\Pi_{1}$ determinations for K2 red giants, obtained from only a single campaign of data, and extend ensemble-level core-structure diagnostics beyond the Kepler field to the ecliptic populations sampled by K2.

\section*{Data availability}
The full version of Table~\ref{tab:k2_red_giants_dpi}, listing the inferred $\nu_{\mathrm{max}}$, $\Delta\nu$, and $\Delta\Pi_{1}$ with their asymmetric uncertainties and the anomalous-inference flag for all 232 K2 red giants, is available in machine-readable form at the CDS. The K2 light curves used in this work are publicly available from MAST, and the Kepler data underlying the validation in Sect.~\ref{sec:results_ekic} are likewise public.

\begin{acknowledgements}
We acknowledge support from the Department of Atomic Energy, Government of India, under Project Identification No. RTI 4002 and Google Research India for providing credits to be used in computational resources. NG would like to thank Shatanik Bhattacharya (TIFR) for all the helpful discussions and Meenakshi Gaira (TIFR) for her help with MCMC analysis. This paper includes data collected by the Kepler and K2 missions and obtained from the MAST data archive at the Space Telescope Science Institute (STScI). Funding to US Institutions for the Kepler mission was provided by the NASA Science Mission Directorate. STScI is operated by the Association of Universities for Research in Astronomy, Inc., under NASA contract NAS 5-26555. This research made use of Lightkurve, a Python package for Kepler, K2 and TESS data analysis \citep{Lightkurve_2018}. This research was supported in part by a generous donation (from the Murty Trust) aimed at enabling advances in astrophysics through the use of machine learning. Murty Trust, an initiative of the Murty Foundation, is a not-for-profit organisation dedicated to the preservation and celebration of culture, science, and knowledge systems born out of India. The Murty Trust is headed by Mrs. Sudha Murty and Mr. Rohan Murty.

We also acknowledge the use of Claude, developed by Anthropic, for assistance in language editing during the preparation of this manuscript.

\textit{Software}: NumPy \citep{harris2020array}, Lightkurve \citep{Lightkurve_2018}, TensorFlow \citep{tensorflow2015-whitepaper}, Pandas \citep{reback_2022_6702671}.
\end{acknowledgements}

\bibliographystyle{aa}
\bibliography{ref}

\appendix
\nolinenumbers
\section{Generating synthetics}
\label{appendix:generating_synthetics}
The detailed formulation for generating synthetic spectra can be found in \cite{dhanpal2022}, which is implemented in the Spectra Simulator code developed by Dr. Othman Benomar \cite{othman_benomar_2023_spectra_simulator}. However, while \cite{dhanpal2022} assumes that all oscillation modes are resolved, we have modified the simulator to account for unresolved modes, improving its applicability to our target systems. Additionally, we adopt a different formalism for computing the heights and widths of mixed modes, following the approaches outlined in \cite{Grosjean_2014A&A...572A..11G} and \cite{Mosser_2018}. Furthermore, we are passing all the small spacing values separately, namely $d_{01}, d_{02}~\&~d_{03}$, unlike \cite{dhanpal2022} where these values are calculated internally as some fraction of a single parameter $d_{0,\ell}$. For the sake of continuity, we provide below a brief summary of the overall methodology for generating synthetic spectra, while also highlighting the key differences introduced in our implementation.

\subsection{Mode frequencies}
The asymptotic theory of oscillations of p modes in red giants may be expressed as \citep{Mosser_2012_evolution}, 
\begin{equation}
\label{eq:rg_universal_p_pattern}
    \frac{\nu_{n_{p},l}}{\Delta\nu} =  n_{p} + \frac{l}{2} + \epsilon_{p} - d_{0l}(\Delta\nu) +\frac{\alpha_{l}}{2}\left[n_{p} - \frac{\nu_{\text{max}}}{\Delta\nu}\right]^2,
\end{equation}
where $\Delta\nu$ is the large separation, $\nu_{\text{max}}$ is the frequency at maximum power, $n_{p}$ is the p-mode radial order, $\ell$ is the angular degree, $\epsilon_{p}$ is the phase offset, $d_{0\ell}$ is the small frequency separation and $\alpha_{\ell}$ term accounts for the curvature of p-modes.

We calculate $\nu_{\text{max}}$ based on $\Delta\nu$ using the relation for solar-like oscillators given in \cite{Stello_2009} as given below and add a 10\% spread on top (see table \ref{tab:parameter_ranges}):
\begin{equation}
    \nu_{\rm{max,scaled}}=(\Delta\nu/0.263)^{1/0.77}
\end{equation}
For $\delta_{02}~\&~\delta_{03}$ we use the $\delta_{0\ell}-\Delta\nu$ relation for Kepler red giants based on \cite{Huber_2010} and add a 20\% spread on top (see table \ref{tab:parameter_ranges}):
\begin{equation}
    \delta_{02, \rm{scaled}} =  0.121~\Delta\nu+0.047
\end{equation} 
\begin{equation}
    \delta_{03,\rm{scaled}} = 0.282~\Delta\nu+0.16
\end{equation}
Meanwhile for $\delta_{01}$ we take uniform values from (-0.6 to 0.6) based on the same study. Note that we need to divide these values of small spacings by $\Delta\nu$ to account for the difference in the definition of $\delta_{0\ell}$ in \cite{Huber_2010} and in equation \ref{eq:rg_universal_p_pattern} (see table \ref{tab:parameter_ranges}).
Mixed-mode frequencies for red giants are given by an implicit equation \citep{Mosser_and_Belkacem_I_2015},
\begin{equation}
\label{eq:mixed_mode_implicit}
\mathrm{tan}~\pi\frac{\nu-\nu_{p}}{\Delta\nu} = q~\mathrm{tan}\frac{\pi}{\Delta\Pi_{1}}\left(\frac{1}{\nu} - \frac{1}{\nu_{g}}\right),
\end{equation}
where $\nu_{g}$ is the asymptotic frequency of pure \textit{g} modes. 
For dipole modes $\nu_{g} = 1/(-n_{g}+\epsilon_{g})~\Delta\Pi_{1}$, where $n_{g}$ is the radial order and $\epsilon_{g}$ is the offset parameter for \textit{g} modes.

We solve equation \ref{eq:rg_universal_p_pattern} to determine frequencies of $\ell$=0, 2 and 3 modes, which are treated as pure pressure modes in our synthetics. $\ell$=2 and 3 are in principle mixed modes, however, as the coupling between g-mode and p-mode cavity for these modes is very weak, only the p-dominated modes are visible. To determine the frequencies of dipole mixed modes, we take solutions of the implicit equation \ref{eq:mixed_mode_implicit} in a range of 1.2 times $\Delta\nu$ for each pure $\ell$ = 1 p mode.

\subsection{Rotational splittings}
Rotation breaks the spherical symmetry of the star and lifts the degeneracy in $m$, splitting each mode of degree $\ell$ into $2\ell +1$ azimuthal components. The frequency of each of these components is given by $\nu_{n,\ell,m} = \nu_{n,\ell}+\delta\nu_{n,\ell}$, where $\delta\nu_{n,\ell}$ is the rotational splitting. For p-modes in solar-like stars, the dependence of rotational splitting on $(n,\ell)$ is weak within the observed frequency range \citep{Lund_2014}. Rotational splittings for p-modes may be approximated as,
\begin{equation}
    \nu_{n,\ell,m} = \nu_{n,\ell}-m~\delta\nu_{n,\ell}~~,
\end{equation}
where $\delta\nu_{n,\ell}=\Omega/2\pi$ is a function only of average internal rotation rate $\Omega$. Furthermore, owing to the large envelopes in red giants and the high sensitivity of p-modes rotational kernels to the outer layers of the star, it is common to approximate average rotation as $\Omega\simeq\Omega_{env}$ and hence $\delta\nu_{n,l}\simeq\Omega_{env}/2\pi$ \citep{Goupil_2013}.

\indent Mixed modes on the other hand are influenced by both the core and the envelope. For $\ell=1$ mixed modes in red giants and early subgiants, a two-zone model of rotation can be used to estimate the rotational splittings as shown by \cite{Goupil_2013}. Moreover, they also demonstrate that the contribution from core and envelope to the rotational splitting depends upon the ratio $\zeta(\nu)$ of kinetic energy of the modes in g-cavity and the total kinetic energy of modes as,
\begin{equation}
\label{eq:rotation_splitting_goupil}
    \delta\nu_{rot} = -\frac{1}{2}\frac{\Omega_{core}}{2\pi}\zeta(\nu)+\frac{\Omega_{env}}{2\pi}(1-\zeta(\nu)).
\end{equation}
Additionally, $\zeta(\nu)$ is well approximated by the properties of pure p and g modes as \citep{Deheuvels_2015,Mosser_and_Belkacem_I_2015},
\begin{equation}
    \zeta(\nu) = \left[1+\frac{1}{q}\frac{\nu^{2}\Delta\Pi_{1}}{\Delta\nu}\frac{cos^{2}\pi\frac{1}{\Delta\Pi_{1}}\left(\frac{1}{\nu}-\frac{1}{\nu_{g}}\right)}{cos^2\pi\frac{\nu-\nu_{p}}{\Delta\nu}}   \right]^{-1}.
\end{equation}

However, this first-order description of rotation does not account for near-degeneracy effects (NDE), which arise when frequency separation between consecutive mixed modes becomes comparable or lower than the rotational splitting \cite{Deheuvels_2017}. NDE leads to avoided crossings of each multiplet component at different age, causing asymmetries in the rotational splittings. There are two formalisms to account for these effects, one is based on decomposing mixed modes in the basis of pure pressure modes ($\pi$ modes) and pure gravity modes ($\gamma$ modes) \citep{Ong_Basu_2020} in which the rotational operator is diagonal \cite{ong2021MixedModesAsteroseismic} and one can solve the eigenvalue problem to get the frequencies \cite{Ong_2022}. The second approach, discussed in \cite{Gang_Li_2024}, applies rotational perturbations to the asymptotic frequencies of pure p- and g-modes and then solve equation \ref{eq:mixed_mode_implicit} separately for each multiplet. Furthermore, authors argue that this approach is similar to the one proposed by \cite{Ong_2022} and adequately accounts for NDE. Thus, for stars showing non-negligible NDE the first-order description of rotation is inadequate and could lead to potential biases in inferences of period spacings and coupling factor. However, \cite{Gang_Li_2024} noted that they found consistent results for $\Delta\Pi_{1}$, $q$, and other parameters using these two approaches, with significant differences appearing only in the inference of the envelope rotation rate, which we do not attempt to infer here. Also, we compared $\Delta\Pi_{1}$ inferences from our model with \cite{Gang_Li_2024} in section \ref{sec:results_ekic_dpi}, where the period spacings agree with a robust dispersion of 0.9\% for stars satisfying the quality thresholds, which confirms that the first-order rotational formalism is generally sufficient to obtain accurate $\Delta\Pi_{1}$ inferences, at least for stars satisfying quality thresholds discussed in section \ref{sec:results_ekic_dpi}. 

Finally, we found the inclusion of this first-order rotational description to be very effective in comparison to not including any rotation at all, as is shown in appendix \ref{appendix:effects_of_rot_norot}.

\subsection{Heights and Widths of modes}
Each oscillation mode is modelled as a Lorentzian centred at $\nu(n,\ell,m)$ with height $H(n,\ell,m)$ and linewidth $\Gamma(n,\ell,m)$. To obtain realistic mode heights and widths for $\ell = 0, 2$ and $3$ \textit{p} modes, we adopt the heights and linewidths of radial modes derived from templates of red giants and sub-giants observed by Kepler (KIC 10147635, KIC 11414712, KIC 12508433, KIC 6144777, KIC 8026226, KIC 11026764, KIC 11771760, KIC 2437976 and KIC 6370489), rescaling them based on $\nu_{\text{max}}$ and $\Delta\nu$, and applying the appropriate geometrical factors for the $\ell =2~\&~3$ modes, following a technique similar to \cite{Kamiaka_2018} (see also \cite{dhanpal2022} for an example). For the dipole mixed modes, we compute amplitudes and linewidths by scaling the radial modes using the ratio of kinetic energy trapped in the g-mode cavity to the total mode kinetic energy, $\zeta(\nu)$. We use the formalism highlighted in \cite{Mosser_2018}; specifically, \cite{Benomar_2014_APJ} estimate the width of mixed modes as $\Gamma_{1}(\nu)= \Gamma_{0}(1-\zeta(\nu))$, and \cite{Belkacem_2015} have shown that the amplitude for resolved dipole mixed modes is $A_{1}^{2}(\nu) = A_{0}^{2}(1-\zeta(\nu))$. As noted in \cite{Mosser_2018}, when the geometrical factor is omitted, these amplitudes correspond to similar heights for resolved dipolar mixed modes since $A^2=\pi\Gamma H/2$.

We treat any two modes with a frequency separation smaller than $\delta f_{\rm{res}}$ as duplicates and retain only one of them in the final calculations, where $\delta f_{\rm{res}}$ is the frequency resolution of the PSD, defined as the inverse of the light curve’s total observation time. Further, we set a minimum observable linewidth ($\Gamma_{min}=\frac{2~\delta f_{\rm{res}}}{\pi}$) for all the modes i.e $\Gamma_n=\rm{max(\Gamma_n, \Gamma_{min})}$. For all unresolved modes, i.e., when $\Gamma_n$ was less than $\Gamma_{min}$, we also apply a dilution factor to the heights \citep{Dupret_2009}, which expresses,
\begin{equation}
    \tilde{H}_n = \frac{\pi}{2~\delta f_{res}} \Gamma_n~H_n
\end{equation}
where $H_{n}$ and $\tilde{H}_n$ are the mode heights before and after applying the dilution factor, so that the total power of the mode is conserved.
Accounting for geometrical and visibility factor, the final heights $H(n,\ell,m)$ of modes are determined as
\begin{equation}
    H(n,\ell,m) = r^{2}_{\ell,m}(\iota)V(\ell)\tilde{H}_n,
\end{equation}
where $V(\ell)$ is the mode visibility, $\tilde{H}_n$ is the height for the mode of radial order $n$ (without geometrical and visibility factor) and $r^{2}_{\ell,m}(\iota)$ is the geometrical factor which depends on the inclination angle $\iota$, determined according to
\begin{equation}
    r^{2}_{\ell,m}(\iota)=\frac{(\ell-|m|)!}{(\ell+|m|)!}[P_{\ell}^{|m|}(cos\iota)]^2,
\end{equation}
where $P_\ell^{|m|}$ is the associated Legendre polynomial.

\subsection{Noise Model}
The noise model comprises two components, a Harvey-like profile generated by surface granulation and white noise which is frequency independent. At low frequencies, the granulation component dominates and at high frequencies, white noise is the primary contributor. We do not consider additional facular signatures in our synthetics.

The noise model is given by
\begin{equation}
    B(\nu) = \frac{H_{g}}{1+(\tau\nu)^{\mathrm{p}}} + N_{0},
\end{equation}
where the first term is the Harvey-like component which depends on the characteristic granulation amplitude $H_{g}$, the granulation timescale $\tau$, and power-law exponent $p$. The second term, $N_{0}$, represents white noise.

Furthermore, granulation amplitude and timescale have been observed to vary with $\nu_{\text{max}}$ \citep{Kallinger_2010,Mathur_2011,Chaplin_2011}. Hence, the granulation amplitude and timescale are modeled as,
\begin{equation}
    H_{g} = A_{g}\nu_{\text{max}}^{B_{g}}~+~C_{g}~,~~\tau = A_{\tau}\nu_{\text{max}}^{B_{\tau}}~+~C_{\tau},
\end{equation}
where ($A_{g},B_{g},C_{g}$) and ($A_{\tau},B_{\tau},C_{\tau}$) are free parameters which control the granulation amplitude and timescale respectively. 

\subsection{Power Spectrum Model}
The power spectrum model comprises two parts, the oscillation signal $S$ and noise profile $B$. The oscillation signal is a sum of Lorentzians centred around the respective mode frequencies, given by,
\begin{equation}
    S(\nu) = \sum_{n}\sum_{\ell=0}^{3}\sum_{m=-\ell}^{\ell}~\frac{H~(n,\ell, m)}{1+4\left(\frac{\nu-\nu_{n\ell m}}{\Gamma(n,\ell, m)}\right)^2}.
\end{equation}

Mode frequencies $\nu_{n\ell m}$, height $H(n,\ell,m)$ and width $\Gamma(n,\ell,m)$ of Lorentzian peaks are determined as explained previously.

For a given set of parameters, the power spectrum model is calculated by adding the signal and noise profile parts as,
\begin{equation}
    M(\nu) = S(\nu)~+~B(\nu).
\end{equation}
We then multiply a random realization of chi-squared noise with two degrees of freedom to the resulting model to obtain the synthetic PSD profile.

\section{Choosing bin size}\label{appendix:choosing_bin_size}
To determine the optimal bin sizes for our network architectures, we conducted convergence exercises by training multiple model instances with progressively finer bins. We compared performance across several diagnostics: Median absolute error (MAE), empirical 1$\sigma$ fraction, mean/median and standard deviation of normalized residuals, median predicted uncertainty, and the confident fraction (inferences satisfying the quality thresholds). Together, these metrics capture the effects of binning resolution on model accuracy, precision, calibration, and outlier behavior. We utilized MAE rather than root mean squared error because it is significantly more robust to extreme outliers, thereby more accurately reflecting the model's baseline performance across diverse bin configurations.

Strictly varying one parameter at a time across our entire multi-dimensional grid would constitute the ideal controlled experiment, but would require training approximately 60 separate models. Our convergence exercises therefore use a multi-variable sweep to map the parameter space efficiently, supplemented by targeted single-parameter control models for the most sensitive parameter -- \dpi -- to demonstrate the empirical robustness of this approach. We trained 10 instances of models with varying bin sizes for the output parameters. The models were trained, validated and tested on a set of 5 million young red giants synthetics (half of the synthetic dataset used for final model training, see section \ref{sec:training_set}), with 82.5\% used for training, 15\% for validation and 2.5\% held back for testing.

\begin{figure*}
    \centering
    \includegraphics[width=0.95\textwidth]{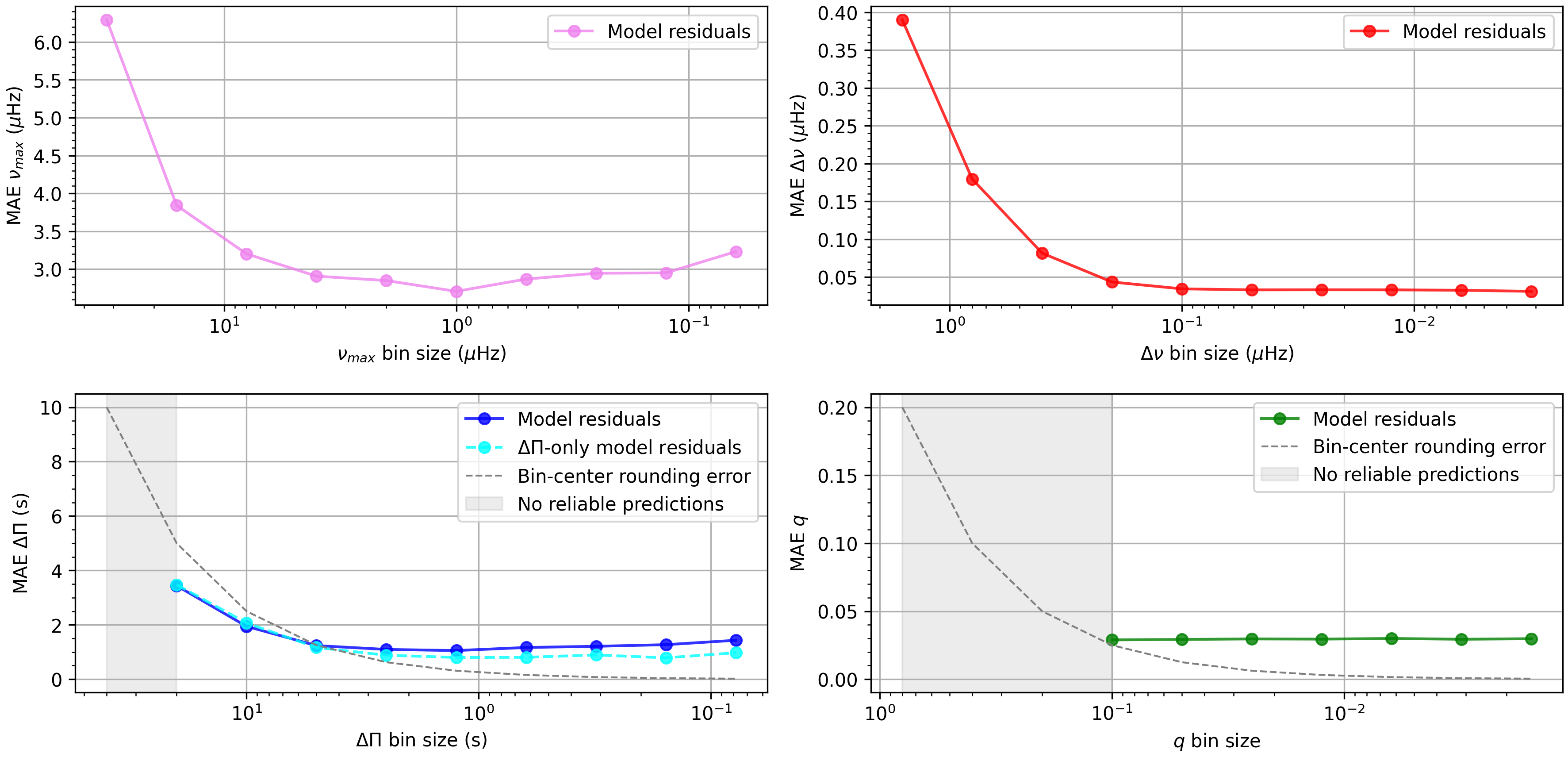}
    \caption{Median absolute error on the synthetic test set as a function of the output bin size, shown for \numax, \dnu, and \dpi. Solid curves correspond to the models which infer all parameters jointly; the dashed aqua curve shows a variant trained to infer $\Delta\Pi_{1}$ alone, with the full loss weighting assigned to that parameter. Final adopted bin sizes are 1~$\mu$Hz in \numax, 0.05~$\mu$Hz in \dnu, and 1.25~s in \dpi and 0.025 for \q.}
    \label{fig:sbins_mse_k2}
\end{figure*}

We use the same uncertainty-based quality cuts as described in section \ref{sec:results_ekic} to select confident inferences; however, we apply the selection cuts to each parameter separately and based only on the conditions for the corresponding parameter -- to minimize dependence on other parameters. Table \ref{tab:sbins_mse_k2} shows the MAE, the normalised residuals and the remaining metrics for 10 models trained with varying bin sizes and figure \ref{fig:sbins_mse_k2} shows the corresponding \textit{MAE} curves for the four output parameters. Again, all the metrics were calculated for the subset of stars satisfying the quality thresholds, out of a total of 79,499 synthetics.

For $\nu_{\text{max}}$ and $\Delta\nu$, MAE first decreases rapidly with decreasing bin sizes, quickly plateauing and becoming relatively constant for successively finer bins. This is the expected behaviour: the accuracy first increases with decreasing bin size and then plateaus for successively finer bins -- signifying the precision limit achievable with the ML model on the synthetic dataset at K2-like resolution. The "knee" occurs at around 4 $\mu$Hz for $\nu_{\text{max}}$ and 0.1 $\mu$Hz for $\Delta\nu$, see figure \ref{fig:sbins_mse_k2}. The same trend is observed in the empirical $1\sigma$ fraction and the median uncertainties, both of which first decrease from coarser to finer bin sizes -- signifying the increasing sharpness of the posteriors -- and then plateau (around 2~$\mu$Hz for $\nu_{\text{max}}$ and 0.05~$\mu$Hz for $\Delta\nu$). Finally, there is no prediction bias in the models, as reflected by the mean/median normalized residual values being close to zero. 

For $\nu_{\text{max}}$, we select 1~$\mu$Hz as the optimal bin size, as it has the lowest MAE, the highest yield, low median uncertainty and well calibrated uncertainties with a $1\sigma$ empirical fraction of 0.676 (very close to the ideal value of 0.683). For $\Delta\nu$, we select the optimal bin size of 0.05~$\mu$Hz, as it has the highest yield, low MAE, well calibrated uncertainties with a $1\sigma$ empirical fraction of 0.717, as well as the low number of outliers compared to models with finer bins (which have higher standard deviation for normalized residuals).

For $q$, for bins coarser than 0.1, none of the inferences satisfy the quality cut of a combined $q$ uncertainty below 0.12 (see table \ref{tab:sbins_mse_k2}). Furthermore, the MAE for $q$ has already saturated at bin size of 0.1 (see table \ref{tab:sbins_mse_k2}) with a value of 0.028, implying a median prediction error of only 0.028. As the high MAE values expected for the coarsest bins are missing due to selection cuts, we have plotted the grey curves for reference, which represent the error expected purely from the binning, i.e. the error that would remain even if the model were to guess the correct bin every time, assuming the stars are uniformly distributed over the full range (which is true here, see table \ref{tab:parameter_ranges} showing ranges and distribution for synthetics). Also, even at finer bins, we obtain significant normalized residuals mean and median values, upto an absolute value of 0.1. However, this corresponds only to a modest shift of approximately $\pm0.01$ in $q$ values. 

We choose an optimal bin size of 0.025 for $q$, which has the highest yield, $1\sigma$ empirical fraction of 0.694 -- close to ideal value 0.683. The median offset in normalized residuals is -0.06 which, corresponds to roughly -0.06$\sigma$ (or a shift of $\sim$0.006 in $q$ values), which is very small and an order of magnitude below the typical uncertainties.

For $\Delta\Pi_{1}$, we again observe a trend in MAE similar to that for $\nu_{\text{max}}$ and $\Delta\nu$ (see figure \ref{fig:sbins_mse_k2}), and the entry corresponding to the coarsest bin size is missing for the same reason as for \q. The "knee" in MAE occurs at a bin size of 2.5 seconds, after which it plateaus -- signifying the precision limit achievable with the model on synthetics at K2-like resolution. We even see a slight increasing trend at the finest bin sizes, and the yield for confident inferences also decreases progressively for bins finer than 1.25 seconds (table \ref{tab:sbins_mse_k2}), which implies a slight deterioration in model performance for finer bins. Additionally, we note that high values of standard deviation for normalized residual are caused by a few outliers, since the trimmed standard deviations (after removing the top and bottom 5\% of residuals) are much lower. Furthermore, the $1\sigma$ empirical fraction stays high -- signifying uncertainties wider than the residuals warrant -- at every bin width tested, and the median reported uncertainty does not decrease below the adopted bin width. The over-coverage in \dpi\ is therefore not set by the output discretisation, a point we return to in Sect.~\ref{sec:results_ekic_dpi}.

Furthermore, to test if $\Delta\Pi_{1}$ is optimally trained in these multi-output models and to confirm that the optimal bin size for $\Delta\Pi_{1}$ is not being affected by other output parameters, we trained models where $\Delta\Pi_{1}$ is accorded a weight of 100\% during training. The MAE for these models is similar to the multi-output models, as shown by the aqua curve in figure \ref{fig:sbins_mse_k2}, which means a similar level of precision is achieved in $\Delta\Pi_{1}$ predictions by multi-output models. This implies that $\Delta\Pi_{1}$ is optimally trained in these multi-output models. We also confirmed that the optimal bin size for $\Delta\Pi_{1}$ is the same for the $\Delta\Pi_{1}$-specific model and is therefore unaffected by the presence of additional output parameters in the multi-output model. However, the mean squared error for these $\Delta\Pi_{1}$-specific models turned out to be higher than multi-output models (80 vs 67 $\text{seconds}^2$ at bin size of 1.25 seconds), which, combined with the fact that they have similar MAE means a larger number of outliers. This highlights the importance of having multi-output models, where the model tries to constrain all these four parameters simultaneously, leading to fewer outliers in $\Delta\Pi_{1}$.

We choose an optimal bin size of 1.25 seconds for $\Delta\Pi_{1}$, since it has the lowest MAE, the highest yield, and the lowest median uncertainty. Also, we do not repeat this exercise for the model covering the full range (i.e. including evolved red giants), and adopt the same bin sizes for consistency.

\begin{table*}[p]
\caption{Performance metrics for 10 model instances with different bin sizes for its four output parameters evaluated on synthetic set of 79,499 young red giants. Here, \textit{MAE} denotes the median absolute error, \textit{conf.} the number of stars satisfying quality thresholds, \textit{1$\sigma$\_frac} the empirical 1$\sigma$ fraction, \textit{nr} the normalized residual, \textit{mdn} the median, \textit{std} the standard deviation, and \textit{std\_trim} the standard deviation after excluding the top and bottom 5\% of residuals.}
\label{tab:sbins_mse_k2}
\small

\begin{tabular*}{\textwidth}{@{\extracolsep{\fill}}lrrrrrrrr@{}}%
\toprule
 & \numax bin & \numax MAE & \numax conf. & \numax 1$\sigma$\_frac & \numax nr\_mean & \numax nr\_std & \numax nr\_mdn & \numax sig\_mdn \\
\midrule
1 & 32.0000 & 6.2904 & 57013 & 0.8612 & 0.0260 & 0.6800 & 0.0423 & 22.8384 \\
2 & 16.0000 & 3.8414 & 77922 & 0.8426 & 0.0044 & 0.7282 & 0.0028 & 17.0350 \\
3 & 8.0000 & 3.2022 & 79063 & 0.7696 & 0.0204 & 0.8719 & 0.0086 & 11.8656 \\
4 & 4.0000 & 2.9081 & 79044 & 0.7193 & 0.0133 & 0.9738 & 0.0030 & 10.2895 \\
5 & 2.0000 & 2.8502 & 79044 & 0.6834 & -0.0272 & 1.0457 & -0.0387 & 10.0025 \\
6 & 1.0000 & 2.7076 & 79076 & 0.6721 & 0.0149 & 1.1008 & -0.0003 & 9.8099 \\
7 & 0.5000 & 2.8698 & 79027 & 0.6688 & 0.0160 & 1.0782 & -0.0004 & 9.9232 \\
8 & 0.2500 & 2.9466 & 78974 & 0.6725 & -0.0065 & 1.0573 & -0.0245 & 9.8348 \\
9 & 0.1250 & 2.9518 & 78953 & 0.6689 & 0.0024 & 1.0702 & -0.0118 & 9.6479 \\
10 & 0.0625 & 3.2325 & 78846 & 0.6722 & 0.0143 & 1.0238 & -0.0052 & 9.8354 \\
\bottomrule
\end{tabular*}%

\vspace{1em}

\noindent

\begin{tabular*}{\textwidth}{@{\extracolsep{\fill}}lrrrrrrrr@{}}%
\toprule
 & \dnu bin & \dnu MAE & \dnu conf. & \dnu 1$\sigma$\_frac & \dnu nr\_mean & \dnu nr\_std & \dnu nr\_mdn & \dnu sig\_mdn \\
\midrule
1 & 1.6000 & 0.3899 & 59771 & 0.7444 & 0.0474 & 0.7962 & 0.0620 & 1.0883 \\
2 & 0.8000 & 0.1796 & 76907 & 0.8269 & 0.0282 & 0.7795 & 0.0368 & 0.5460 \\
3 & 0.4000 & 0.0817 & 77803 & 0.8611 & 0.0056 & 0.8229 & 0.0002 & 0.2821 \\
4 & 0.2000 & 0.0436 & 77663 & 0.8525 & 0.0240 & 0.9041 & 0.0164 & 0.1749 \\
5 & 0.1000 & 0.0346 & 77763 & 0.7804 & 0.0072 & 1.1109 & 0.0205 & 0.1369 \\
6 & 0.0500 & 0.0332 & 77739 & 0.7167 & -0.0134 & 1.2766 & 0.0023 & 0.1050 \\
7 & 0.0250 & 0.0334 & 77705 & 0.6928 & -0.0052 & 1.3541 & 0.0086 & 0.0986 \\
8 & 0.0125 & 0.0333 & 77623 & 0.6881 & 0.0024 & 1.3381 & 0.0175 & 0.0978 \\
9 & 0.0063 & 0.0327 & 77611 & 0.6888 & 0.0124 & 1.3075 & 0.0202 & 0.0983 \\
10 & 0.0031 & 0.0313 & 77841 & 0.6804 & 0.0055 & 1.4665 & 0.0069 & 0.0963 \\
\bottomrule
\end{tabular*}%

\vspace{1em}

\begin{tabular*}{\textwidth}{@{\extracolsep{\fill}}lrrrrrrrr@{}}%
\toprule
 & \q bin & \q MAE & \q conf. & \q 1$\sigma$\_frac & \q nr\_mean & \q nr\_std & \q nr\_mdn & \q sig\_mdn \\
\midrule
1 & 0.8000 & NaN & 0 & NaN & NaN & NaN & NaN & NaN \\
2 & 0.4000 & NaN & 0 & NaN & NaN & NaN & NaN & NaN \\
3 & 0.2000 & NaN & 0 & NaN & NaN & NaN & NaN & NaN \\
4 & 0.1000 & 0.0289 & 9176 & 0.6707 & 0.1878 & 0.9549 & 0.3130 & 0.0764 \\
5 & 0.0500 & 0.0293 & 9077 & 0.7234 & -0.0825 & 0.9395 & -0.0236 & 0.1020 \\
6 & 0.0250 & 0.0296 & 10258 & 0.6954 & -0.1049 & 0.9953 & -0.0664 & 0.0972 \\
7 & 0.0125 & 0.0295 & 10181 & 0.6976 & -0.0089 & 0.9838 & 0.0076 & 0.0971 \\
8 & 0.0063 & 0.0300 & 9603 & 0.7027 & -0.0842 & 0.9928 & -0.0638 & 0.0974 \\
9 & 0.0031 & 0.0294 & 9577 & 0.6949 & -0.0322 & 0.9886 & -0.0127 & 0.0977 \\
10 & 0.0016 & 0.0298 & 9603 & 0.6933 & -0.1189 & 1.0029 & -0.0814 & 0.0970 \\
\bottomrule
\end{tabular*}%

\vspace{1em}

\begin{tabular*}{\textwidth}{@{\extracolsep{\fill}}lrrrrrrrrr@{}}%

\toprule
 & \dpi bin & \dpi MAE & \dpi conf. & \dpi 1$\sigma$\_frac & \dpi nr\_mean & \dpi nr\_std & \dpi nr\_mdn & \dpi sig\_mdn & \dpi nr\_std\_trim \\
\midrule
1 & 40.0000 & NaN & 0 & NaN & NaN & NaN & NaN & NaN & NaN \\
2 & 20.0000 & 3.4386 & 734 & 0.8747 & 0.6086 & 1.0013 & 0.4899 & 13.8588 & 0.3564 \\
3 & 10.0000 & 1.9448 & 6051 & 0.8498 & 0.0132 & 1.5114 & -0.0326 & 7.4651 & 0.6458 \\
4 & 5.0000 & 1.2357 & 8633 & 0.8657 & 0.0843 & 1.6990 & 0.0005 & 6.7909 & 0.6062 \\
5 & 2.5000 & 1.0950 & 8514 & 0.8410 & 0.0528 & 2.1734 & -0.0192 & 5.4406 & 0.6429 \\
6 & 1.2500 & 1.0541 & 9144 & 0.8234 & 0.0790 & 2.4216 & -0.0033 & 4.9719 & 0.6612 \\
7 & 0.6250 & 1.1671 & 7841 & 0.8325 & 0.0684 & 1.9028 & 0.0095 & 5.6167 & 0.6582 \\
8 & 0.3125 & 1.2104 & 6737 & 0.8421 & 0.0469 & 1.7517 & -0.0352 & 5.8871 & 0.6374 \\
9 & 0.1562 & 1.2703 & 6673 & 0.8358 & 0.0012 & 1.6466 & -0.0408 & 6.1441 & 0.6467 \\
10 & 0.0781 & 1.4321 & 6026 & 0.8428 & 0.0522 & 1.5211 & -0.0246 & 6.7797 & 0.6486 \\
\bottomrule
\end{tabular*}%
\end{table*}

\section{Period Spacings Selection Cut}
\label{appendix:dpi_selection_cut}
To determine the selection threshold for confident $\Delta\Pi_{1}$ inferences, we evaluate the performance of the g-mode model on the Kepler data at K2-like resolution over a range of threshold values. The threshold, defined as the maximum permitted fractional uncertainty on the inferred $\Delta\Pi_{1}$, is varied from 1 to 30\% in steps of 1\%, and the resulting coverage, accuracy, and outlier behaviour are recorded. No cuts on $q$ or on any other model output are applied, so that this analysis is independent of the remaining selection criteria.

Here coverage denotes the fraction of the full test set satisfying the cut; accuracy the fraction of the selected subset whose $\Delta\Pi_{1}$ values agree with the reference to within 5\%; and outliers the complement of the latter. We additionally track the fraction of the selected subset classified as anomalous inferences under the criteria of Sect.~\ref{sec:results_ekic_dpi}. The difference between the outlier and anomalous fractions gives the proportion of discrepant inferences that the anomalous criterion fails to identify, and it is this quantity, rather than the outlier fraction itself, that limits the usable threshold. A suitable threshold maximises coverage while maintaining high accuracy and a low fraction of unidentified outliers.

\begin{figure*}
    \centering
    \includegraphics[width=0.95\textwidth]{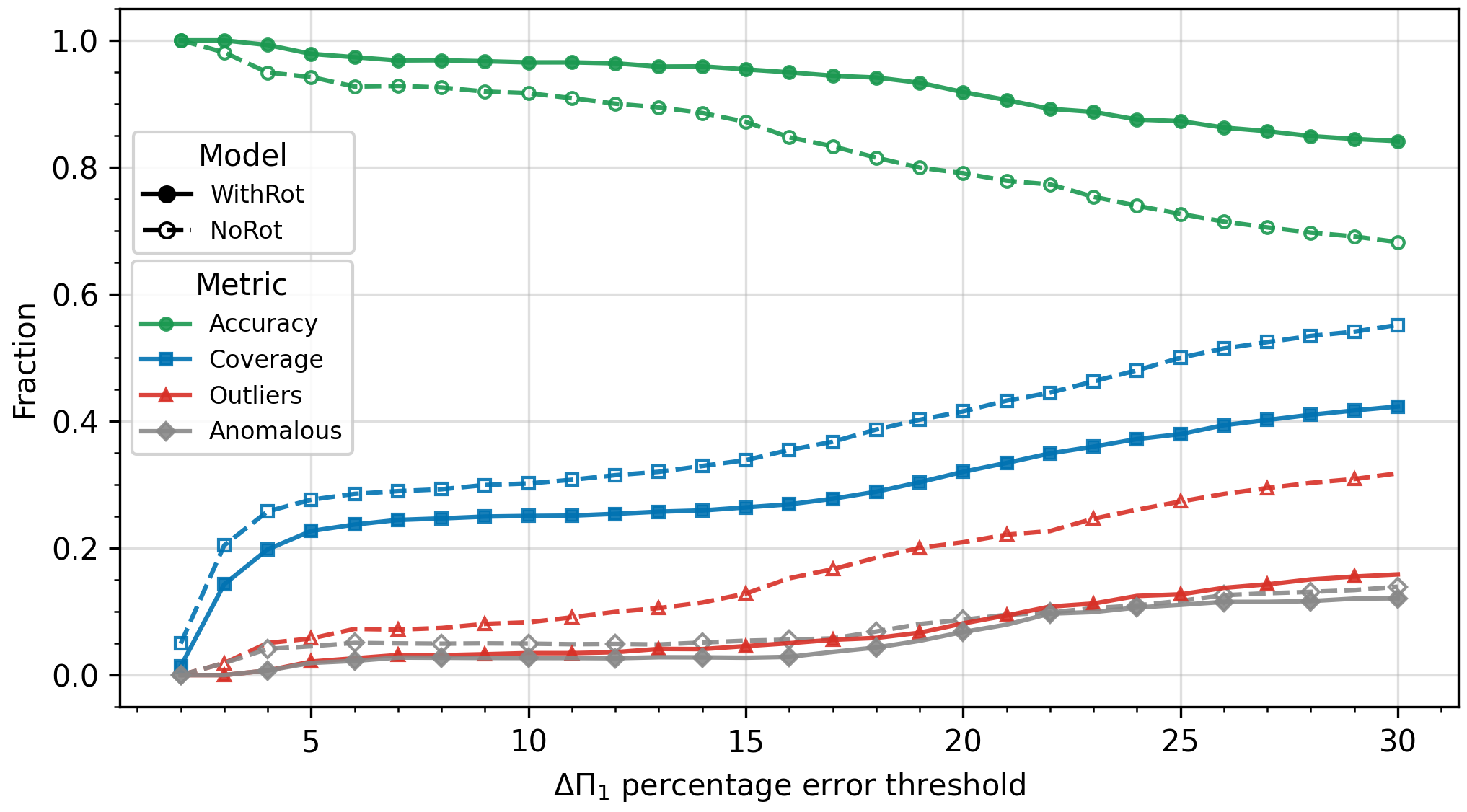}
    \caption{Coverage, accuracy, outlier fraction, and anomalous fraction as functions of the threshold on the fractional $\Delta\Pi_{1}$ uncertainty, evaluated on the $\sim$2,000 Kepler stars at K2-like resolution in common with \citet{Gang_Li_2024}. Coverage is the fraction of the full sample satisfying the threshold; accuracy, outlier fraction, and anomalous fraction are all computed within the selected subset. Solid curves and filled symbols show the model trained with rotational splittings included (the model adopted in this work); dashed curves and open symbols show the model trained without them.}
    \label{fig:dpi_cut_accuracy_coverage_outliers}
\end{figure*}

The solid curves and closed symbols in Figure~\ref{fig:dpi_cut_accuracy_coverage_outliers} show these quantities for the 2,070 stars with $\Delta\Pi_{1}$ values reported by \citet{Gang_Li_2024} (Sect.~\ref{sec:results_ekic_dpi}). Coverage rises steeply to 5\% and flattens by 10\%, beyond which further gains are offset by a decline in accuracy and, more importantly, by a growing population of discrepant inferences that the anomalous criterion does not flag. We therefore adopt a threshold of 10\% on the fractional $\Delta\Pi_{1}$ uncertainty. The dashed curves and open symbols in the same figure show the corresponding performance of the \texttt{g-mode no-rot model}, trained without rotation, discussed in Appendix~\ref{appendix:effects_of_rot_norot}.

\section{Effectiveness of first-order rotation formalism}
\label{appendix:effects_of_rot_norot}
To verify the effectiveness of our currently implemented first-order rotation formalism (see Appendix~\ref{appendix:generating_synthetics}), we created a synthetic dataset with core and envelope rotation rates set to zero, while keeping all other parameter ranges the same as in Table~\ref{tab:parameter_ranges}. We trained an instance of the g-mode model, termed the \texttt{g-mode no-rot model}, on this dataset to compare its performance with the g-mode model trained on synthetics with rotation included. We use the Kepler-as-K2 test set (see Sect.~\ref{sec:results_ekic_dpi}) to assess model performance on observations directly, evaluating accuracy, coverage, outlier, and anomalous fraction as in Appendix~\ref{appendix:dpi_selection_cut}, to allow a full comparison between the two. The dashed curves and open symbols in Fig.~\ref{fig:dpi_cut_accuracy_coverage_outliers} show these values for the \texttt{g-mode no-rot model} at different $\Delta\Pi_{1}$ percentage-error thresholds: compared to the g-mode model, its accuracy drops rapidly with increasing threshold, and the outlier fraction rises correspondingly. Furthermore, the number of unidentified outliers -- the difference between the outlier and anomalous fractions -- is non-negligible even at very low uncertainties and grows substantial at higher thresholds. All the gains in coverage from the \texttt{g-mode no-rot model} are therefore offset by its steep decline in accuracy.

The poor performance of the \texttt{g-mode no-rot model} relative to the g-mode model is likely attributable to the mismatch between the synthetic dataset, which has no rotational splittings, and real observations, which do show such splittings. This indicates that including even a first-order description of rotation -- despite it not accounting for near-degeneracy effects -- helps the models infer $\Delta\Pi_{1}$ more accurately. Nonetheless, the great majority of the common confident, non-anomalous inferences from the two models agree with each other, indicating that for most stars in the confident set, adding the first-order rotation formalism does not change the $\Delta\Pi_{1}$ inference.

The comparison of the g-mode model's $\Delta\Pi_{1}$ inferences against \citet{Gang_Li_2024}, who employed a rotational formalism that accounts for near-degeneracy effects, is presented in Sect.~\ref{sec:results_ekic_dpi} (Fig.~\ref{fig:eKIC_dpi_comparison_scatter}). We repeated the same comparison for the \texttt{g-mode no-rot model} (not shown here for brevity) and find that it too recovers $\Delta\Pi_{1}$ well, giving values consistent with the g-mode model for the majority of stars, though with a higher fraction of outliers. This further highlights the importance of incorporating rotation -- even at the level of a first-order approximation -- into the synthetic training set, in order to reduce outliers and obtain more reliable $\Delta\Pi_{1}$ estimates.

\section{Model performance on synthetics}
\label{sec:results_on_synthetics}
Both networks are trained on synthetic power spectra (Appendix~\ref{sec:training_set}), so we first characterise their performance on a held-out synthetic test set. This test benchmarks the model: it measures how well each network recovers parameters under the same generative model that produced its training data, and therefore isolates the intrinsic difficulty of the inference from any mismatch between the synthetics and real observations. The latter is addressed by the comparisons with observed data in Sect.~\ref{sec:results_ekic} and~\ref{sec:results_k2gap}. Because the true parameters are known exactly here, the reference uncertainty vanishes and the fraction of stars agreeing within the combined uncertainties coincides identically with the empirical $1\sigma$ fraction. We apply the same selection criteria as in the observational benchmarks throughout.

\begin{figure*}
\centering
    \subfloat[\label{fig:synthetics_numax_comparison_scatter}]{
        \includegraphics[width=0.48\textwidth]{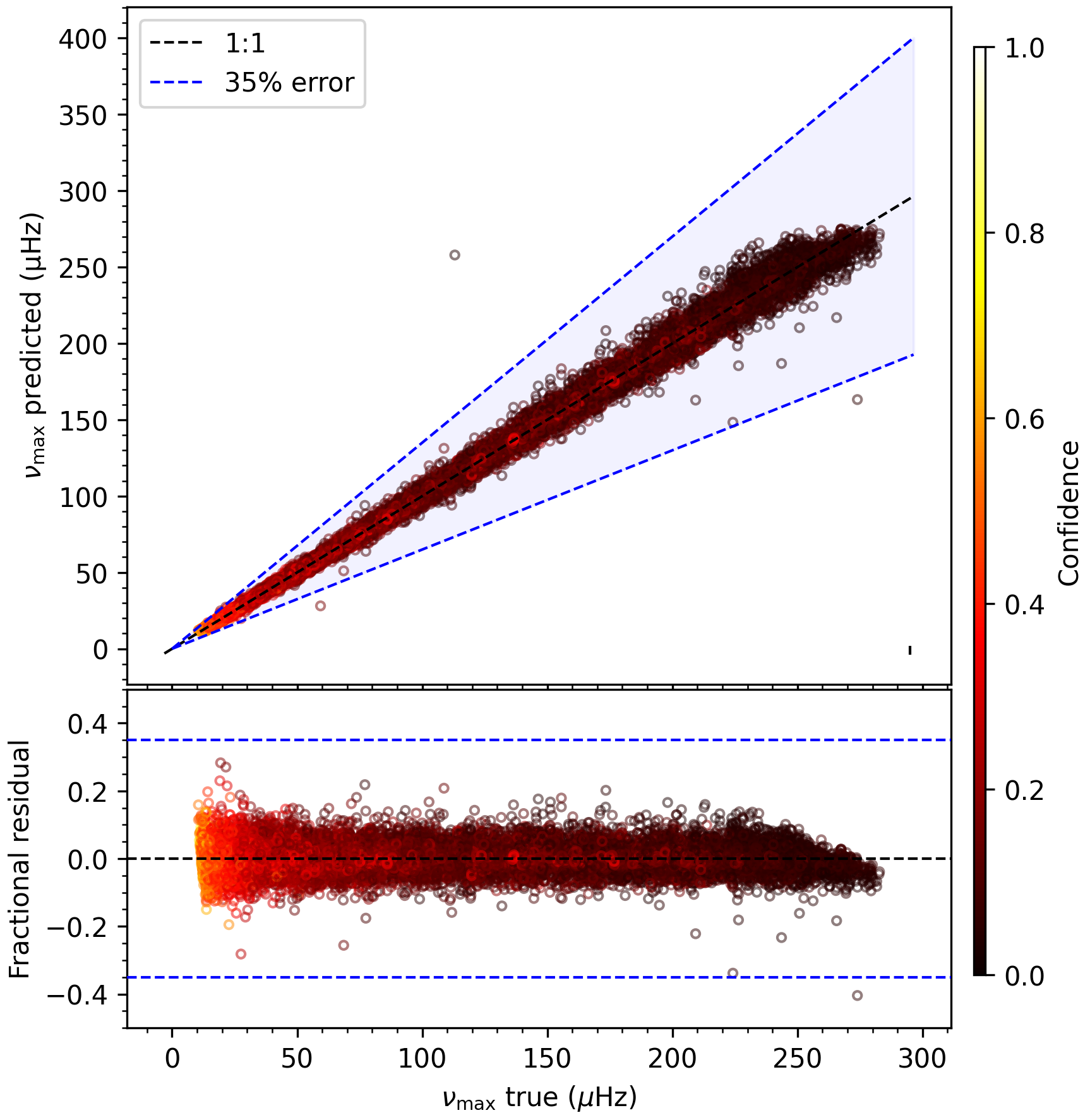}
    }
    \hfill
    \subfloat[\label{fig:synthetics_dnu_comparison_scatter}]{
        \includegraphics[width=0.48\textwidth]{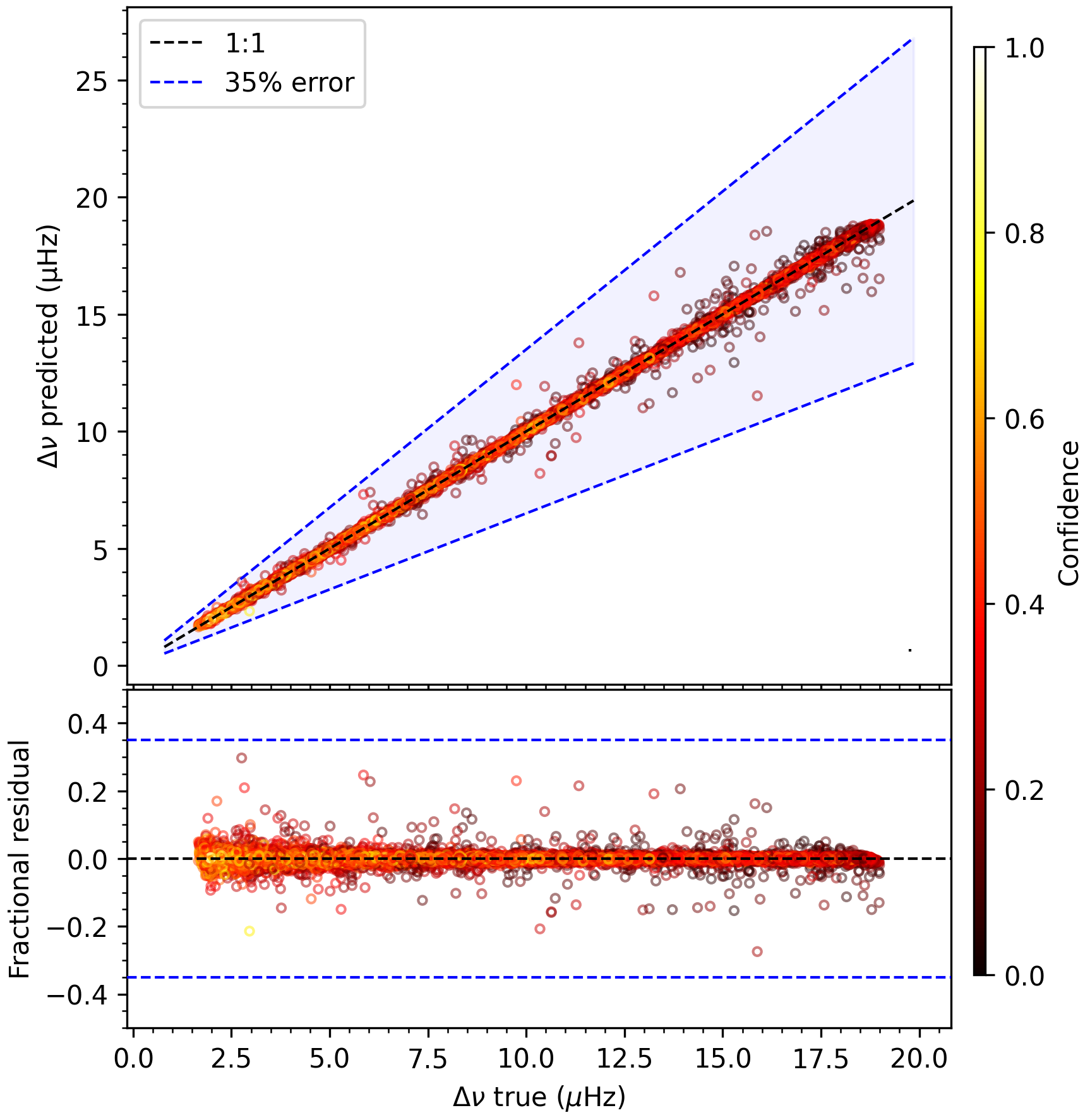}
    } \\ 

    \subfloat[\label{fig:synthetics_numax_comparison_hist}]{
        \includegraphics[width=0.45\textwidth]{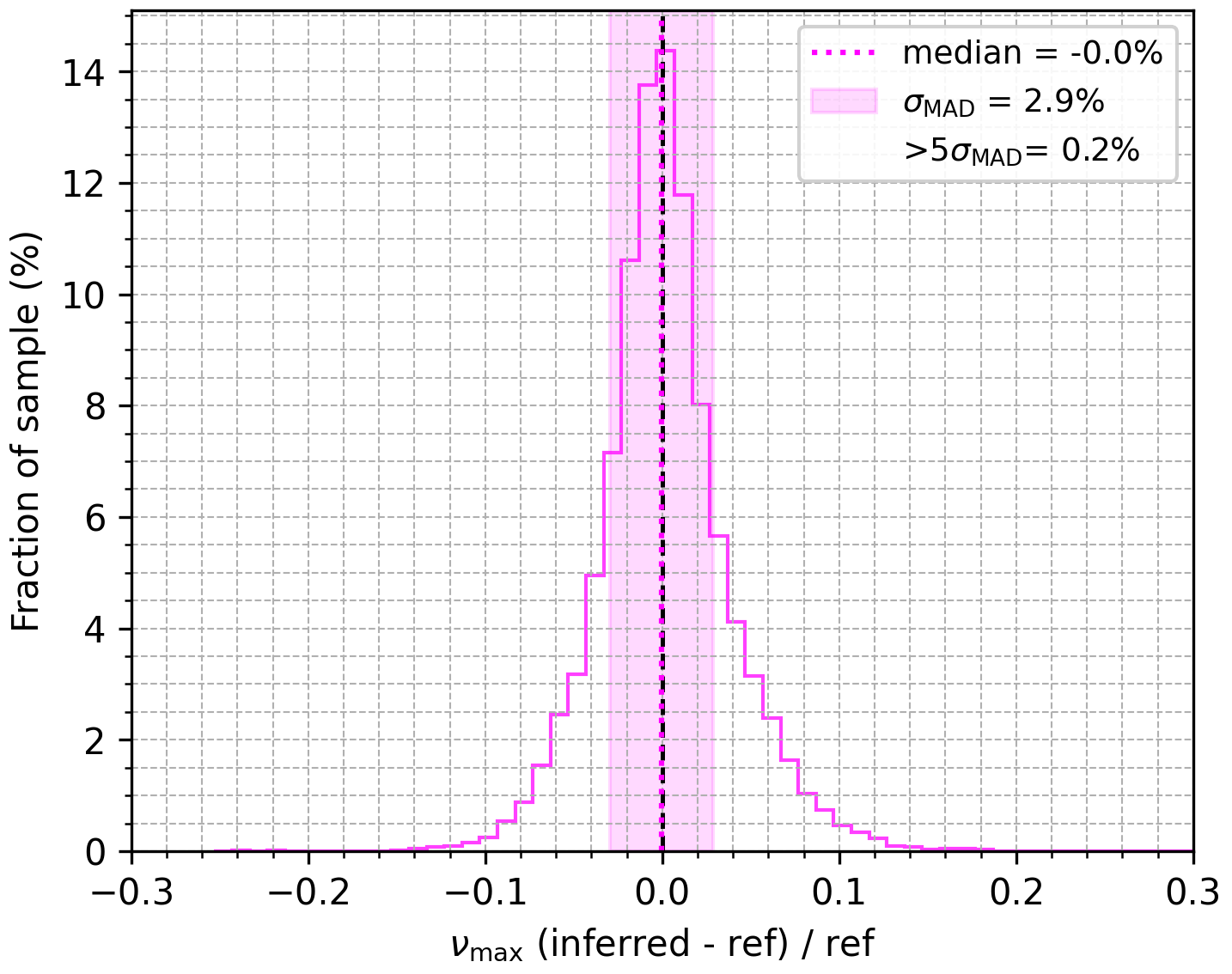}
    }
    \hfill
    \subfloat[\label{fig:synthetics_dnu_comparison_hist}]{
        \includegraphics[width=0.45\textwidth]{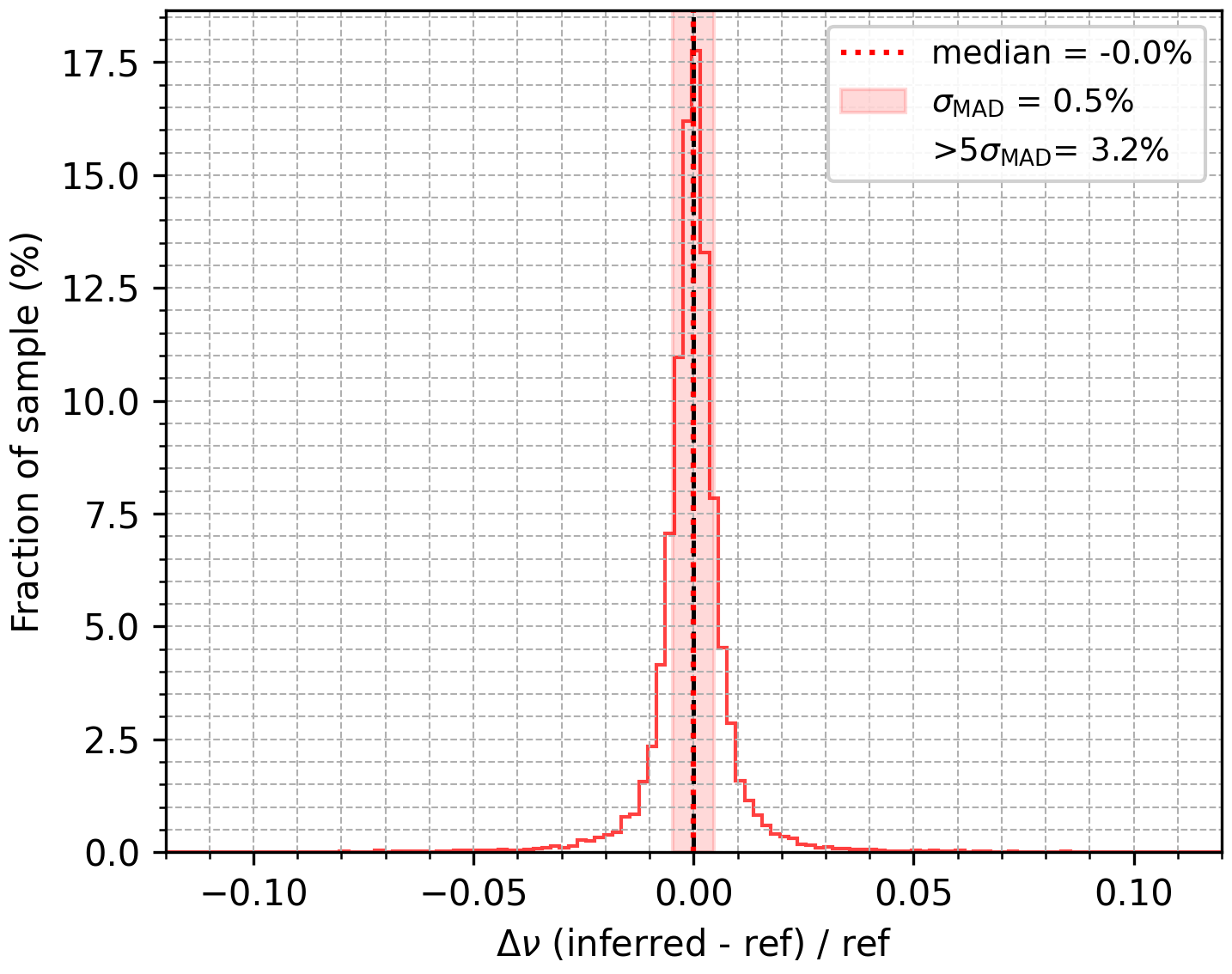}
    }
    \caption{Comparison of the model predictions on a set of unseen synthetic stars. Panel (a): confident $\nu_{\mathrm{max}}$ values. Panel (b): confident $\Delta\nu$ values. Panels (c--d): histograms of relative errors corresponding to panels (a) and (b), respectively.}
    \label{fig:synthetics_comparison}
\end{figure*}

Of the 20,000 test spectra, 19,224 (96.1\%) satisfy the \numax\ selection criterion and 18,630 (93.2\%) the \dnu\ threshold, yields close to those obtained on real Kepler data of the same baseline (Sect.~\ref{sec:results_ekic}). The p-mode model recovers both parameters without measurable bias, the median fractional offsets being below $0.01\%$ in each case, and with robust dispersions of $\sigma_{\mathrm{MAD}} = 2.9\%$ in \numax\ and $0.5\%$ in \dnu\ (Figs.~\ref{fig:synthetics_numax_comparison_scatter} and~\ref{fig:synthetics_dnu_comparison_scatter}). As on real data, \dnu\ has the heavier tails of the two: 3.2\% of the confident sample lies beyond $5\sigma_{\mathrm{MAD}}$, against 0.2\% for \numax, as the distributions of fractional residuals in Figs.~\ref{fig:synthetics_numax_comparison_hist} and~\ref{fig:synthetics_dnu_comparison_hist} show. Outright failures are practically absent, with fewer than 0.02\% of either sample differing from the true value by more than 35\%.

\begin{figure*}
\centering
    \subfloat[\label{fig:synthetics_dpi_comparison_density}]{
        \includegraphics[width=0.48\textwidth]{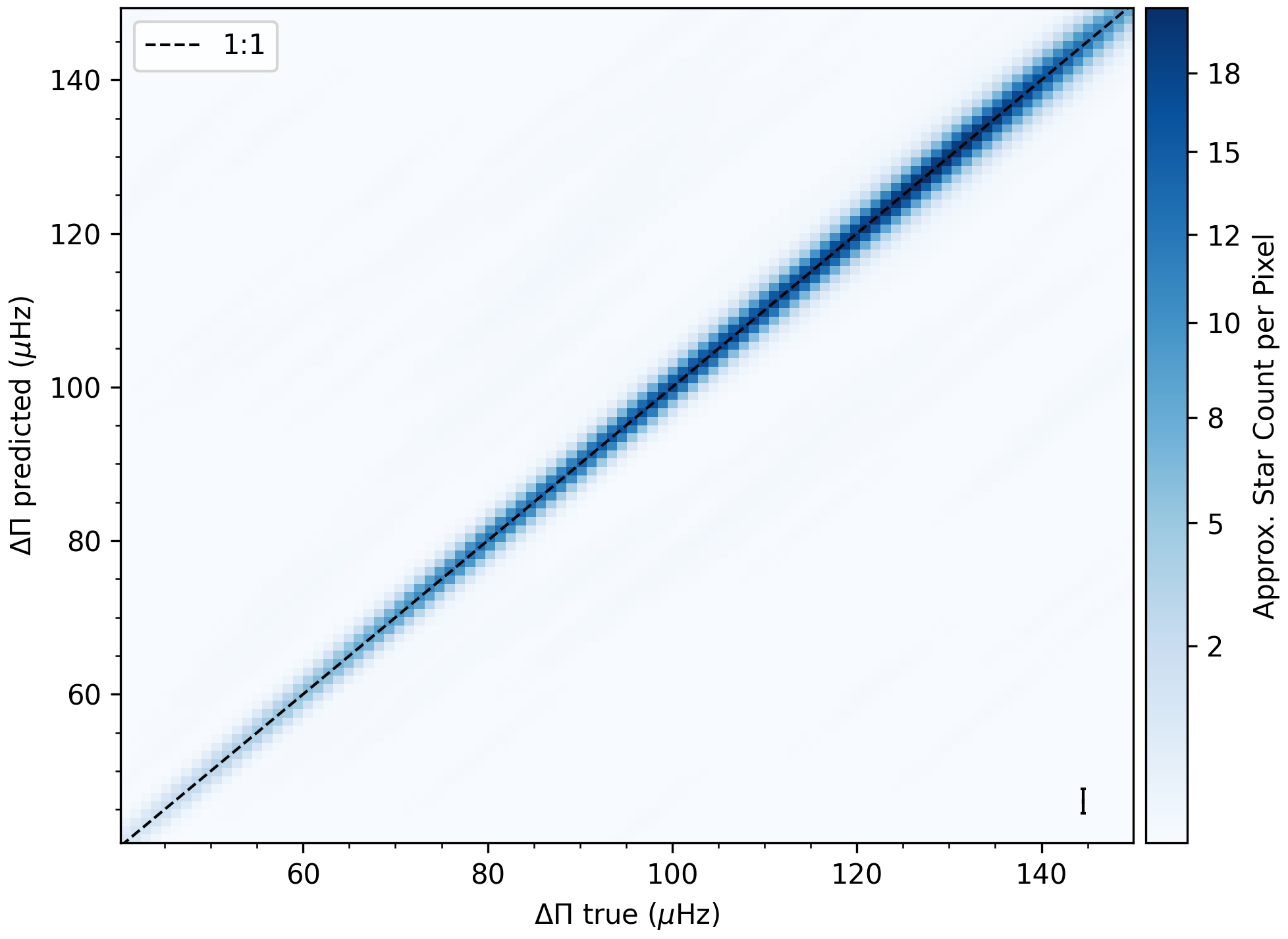}
    }
    \hfill
    \subfloat[\label{fig:synthetics_dpi_comparison_hist}]{
        \includegraphics[width=0.45\textwidth]{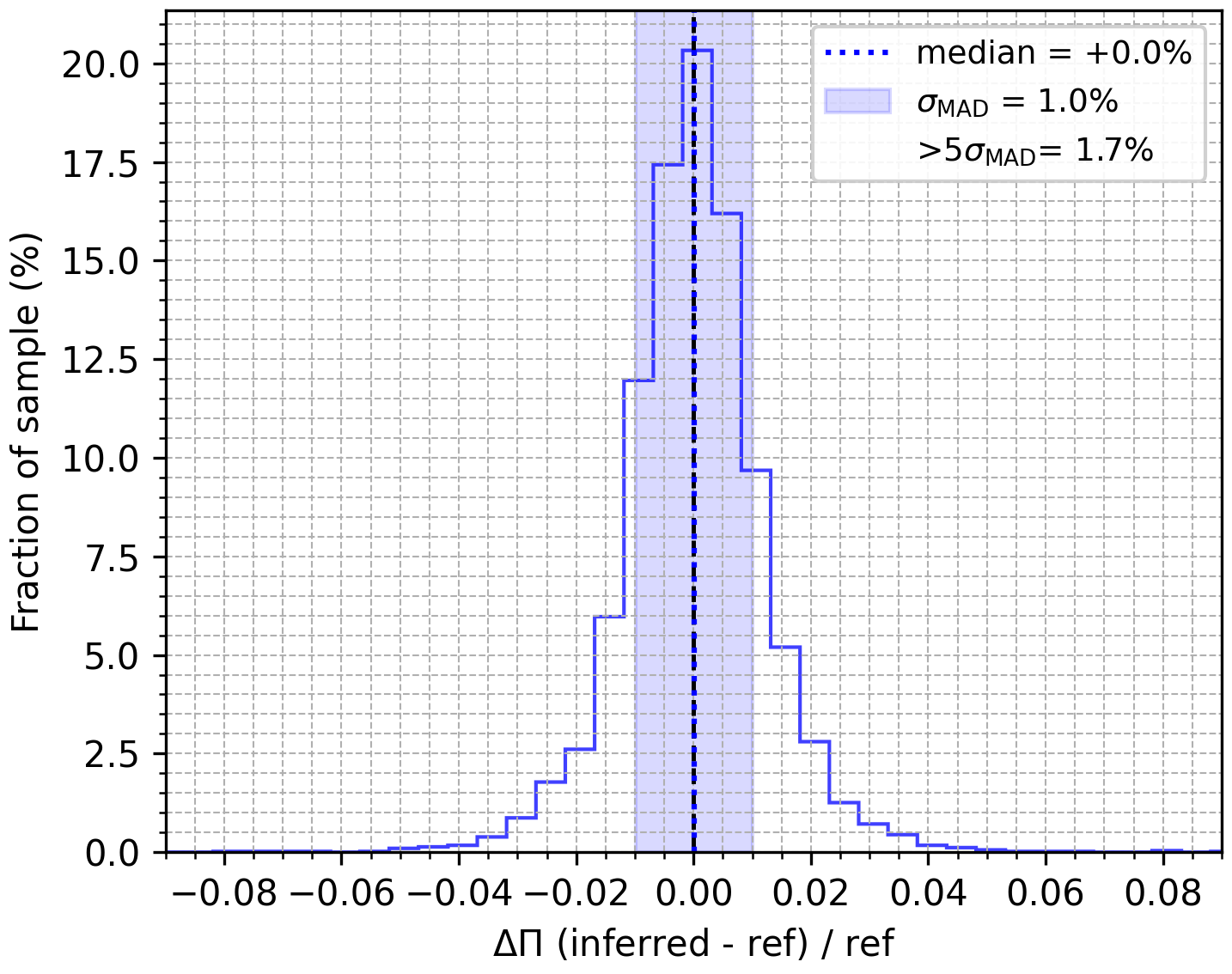}
    }
    \caption{\textbf{(a)} Density of confident $\Delta\Pi_{1}$ predictions against the true values for the synthetic test set. \textbf{(b)} Histogram of the corresponding fractional residuals.}
    \label{fig:k2_synthetics_dpi_comparison_matrix_hist}
\end{figure*}

For the g-mode model, only 5,421 of 79,616 synthetic young red giants (6.8\%) satisfy the selection criteria. This low yield is intrinsic to the problem: at K2-like resolution the mixed-mode pattern is resolved sufficiently to constrain \dpi\ only for a minority of stars, depending on the coupling strength, the inclination, the signal-to-noise ratio, and where the oscillation power falls relative to the frequency resolution. The selection criteria are precisely the mechanism by which we identify this minority, and we emphasise that the statistics below describe the selected subset and not the population as a whole. Within that subset the recovery is accurate, with a median fractional offset below $0.01\%$ and $\sigma_{\mathrm{MAD}} = 1.0\%$; the inferred values track the true ones along the 1:1 relation over the full range (Fig.~\ref{fig:synthetics_dpi_comparison_density}). The tails are again non-negligible, 1.7\% of the subset lying beyond $5\sigma_{\mathrm{MAD}}$ and 0.5\% differing from the true value by more than 35\% (Fig.~\ref{fig:synthetics_dpi_comparison_hist}), so the selection is effective but not infallible.

Since the reference values are exact, the uncertainties can be assessed directly. The empirical $1\sigma$ fractions are 0.681 for \numax, 0.720 for \dnu, and 0.827 for \dpi, against the 0.683 expected for well-calibrated Gaussian uncertainties. The corresponding robust dispersions of the normalized residuals are 1.02, 0.89, and 0.72, so the predictive distributions have near-Gaussian cores whose widths are accurate to better than 2\% for \numax, wider than the residuals warrant by roughly 10\% for \dnu, and by some 40\% for \dpi. For \dpi\ the effect is also visible in the reported values themselves, the median fractional uncertainty of $1.40\%$ exceeding the observed dispersion of $1.0\%$ by a similar factor. This over-coverage in \dpi\ is not an artefact of the discretisation of the network output: the median reported uncertainty does not decrease as the bin width is reduced below 1.25~s, and the over-coverage persists essentially unchanged at every bin width tested (Appendix~\ref{appendix:choosing_bin_size}). It appears instead to be a property of the predictive distributions themselves, which are broader than the precision the model ultimately achieves at this resolution. Multimodality contributes where it occurs, since for some stars the output carries secondary probability mass at candidate period spacings displaced from the primary peak, but it does not account for the effect in full. These uncertainties are therefore conservative by roughly 40\%.

\section{Period spacings for K2 red giants}
\label{appendix:k2_red_giants_dpi}
Table~\ref{tab:k2_red_giants_dpi} lists the confident period-spacing inferences for the young
red giants, obtained from single-campaign K2 observations. For each parameter the quoted value
is the median of the output probability distribution and the uncertainties are the 16th and
84th percentiles about that median (Sect.~\ref{sec:results}). Stars flagged as anomalous
inferences (Sect.~\ref{sec:results_ekic_dpi}) are marked with a dagger; their $\Delta\Pi_1$
distributions carry a secondary peak lying on the $\Delta\nu$--$\Delta\Pi_1$ sequence, and
these values should be treated with caution.

\onecolumn
\footnotesize
\setlength{\tabcolsep}{4pt}
\renewcommand{\arraystretch}{1.45}
\begin{longtable}{r r r r c @{\hspace{1.2em}}|@{\hspace{1.2em}} r r r r c}
\caption{Global seismic parameters and gravity-mode period spacings for the 232 K2 red giants
with confident $\Delta\Pi_1$ inferences. The table is continued across the two halves of each row.
The full table is also available in machine-readable form at the CDS.}
\label{tab:k2_red_giants_dpi} \\
\hline\hline
EPIC & $\nu_{\rm max}$ ($\mu$Hz) & $\Delta\nu$ ($\mu$Hz) & $\Delta\Pi_1$ (s) & Flag & EPIC & $\nu_{\rm max}$ ($\mu$Hz) & $\Delta\nu$ ($\mu$Hz) & $\Delta\Pi_1$ (s) & Flag \\
\hline
\endfirsthead
\caption[]{\textit{continued.}} \\
\hline\hline
EPIC & $\nu_{\rm max}$ ($\mu$Hz) & $\Delta\nu$ ($\mu$Hz) & $\Delta\Pi_1$ (s) & Flag & EPIC & $\nu_{\rm max}$ ($\mu$Hz) & $\Delta\nu$ ($\mu$Hz) & $\Delta\Pi_1$ (s) & Flag \\
\hline
\endhead
\hline
\multicolumn{10}{r}{\textit{Continued on next page}} \\
\endfoot
\hline
\endlastfoot
201162894 & $165.03^{+4.89}_{-3.27}$ & $13.879^{+0.043}_{-0.035}$ & $83.82^{+0.94}_{-1.14}$ & & 201184067 & $195.32^{+6.20}_{-10.15}$ & $15.647^{+0.046}_{-0.041}$ & $88.17^{+2.34}_{-1.62}$ & \\
201188911 & $136.54^{+5.89}_{-1.82}$ & $12.349^{+0.042}_{-0.049}$ & $82.10^{+2.56}_{-2.01}$ & & 201211472 & $235.91^{+5.54}_{-6.30}$ & $18.244^{+0.070}_{-0.040}$ & $91.45^{+1.41}_{-1.35}$ & \\
201260990 & $215.99^{+9.12}_{-6.66}$ & $16.745^{+0.054}_{-0.041}$ & $88.22^{+1.03}_{-0.79}$ & & 201420000 & $185.00^{+3.23}_{-3.62}$ & $15.222^{+0.045}_{-0.048}$ & $86.14^{+1.00}_{-0.89}$ & \\
201494732 & $186.58^{+3.64}_{-3.58}$ & $15.390^{+0.061}_{-0.057}$ & $86.10^{+1.30}_{-1.43}$ & & 201500806 & $191.40^{+8.00}_{-5.62}$ & $15.606^{+0.037}_{-0.048}$ & $85.69^{+1.01}_{-0.86}$ & \\
201535193 & $203.62^{+9.68}_{-2.53}$ & $16.071^{+0.054}_{-0.053}$ & $88.80^{+1.21}_{-1.20}$ & & 201571570 & $142.43^{+3.07}_{-3.56}$ & $12.384^{+0.064}_{-0.060}$ & $81.24^{+2.96}_{-2.36}$ & \\
201583796 & $173.36^{+3.69}_{-7.11}$ & $13.580^{+0.054}_{-0.062}$ & $84.84^{+1.10}_{-0.92}$ & & 201584014 & $188.49^{+9.48}_{-3.20}$ & $15.226^{+0.051}_{-0.048}$ & $85.22^{+0.93}_{-1.30}$ & \\
201615261 & $148.89^{+2.79}_{-3.37}$ & $12.305^{+0.049}_{-0.050}$ & $80.45^{+1.47}_{-1.52}$ & & 201626832 & $152.24^{+2.02}_{-10.08}$ & $12.697^{+0.048}_{-0.042}$ & $81.81^{+1.47}_{-1.41}$ & \\
201668891 & $178.46^{+6.32}_{-5.12}$ & $14.498^{+0.068}_{-0.049}$ & $84.78^{+1.05}_{-0.75}$ & & 201696302 & $183.27^{+7.11}_{-5.48}$ & $14.712^{+0.063}_{-0.053}$ & $85.69^{+1.31}_{-1.60}$ & \\
201705355 & $183.63^{+3.15}_{-2.28}$ & $14.986^{+0.072}_{-0.060}$ & $85.61^{+0.79}_{-0.84}$ & & 201717672 & $169.24^{+3.37}_{-2.46}$ & $13.443^{+0.049}_{-0.048}$ & $81.82^{+1.20}_{-1.17}$ & \\
201722849 & $203.79^{+5.19}_{-4.02}$ & $15.970^{+0.047}_{-0.051}$ & $85.35^{+0.81}_{-1.07}$ & & 201764418 & $191.10^{+8.82}_{-2.85}$ & $15.703^{+0.046}_{-0.046}$ & $87.03^{+1.10}_{-0.76}$ & \\
201785415 & $197.63^{+6.29}_{-7.70}$ & $15.337^{+0.056}_{-0.066}$ & $84.45^{+1.53}_{-1.30}$ & & 201843617 & $173.33^{+6.33}_{-4.03}$ & $14.421^{+0.059}_{-0.053}$ & $86.03^{+1.18}_{-0.96}$ & \\
201853779 & $164.89^{+2.49}_{-3.13}$ & $13.743^{+0.069}_{-0.056}$ & $84.85^{+1.29}_{-1.04}$ & & 201920393 & $221.77^{+5.60}_{-5.66}$ & $16.823^{+0.057}_{-0.059}$ & $88.34^{+1.08}_{-0.77}$ & \\
201928170 & $177.86^{+7.05}_{-5.41}$ & $14.228^{+0.052}_{-0.050}$ & $83.97^{+0.80}_{-1.08}$ & & 205994284 & $164.26^{+2.34}_{-2.16}$ & $13.654^{+0.040}_{-0.040}$ & $83.75^{+0.94}_{-1.06}$ & \\
206015475 & $160.43^{+4.52}_{-4.19}$ & $13.143^{+0.048}_{-0.037}$ & $82.78^{+2.32}_{-1.45}$ & & 206023175 & $184.86^{+4.68}_{-10.31}$ & $14.945^{+0.052}_{-0.046}$ & $86.57^{+1.98}_{-1.72}$ & \\
206049476 & $203.73^{+6.30}_{-5.00}$ & $16.666^{+0.059}_{-0.055}$ & $88.35^{+1.12}_{-0.82}$ & & 206062898 & $192.33^{+5.41}_{-8.08}$ & $15.425^{+0.050}_{-0.048}$ & $87.31^{+1.16}_{-0.85}$ & \\
206075005 & $190.25^{+2.66}_{-3.32}$ & $14.610^{+0.037}_{-0.049}$ & $83.67^{+1.31}_{-1.05}$ & & 206088822 & $175.29^{+2.88}_{-3.25}$ & $13.821^{+0.048}_{-0.053}$ & $81.63^{+0.72}_{-1.13}$ & \\
206100060 & $220.44^{+5.48}_{-5.48}$ & $17.012^{+0.036}_{-0.044}$ & $89.22^{+1.12}_{-1.20}$ & & 206107394 & $233.63^{+2.42}_{-2.77}$ & $17.269^{+0.054}_{-0.049}$ & $85.36^{+1.90}_{-1.39}$ & \\
206136293 & $190.51^{+10.16}_{-3.24}$ & $15.979^{+0.049}_{-0.047}$ & $87.57^{+0.91}_{-0.93}$ & & 206139567 & $215.30^{+3.29}_{-5.87}$ & $17.117^{+0.042}_{-0.056}$ & $89.04^{+0.89}_{-1.13}$ & \\
206141983 & $227.33^{+4.14}_{-6.24}$ & $16.797^{+0.055}_{-0.051}$ & $87.84^{+1.67}_{-1.47}$ & & 206166135 & $186.39^{+7.86}_{-6.13}$ & $15.248^{+0.070}_{-0.049}$ & $86.88^{+0.68}_{-0.68}$ & \\
206187222 & $194.57^{+5.32}_{-5.88}$ & $15.502^{+0.075}_{-0.063}$ & $85.22^{+1.11}_{-1.18}$ & & 206188223 & $179.61^{+4.75}_{-8.97}$ & $14.295^{+0.053}_{-0.048}$ & $85.04^{+2.11}_{-1.07}$ & \\
206206667 & $150.11^{+3.63}_{-2.80}$ & $13.217^{+0.065}_{-0.060}$ & $81.50^{+1.44}_{-1.28}$ & & 206211295 & $233.77^{+5.61}_{-8.30}$ & $17.696^{+0.045}_{-0.058}$ & $90.56^{+0.68}_{-0.94}$ & \\
206223294 & $150.31^{+3.05}_{-4.09}$ & $12.077^{+0.044}_{-0.042}$ & $79.93^{+2.96}_{-2.01}$ & & 206283558 & $150.43^{+2.95}_{-4.09}$ & $12.701^{+0.078}_{-0.073}$ & $81.56^{+1.57}_{-1.60}$ & \\
206348556 & $145.78^{+3.21}_{-8.34}$ & $12.374^{+0.045}_{-0.048}$ & $83.54^{+1.77}_{-2.28}$ & & 206351132 & $146.57^{+4.57}_{-4.39}$ & $12.743^{+0.051}_{-0.045}$ & $81.39^{+1.56}_{-1.60}$ & \\
206375929 & $161.22^{+4.17}_{-6.06}$ & $13.580^{+0.059}_{-0.079}$ & $82.81^{+1.51}_{-1.22}$ & & 206434608 & $176.45^{+2.05}_{-2.31}$ & $13.532^{+0.055}_{-0.055}$ & $81.70^{+1.05}_{-1.45}$ & \\
206452199 & $229.25^{+8.10}_{-4.27}$ & $17.863^{+0.037}_{-0.057}$ & $91.90^{+1.36}_{-1.27}$ & & 206469672 & $186.54^{+8.66}_{-4.63}$ & $14.539^{+0.053}_{-0.059}$ & $85.84^{+1.21}_{-1.28}$ & \\
206476223 & $167.59^{+3.07}_{-3.08}$ & $14.038^{+0.055}_{-0.052}$ & $84.60^{+1.16}_{-1.06}$ & & 206515124 & $185.11^{+6.50}_{-2.35}$ & $15.627^{+0.045}_{-0.042}$ & $87.31^{+1.08}_{-0.82}$ & \\
210435152 & $144.95^{+2.24}_{-5.58}$ & $12.469^{+0.059}_{-0.055}$ & $81.36^{+2.33}_{-1.89}$ & & 210436932 & $168.04^{+3.14}_{-5.43}$ & $13.052^{+0.040}_{-0.049}$ & $67.34^{+1.15}_{-1.06}$ & \\
210483090 & $182.41^{+4.21}_{-4.62}$ & $13.844^{+0.047}_{-0.059}$ & $86.60^{+3.08}_{-2.74}$ & & 210516993 & $161.35^{+3.76}_{-6.26}$ & $13.652^{+0.045}_{-0.041}$ & $71.45^{+2.61}_{-1.23}$ & $\dagger$ \\
210548781 & $167.79^{+8.27}_{-4.20}$ & $13.598^{+0.041}_{-0.045}$ & $82.82^{+1.21}_{-1.37}$ & & 210562121 & $213.24^{+4.47}_{-11.24}$ & $15.862^{+0.035}_{-0.045}$ & $84.89^{+1.25}_{-1.01}$ & \\
210563947 & $147.85^{+2.66}_{-2.15}$ & $12.315^{+0.052}_{-0.055}$ & $81.65^{+0.88}_{-1.16}$ & & 210568587 & $170.52^{+8.30}_{-3.25}$ & $13.877^{+0.047}_{-0.046}$ & $81.70^{+0.95}_{-1.14}$ & \\
210573512 & $135.30^{+5.04}_{-3.69}$ & $11.457^{+0.040}_{-0.050}$ & $80.71^{+1.69}_{-1.90}$ & & 210634677 & $169.02^{+2.34}_{-1.97}$ & $14.096^{+0.041}_{-0.042}$ & $84.77^{+1.32}_{-1.03}$ & \\
210653298 & $176.21^{+4.79}_{-4.74}$ & $14.462^{+0.038}_{-0.050}$ & $84.67^{+1.10}_{-0.82}$ & & 210665262 & $138.01^{+2.42}_{-6.76}$ & $11.513^{+0.055}_{-0.052}$ & $80.25^{+3.23}_{-2.71}$ & \\
210683150 & $180.17^{+4.23}_{-2.86}$ & $14.938^{+0.053}_{-0.046}$ & $85.86^{+1.37}_{-1.10}$ & & 210733885 & $166.71^{+2.90}_{-3.86}$ & $13.968^{+0.079}_{-0.072}$ & $84.46^{+1.06}_{-0.83}$ & \\
210749402 & $145.31^{+4.52}_{-6.47}$ & $12.438^{+0.061}_{-0.056}$ & $81.30^{+1.05}_{-1.08}$ & & 210790944 & $144.65^{+10.04}_{-1.54}$ & $12.448^{+0.055}_{-0.046}$ & $81.25^{+1.15}_{-1.03}$ & \\
210791216 & $190.10^{+4.14}_{-6.04}$ & $15.232^{+0.060}_{-0.054}$ & $85.99^{+1.04}_{-0.78}$ & & 210845917 & $231.59^{+5.57}_{-6.96}$ & $17.201^{+0.044}_{-0.053}$ & $87.28^{+1.25}_{-1.06}$ & \\
210858949 & $173.87^{+6.23}_{-5.27}$ & $14.073^{+0.072}_{-0.090}$ & $83.25^{+1.28}_{-1.21}$ & & 210870657 & $203.78^{+6.37}_{-4.96}$ & $16.722^{+0.065}_{-0.074}$ & $88.20^{+1.40}_{-1.34}$ & \\
210873190 & $168.84^{+5.88}_{-3.74}$ & $13.265^{+0.057}_{-0.053}$ & $81.71^{+0.69}_{-1.02}$ & & 210968089 & $163.76^{+4.82}_{-4.07}$ & $13.535^{+0.056}_{-0.059}$ & $82.92^{+1.05}_{-1.15}$ & \\
211404786 & $195.35^{+7.44}_{-4.42}$ & $15.672^{+0.053}_{-0.051}$ & $87.77^{+1.33}_{-1.19}$ & & 211417815 & $200.74^{+3.05}_{-2.83}$ & $16.459^{+0.038}_{-0.053}$ & $88.26^{+1.01}_{-0.71}$ & \\
211505743 & $198.76^{+2.95}_{-2.26}$ & $16.471^{+0.058}_{-0.065}$ & $88.17^{+1.34}_{-1.55}$ & & 211513489 & $206.81^{+7.13}_{-5.45}$ & $16.555^{+0.075}_{-0.068}$ & $87.86^{+1.58}_{-1.38}$ & \\
211528211 & $171.85^{+2.52}_{-2.09}$ & $14.496^{+0.062}_{-0.045}$ & $84.94^{+1.02}_{-0.88}$ & & 211540713 & $169.99^{+3.52}_{-2.52}$ & $14.055^{+0.050}_{-0.052}$ & $85.58^{+0.98}_{-1.05}$ & \\
211568691 & $184.90^{+8.65}_{-2.62}$ & $15.638^{+0.056}_{-0.051}$ & $88.24^{+4.13}_{-1.57}$ & & 211609177 & $196.72^{+3.59}_{-10.59}$ & $15.605^{+0.038}_{-0.048}$ & $86.80^{+0.59}_{-0.72}$ & \\
211614234 & $186.65^{+3.21}_{-4.59}$ & $14.160^{+0.049}_{-0.046}$ & $82.23^{+1.11}_{-0.82}$ & & 211614725 & $176.08^{+10.48}_{-3.30}$ & $14.376^{+0.050}_{-0.050}$ & $70.90^{+1.30}_{-0.85}$ & $\dagger$ \\
211654531 & $201.31^{+7.03}_{-8.61}$ & $16.083^{+0.051}_{-0.059}$ & $87.01^{+1.00}_{-0.66}$ & & 211678470 & $162.44^{+3.61}_{-7.51}$ & $13.217^{+0.054}_{-0.053}$ & $84.66^{+1.18}_{-0.84}$ & \\
211692043 & $227.30^{+4.96}_{-8.57}$ & $16.488^{+0.051}_{-0.039}$ & $85.28^{+1.24}_{-1.17}$ & & 211698152 & $150.24^{+4.28}_{-5.39}$ & $12.230^{+0.046}_{-0.033}$ & $81.16^{+1.59}_{-1.08}$ & \\
211704166 & $172.12^{+8.63}_{-2.54}$ & $14.324^{+0.053}_{-0.058}$ & $84.09^{+0.74}_{-1.07}$ & & 211704574 & $194.54^{+3.29}_{-7.17}$ & $15.713^{+0.054}_{-0.050}$ & $87.76^{+1.90}_{-1.31}$ & \\
211706751 & $149.10^{+6.78}_{-4.15}$ & $12.128^{+0.047}_{-0.042}$ & $79.19^{+2.48}_{-2.18}$ & & 211707086 & $217.17^{+3.67}_{-10.00}$ & $15.672^{+0.060}_{-0.056}$ & $81.33^{+0.92}_{-1.02}$ & \\
211732416 & $234.60^{+5.20}_{-11.44}$ & $17.515^{+0.079}_{-0.065}$ & $87.96^{+0.94}_{-1.12}$ & & 211732772 & $189.42^{+4.31}_{-5.84}$ & $14.736^{+0.049}_{-0.043}$ & $83.94^{+0.79}_{-1.06}$ & \\
211739709 & $219.49^{+6.41}_{-7.80}$ & $16.714^{+0.040}_{-0.056}$ & $88.11^{+1.25}_{-1.21}$ & & 211741853 & $221.53^{+4.22}_{-3.39}$ & $17.475^{+0.051}_{-0.050}$ & $89.64^{+1.08}_{-0.87}$ & \\
211764055 & $224.44^{+4.55}_{-3.90}$ & $17.725^{+0.045}_{-0.048}$ & $90.14^{+0.93}_{-1.32}$ & & 211779384 & $197.71^{+5.86}_{-6.43}$ & $15.489^{+0.053}_{-0.038}$ & $85.08^{+0.91}_{-0.91}$ & \\
211793455 & $168.66^{+2.09}_{-1.98}$ & $13.946^{+0.052}_{-0.044}$ & $83.87^{+1.02}_{-1.23}$ & & 211793638 & $214.48^{+8.13}_{-8.86}$ & $16.812^{+0.060}_{-0.054}$ & $89.35^{+1.51}_{-1.51}$ & \\
211796103 & $159.94^{+7.53}_{-3.40}$ & $12.981^{+0.049}_{-0.037}$ & $73.25^{+1.07}_{-0.86}$ & & 211806774 & $155.33^{+3.54}_{-2.83}$ & $13.019^{+0.058}_{-0.053}$ & $82.39^{+1.53}_{-1.86}$ & \\
211830799 & $160.38^{+4.67}_{-4.37}$ & $13.604^{+0.053}_{-0.075}$ & $83.78^{+2.23}_{-1.88}$ & & 211833958 & $153.26^{+3.55}_{-4.79}$ & $12.237^{+0.055}_{-0.045}$ & $79.82^{+1.37}_{-1.34}$ & \\
211897908 & $187.73^{+3.60}_{-5.13}$ & $15.327^{+0.049}_{-0.055}$ & $86.08^{+1.05}_{-0.93}$ & & 211906415 & $161.84^{+3.05}_{-2.27}$ & $12.913^{+0.045}_{-0.046}$ & $89.94^{+1.06}_{-1.03}$ & \\
211906830 & $188.03^{+5.21}_{-5.35}$ & $14.813^{+0.044}_{-0.055}$ & $83.82^{+0.82}_{-0.94}$ & & 211929298 & $170.68^{+2.31}_{-2.33}$ & $14.381^{+0.048}_{-0.034}$ & $84.18^{+0.71}_{-1.17}$ & \\
211947009 & $160.53^{+6.62}_{-2.30}$ & $13.895^{+0.049}_{-0.041}$ & $83.77^{+1.03}_{-1.54}$ & & 211954121 & $170.98^{+2.54}_{-2.51}$ & $14.163^{+0.054}_{-0.048}$ & $84.23^{+0.78}_{-1.11}$ & \\
211962883 & $211.58^{+4.83}_{-7.07}$ & $15.627^{+0.058}_{-0.059}$ & $83.14^{+0.81}_{-0.76}$ & & 211974782 & $209.40^{+5.48}_{-6.98}$ & $16.372^{+0.048}_{-0.051}$ & $86.87^{+1.55}_{-2.77}$ & \\
211977001 & $188.32^{+3.35}_{-2.95}$ & $14.916^{+0.053}_{-0.054}$ & $84.67^{+1.09}_{-0.74}$ & & 211977346 & $182.72^{+2.28}_{-6.07}$ & $14.746^{+0.071}_{-0.055}$ & $86.53^{+0.75}_{-1.06}$ & \\
211982071 & $192.13^{+6.52}_{-6.47}$ & $15.628^{+0.049}_{-0.050}$ & $86.80^{+0.69}_{-0.94}$ & & 211987214 & $151.92^{+3.66}_{-4.30}$ & $12.641^{+0.046}_{-0.038}$ & $80.98^{+1.28}_{-1.24}$ & \\
211993851 & $236.37^{+5.32}_{-7.86}$ & $17.334^{+0.056}_{-0.065}$ & $89.78^{+1.10}_{-1.24}$ & & 211994196 & $187.32^{+5.62}_{-7.09}$ & $14.931^{+0.054}_{-0.049}$ & $84.32^{+0.77}_{-0.96}$ & \\
212010612 & $185.95^{+2.59}_{-5.92}$ & $14.849^{+0.045}_{-0.047}$ & $86.19^{+0.94}_{-0.85}$ & & 212023032 & $163.05^{+4.74}_{-5.87}$ & $13.437^{+0.047}_{-0.036}$ & $82.98^{+0.83}_{-1.07}$ & \\
212073752 & $202.28^{+7.48}_{-3.74}$ & $15.275^{+0.056}_{-0.056}$ & $84.45^{+1.20}_{-1.06}$ & & 212106017 & $183.64^{+3.27}_{-3.46}$ & $15.089^{+0.045}_{-0.040}$ & $87.01^{+1.14}_{-0.92}$ & \\
212136615 & $156.12^{+3.88}_{-3.01}$ & $13.110^{+0.034}_{-0.047}$ & $82.12^{+1.61}_{-0.97}$ & & 212161576 & $173.23^{+3.77}_{-6.20}$ & $14.366^{+0.033}_{-0.050}$ & $85.64^{+1.08}_{-1.06}$ & \\
212177247 & $204.14^{+9.87}_{-3.17}$ & $16.125^{+0.050}_{-0.051}$ & $88.27^{+1.13}_{-0.93}$ & & 212178875 & $233.04^{+4.97}_{-5.51}$ & $17.221^{+0.046}_{-0.050}$ & $89.33^{+1.34}_{-1.31}$ & \\
212207491 & $198.68^{+2.94}_{-2.44}$ & $16.338^{+0.052}_{-0.064}$ & $87.61^{+0.97}_{-1.02}$ & & 212207638 & $220.24^{+4.42}_{-4.15}$ & $16.737^{+0.055}_{-0.045}$ & $87.08^{+1.28}_{-1.10}$ & \\
212297049 & $214.20^{+2.72}_{-2.71}$ & $17.017^{+0.048}_{-0.048}$ & $89.22^{+1.17}_{-1.27}$ & & 212329497 & $154.09^{+2.96}_{-8.78}$ & $12.948^{+0.042}_{-0.039}$ & $79.79^{+1.68}_{-1.64}$ & \\
212366315 & $180.83^{+2.12}_{-2.53}$ & $14.398^{+0.044}_{-0.035}$ & $82.88^{+0.67}_{-0.99}$ & & 212392830 & $168.65^{+3.58}_{-3.30}$ & $13.606^{+0.063}_{-0.076}$ & $83.09^{+1.32}_{-1.33}$ & \\
212426865 & $184.09^{+4.26}_{-3.81}$ & $14.235^{+0.048}_{-0.038}$ & $82.42^{+0.97}_{-0.86}$ & & 212452639 & $153.53^{+1.96}_{-2.07}$ & $12.833^{+0.050}_{-0.054}$ & $81.38^{+1.41}_{-1.34}$ & \\
212452985 & $178.10^{+3.21}_{-6.84}$ & $14.673^{+0.053}_{-0.052}$ & $85.34^{+1.87}_{-2.21}$ & & 212457945 & $168.36^{+4.19}_{-3.56}$ & $13.545^{+0.044}_{-0.052}$ & $84.02^{+0.83}_{-1.16}$ & \\
212470043 & $192.68^{+9.47}_{-4.26}$ & $15.260^{+0.058}_{-0.048}$ & $85.82^{+1.01}_{-0.70}$ & & 212481465 & $201.53^{+9.97}_{-4.05}$ & $16.436^{+0.049}_{-0.055}$ & $88.39^{+1.24}_{-1.12}$ & \\
212502566 & $163.50^{+2.68}_{-6.20}$ & $13.809^{+0.044}_{-0.058}$ & $84.99^{+1.67}_{-1.12}$ & & 212557162 & $126.57^{+6.13}_{-2.68}$ & $11.532^{+0.053}_{-0.056}$ & $80.65^{+2.98}_{-2.22}$ & \\
212559650 & $151.65^{+4.50}_{-7.28}$ & $12.723^{+0.057}_{-0.054}$ & $81.90^{+2.20}_{-5.74}$ & & 212562020 & $182.51^{+2.34}_{-3.09}$ & $15.113^{+0.032}_{-0.048}$ & $86.00^{+1.07}_{-0.87}$ & \\
212585386 & $163.15^{+5.13}_{-5.18}$ & $13.208^{+0.060}_{-0.055}$ & $81.82^{+0.66}_{-0.87}$ & & 212610492 & $195.16^{+2.82}_{-7.19}$ & $14.698^{+0.081}_{-0.079}$ & $67.91^{+0.76}_{-1.05}$ & $\dagger$ \\
212633100 & $145.03^{+5.18}_{-2.60}$ & $12.866^{+0.032}_{-0.045}$ & $84.02^{+2.22}_{-2.36}$ & & 212642718 & $187.18^{+2.20}_{-8.19}$ & $14.823^{+0.075}_{-0.081}$ & $84.90^{+1.10}_{-0.94}$ & \\
212653037 & $152.43^{+5.95}_{-3.46}$ & $12.572^{+0.059}_{-0.064}$ & $81.39^{+1.62}_{-1.41}$ & & 212698870 & $160.80^{+3.88}_{-2.91}$ & $13.109^{+0.034}_{-0.048}$ & $83.19^{+1.77}_{-1.63}$ & \\
212703382 & $168.88^{+6.50}_{-5.13}$ & $13.446^{+0.048}_{-0.047}$ & $82.07^{+0.94}_{-0.61}$ & & 212706376 & $152.41^{+5.01}_{-3.69}$ & $12.959^{+0.049}_{-0.051}$ & $80.69^{+1.79}_{-2.10}$ & \\
212750375 & $205.99^{+3.63}_{-2.78}$ & $16.658^{+0.068}_{-0.053}$ & $88.82^{+1.59}_{-1.57}$ & & 212795843 & $195.08^{+3.05}_{-5.30}$ & $15.368^{+0.044}_{-0.049}$ & $85.96^{+1.18}_{-0.91}$ & \\
212808788 & $219.25^{+4.32}_{-4.62}$ & $16.204^{+0.042}_{-0.054}$ & $85.17^{+0.83}_{-0.98}$ & & 220224794 & $164.23^{+4.94}_{-5.59}$ & $13.386^{+0.049}_{-0.044}$ & $80.89^{+1.06}_{-0.77}$ & \\
220318037 & $180.45^{+6.14}_{-3.36}$ & $14.732^{+0.053}_{-0.047}$ & $86.07^{+1.58}_{-1.27}$ & & 220327551 & $203.37^{+4.94}_{-7.58}$ & $15.416^{+0.052}_{-0.049}$ & $83.91^{+0.82}_{-1.06}$ & \\
220379656 & $168.76^{+5.36}_{-4.41}$ & $14.228^{+0.056}_{-0.055}$ & $84.13^{+1.34}_{-1.34}$ & & 220417297 & $176.58^{+3.95}_{-8.84}$ & $13.699^{+0.043}_{-0.042}$ & $86.92^{+1.13}_{-1.04}$ & \\
220452204 & $191.29^{+8.56}_{-2.65}$ & $15.880^{+0.050}_{-0.044}$ & $88.30^{+1.30}_{-1.29}$ & & 220489190 & $154.78^{+7.07}_{-4.17}$ & $12.544^{+0.048}_{-0.062}$ & $81.41^{+2.01}_{-2.34}$ & \\
220547602 & $128.96^{+2.55}_{-1.66}$ & $11.792^{+0.054}_{-0.075}$ & $79.34^{+3.65}_{-3.56}$ & & 220635468 & $177.57^{+8.15}_{-3.33}$ & $14.453^{+0.044}_{-0.047}$ & $85.37^{+1.05}_{-1.15}$ & \\
220648976 & $166.23^{+3.67}_{-3.71}$ & $13.619^{+0.030}_{-0.046}$ & $81.94^{+1.00}_{-0.81}$ & & 228788585 & $173.75^{+4.26}_{-4.86}$ & $14.673^{+0.145}_{-0.196}$ & $83.15^{+2.15}_{-1.63}$ & \\
229039250 & $160.61^{+4.05}_{-2.94}$ & $13.938^{+0.056}_{-0.048}$ & $83.11^{+1.53}_{-1.46}$ & & 235058736 & $159.82^{+1.91}_{-1.76}$ & $13.284^{+0.048}_{-0.051}$ & $81.32^{+1.12}_{-1.04}$ & \\
246009840 & $159.86^{+1.82}_{-2.42}$ & $12.810^{+0.037}_{-0.049}$ & $80.86^{+1.22}_{-1.07}$ & & 246050536 & $183.15^{+2.53}_{-3.77}$ & $14.233^{+0.047}_{-0.032}$ & $82.80^{+1.08}_{-1.15}$ & \\
246075387 & $169.93^{+2.95}_{-2.08}$ & $13.559^{+0.040}_{-0.056}$ & $81.69^{+1.39}_{-1.40}$ & & 246079566 & $195.94^{+4.79}_{-6.27}$ & $15.626^{+0.045}_{-0.044}$ & $88.27^{+1.84}_{-1.53}$ & \\
246112343 & $158.47^{+2.21}_{-2.23}$ & $13.737^{+0.071}_{-0.063}$ & $83.89^{+1.68}_{-1.75}$ & & 246139560 & $185.10^{+6.59}_{-4.41}$ & $14.387^{+0.051}_{-0.035}$ & $86.46^{+0.87}_{-1.06}$ & \\
246233436 & $198.29^{+3.66}_{-2.42}$ & $15.861^{+0.036}_{-0.045}$ & $87.68^{+0.85}_{-0.97}$ & & 246233563 & $217.58^{+3.25}_{-2.64}$ & $17.232^{+0.052}_{-0.045}$ & $90.04^{+0.86}_{-1.04}$ & \\
246238698 & $191.68^{+7.92}_{-3.37}$ & $15.627^{+0.047}_{-0.046}$ & $87.22^{+1.04}_{-0.72}$ & & 246299719 & $150.00^{+3.71}_{-6.97}$ & $12.576^{+0.069}_{-0.102}$ & $81.07^{+1.34}_{-1.28}$ & \\
246347334 & $154.41^{+2.12}_{-3.33}$ & $13.074^{+0.051}_{-0.057}$ & $82.24^{+1.29}_{-1.15}$ & & 246416522 & $220.90^{+7.76}_{-5.54}$ & $16.379^{+0.057}_{-0.048}$ & $86.90^{+0.83}_{-0.71}$ & \\
246964021 & $169.88^{+8.87}_{-4.78}$ & $13.694^{+0.076}_{-0.063}$ & $104.18^{+1.81}_{-1.56}$ & $\dagger$ & 246967235 & $211.01^{+4.38}_{-4.58}$ & $16.153^{+0.055}_{-0.042}$ & $85.81^{+1.30}_{-1.34}$ & \\
247136733 & $180.64^{+4.67}_{-4.83}$ & $14.623^{+0.042}_{-0.046}$ & $84.77^{+1.03}_{-0.74}$ & & 247141427 & $214.49^{+3.07}_{-4.84}$ & $15.906^{+0.042}_{-0.044}$ & $84.39^{+0.51}_{-0.51}$ & \\
247278817 & $147.85^{+8.67}_{-2.98}$ & $12.175^{+0.047}_{-0.047}$ & $81.03^{+2.62}_{-1.83}$ & & 247279992 & $187.70^{+10.97}_{-3.19}$ & $15.240^{+0.064}_{-0.047}$ & $84.80^{+1.34}_{-0.99}$ & \\
247298530 & $196.26^{+4.70}_{-5.47}$ & $15.691^{+0.047}_{-0.040}$ & $87.55^{+0.96}_{-0.94}$ & & 247361881 & $163.06^{+2.28}_{-1.96}$ & $13.470^{+0.043}_{-0.047}$ & $83.06^{+1.04}_{-1.10}$ & \\
247364531 & $160.64^{+4.94}_{-6.91}$ & $13.319^{+0.046}_{-0.052}$ & $80.81^{+1.39}_{-1.37}$ & & 247376509 & $146.67^{+5.07}_{-3.79}$ & $12.704^{+0.046}_{-0.045}$ & $81.79^{+1.08}_{-1.15}$ & \\
247387736 & $154.88^{+3.55}_{-5.09}$ & $13.012^{+0.062}_{-0.051}$ & $81.32^{+1.86}_{-1.64}$ & & 248484584 & $157.65^{+3.24}_{-3.06}$ & $12.834^{+0.050}_{-0.054}$ & $83.42^{+1.32}_{-1.55}$ & \\
248599469 & $192.70^{+4.33}_{-8.74}$ & $14.818^{+0.055}_{-0.057}$ & $84.42^{+0.57}_{-0.57}$ & & 248620720 & $222.35^{+1.99}_{-5.33}$ & $15.870^{+0.050}_{-0.050}$ & $114.47^{+1.95}_{-1.74}$ & $\dagger$ \\
248628691 & $187.72^{+10.90}_{-2.55}$ & $15.326^{+0.047}_{-0.051}$ & $86.60^{+0.70}_{-1.02}$ & & 248645998 & $165.02^{+3.45}_{-5.73}$ & $13.607^{+0.036}_{-0.046}$ & $82.98^{+1.43}_{-1.67}$ & \\
248657894 & $154.36^{+2.81}_{-2.40}$ & $13.404^{+0.045}_{-0.051}$ & $83.99^{+1.43}_{-1.40}$ & & 248760946 & $183.67^{+4.53}_{-3.08}$ & $15.467^{+0.058}_{-0.059}$ & $87.08^{+1.29}_{-0.93}$ & \\
248769803 & $168.78^{+2.70}_{-2.82}$ & $14.083^{+0.055}_{-0.064}$ & $84.00^{+1.49}_{-1.48}$ & & 248798639 & $160.39^{+6.91}_{-3.62}$ & $13.685^{+0.064}_{-0.058}$ & $84.32^{+2.06}_{-1.44}$ & \\
248824042 & $238.59^{+5.35}_{-4.32}$ & $17.426^{+0.059}_{-0.056}$ & $88.14^{+1.78}_{-1.34}$ & & 248867024 & $182.69^{+5.22}_{-7.31}$ & $14.632^{+0.048}_{-0.041}$ & $85.48^{+0.92}_{-1.10}$ & \\
248879557 & $177.11^{+2.83}_{-3.53}$ & $14.862^{+0.034}_{-0.048}$ & $85.73^{+1.52}_{-1.47}$ & & 249251825 & $200.09^{+7.74}_{-8.58}$ & $15.997^{+0.047}_{-0.047}$ & $85.53^{+1.12}_{-1.15}$ & \\
249344166 & $200.35^{+4.43}_{-6.79}$ & $15.847^{+0.053}_{-0.053}$ & $86.69^{+2.01}_{-1.49}$ & & 249369033 & $151.01^{+3.96}_{-2.60}$ & $12.730^{+0.060}_{-0.053}$ & $82.53^{+1.57}_{-1.25}$ & \\
249379090 & $173.21^{+2.58}_{-4.94}$ & $14.621^{+0.051}_{-0.053}$ & $86.20^{+1.30}_{-1.54}$ & & 249383613 & $171.22^{+4.16}_{-8.26}$ & $13.440^{+0.046}_{-0.037}$ & $81.60^{+0.77}_{-1.08}$ & \\
249417643 & $215.44^{+4.97}_{-7.03}$ & $15.873^{+0.053}_{-0.051}$ & $84.42^{+1.16}_{-1.06}$ & & 249434061 & $183.22^{+4.23}_{-7.10}$ & $14.909^{+0.059}_{-0.053}$ & $87.45^{+3.77}_{-1.90}$ & \\
249438765 & $205.00^{+8.23}_{-2.22}$ & $16.684^{+0.053}_{-0.063}$ & $88.28^{+1.21}_{-1.12}$ & & 249462150 & $199.81^{+4.84}_{-7.48}$ & $16.303^{+0.065}_{-0.050}$ & $89.62^{+1.41}_{-1.75}$ & \\
249480797 & $187.69^{+3.23}_{-7.37}$ & $14.571^{+0.049}_{-0.058}$ & $81.75^{+1.46}_{-1.17}$ & & 249563747 & $156.77^{+6.27}_{-2.17}$ & $13.001^{+0.048}_{-0.041}$ & $81.00^{+2.75}_{-1.82}$ & \\
249583241 & $182.07^{+7.04}_{-6.36}$ & $14.615^{+0.037}_{-0.048}$ & $76.87^{+0.51}_{-0.51}$ & & 249599419 & $46.94^{+120.85}_{-0.91}$ & $14.461^{+0.041}_{-0.051}$ & $85.73^{+1.42}_{-1.35}$ & \\
249599650 & $195.98^{+2.88}_{-5.22}$ & $15.550^{+0.052}_{-0.058}$ & $87.00^{+0.87}_{-0.62}$ & & 249615561 & $193.06^{+3.37}_{-3.96}$ & $14.869^{+0.039}_{-0.049}$ & $83.57^{+1.02}_{-0.86}$ & \\
249618659 & $172.70^{+6.04}_{-6.26}$ & $14.176^{+0.055}_{-0.051}$ & $82.80^{+0.71}_{-1.00}$ & & 249621972 & $203.94^{+6.29}_{-6.71}$ & $15.564^{+0.053}_{-0.057}$ & $82.76^{+0.83}_{-1.04}$ & \\
249632579 & $172.13^{+5.12}_{-2.67}$ & $14.263^{+0.056}_{-0.054}$ & $85.48^{+0.68}_{-0.96}$ & & 249635714 & $163.77^{+8.60}_{-2.70}$ & $13.436^{+0.046}_{-0.037}$ & $81.39^{+1.05}_{-1.14}$ & \\
249636998 & $183.81^{+7.93}_{-5.37}$ & $14.816^{+0.048}_{-0.052}$ & $86.93^{+1.01}_{-0.71}$ & & 249662481 & $184.65^{+3.10}_{-3.06}$ & $15.462^{+0.058}_{-0.060}$ & $86.43^{+1.85}_{-1.34}$ & \\
249707852 & $168.61^{+8.68}_{-3.94}$ & $13.757^{+0.044}_{-0.044}$ & $85.54^{+1.30}_{-1.25}$ & & 249812189 & $160.15^{+3.63}_{-7.65}$ & $13.520^{+0.053}_{-0.051}$ & $84.63^{+1.44}_{-1.40}$ & \\
249885882 & $162.83^{+2.02}_{-1.77}$ & $13.529^{+0.050}_{-0.049}$ & $83.79^{+1.04}_{-1.18}$ & & 249938143 & $226.30^{+7.09}_{-5.45}$ & $17.601^{+0.055}_{-0.073}$ & $90.50^{+1.91}_{-1.72}$ & \\
249943706 & $147.41^{+7.47}_{-3.03}$ & $12.969^{+0.031}_{-0.043}$ & $81.55^{+0.91}_{-1.19}$ & & 249947084 & $186.84^{+2.90}_{-1.64}$ & $14.825^{+0.048}_{-0.052}$ & $84.54^{+1.94}_{-3.19}$ & \\
250052020 & $142.05^{+5.13}_{-2.31}$ & $11.811^{+0.038}_{-0.051}$ & $98.09^{+1.68}_{-1.69}$ & $\dagger$ & 250124218 & $159.41^{+4.30}_{-2.22}$ & $13.220^{+0.057}_{-0.054}$ & $84.61^{+1.24}_{-0.89}$ & \\
251310765 & $157.70^{+4.95}_{-6.91}$ & $12.747^{+0.076}_{-0.066}$ & $82.15^{+1.48}_{-1.40}$ & & 251358208 & $161.27^{+3.05}_{-7.38}$ & $13.642^{+0.074}_{-0.062}$ & $82.49^{+2.23}_{-1.22}$ & \\
\end{longtable}

\end{document}